\newif\ifeleven
\elevenfalse
\eleventrue      

\newif\ifpdf
\ifx\pdfoutput\undefined
\pdffalse 
\else
\pdfoutput=1 
\pdftrue \fi

\ifeleven
\documentclass[reqno,twoside,11pt]{amsart}
\else
\documentclass[reqno,twoside,12pt]{amsart}
\fi
\usepackage{amsmath}
\usepackage{amsfonts}
\usepackage{amssymb}
\usepackage{graphicx}
\usepackage{graphics}
\usepackage{verbatim}
\IfFileExists{epsf.def}{\input epsf.def}{\usepackage{epsf}}

\ifeleven
\IfFileExists{myowntimes.sty}{\usepackage{myowntimes}}
{\usepackage{times}\usepackage{mathrsfs}}
\else
\usepackage{mathrsfs}
\renewcommand{\mathscr}{\mathcal}
\fi

\DeclareFontFamily{OT1}{eusb}{} \DeclareFontShape{OT1}{eusb}{m}{n}
{<5> <6> <7> <8> <9> <10> <11> <12> <14.4> eusb10}{}
\DeclareMathAlphabet{\eusb}{OT1}{eusb}{m}{n}

\DeclareFontFamily{OT1}{eusm}{} \DeclareFontShape{OT1}{eusm}{m}{n}
{<5> <6> <7> <8> <9> <10> <11> <12> <14.4> [1.0] eusm10}{}
\DeclareMathAlphabet{\eusmCMD}{OT1}{eusm}{m}{n}
\newcommand{\eusm}[1]{\mathchoice%
    {\lower0.7pt\hbox{$\displaystyle\eusmCMD #1$}}
    {\lower0.5pt\hbox{$\textstyle\eusmCMD #1$}}
    {\hbox{$\scriptstyle\eusmCMD #1$}}
    {\hbox{$\scriptscriptstyle\eusmCMD #1$}}}

\DeclareFontFamily{OT1}{eufm}{} \DeclareFontShape{OT1}{eufm}{m}{n}
{<5> <6> <7> <8> <9> <10> <11> <12> <14.4> eufm10}{}
\DeclareMathAlphabet{\mathfrak}{OT1}{eufm}{m}{n}

\DeclareFontFamily{OT1}{fraktura}{}
\DeclareFontShape{OT1}{fraktura}{m}{n} {<5> <6> <7> <8> <9> <10>
<11> <12> <13> <14.4> [1.1] eufm10}{}
\DeclareMathAlphabet{\fraktura}{OT1}{fraktura}{m}{n}

\DeclareFontFamily{OT1}{cmfi}{} \DeclareFontShape{OT1}{cmfi}{m}{n}
{<5> <6> <7> <8> <9> <10> <11> <12> <13> <14.4> [0.9] cmfi10}{}
\DeclareMathAlphabet{\cmfi}{OT1}{cmfi}{b}{n}

\DeclareFontFamily{OT1}{cmss}{} \DeclareFontShape{OT1}{cmss}{m}{n}
{<5> <6> <7> <8> <9> <10> <11> <12> <13> <14.4> [0.95] cmss10}{}
\DeclareMathAlphabet{\cmssCMD}{OT1}{cmss}{m}{n}
\newcommand{\cmss}[1]{\mathchoice%
    {\lower0.5pt\hbox{$\displaystyle\cmssCMD #1$}}
    {\lower0.3pt\hbox{$\textstyle\cmssCMD #1$}}
    {\lower0.5pt\hbox{$\scriptstyle\cmssCMD #1$}}
    {\lower0.5pt\hbox{$\scriptscriptstyle\cmssCMD #1$}}}

\newtheoremstyle{thm}{1.5ex}{1.5ex}{\itshape\rmfamily}{}
{\bfseries\rmfamily}{}{2ex}{}

\newtheoremstyle{def}{1.5ex}{1.5ex}{\rmfamily}{}
{\bfseries\rmfamily}{}{2ex}{}

\newtheoremstyle{rem}{1.3ex}{1.3ex}{\rmfamily}{}
{\itshape}
{} {1.5ex}{}

\newenvironment{proofsect}[1]
{\vskip0.1cm\noindent{\rmfamily\itshape#1.}}{\qed\vspace{0.15cm}}

\theoremstyle{thm}
\newtheorem{theorem}{Theorem}[section]
\newtheorem{lemma}[theorem]{Lemma}
\newtheorem{proposition}[theorem]{Proposition}

\newtheorem*{Main Theorem}{Main Theorem.}
\newtheorem{corollary}[theorem]{Corollary}

\theoremstyle{def}
\newtheorem{definition}{Definition}

\theoremstyle{rem}
\newtheorem{remark}{{\itshape Remark}}[]

\numberwithin{equation}{section}

\renewcommand{\section}{\secdef\sct\sect}
\newcommand{\sct}[2][default]{\refstepcounter{section}
\addcontentsline{toc}{section}
{{\tocsection {}{\thesection}{\!\!\!\!#1\dotfill}}{}}
\vspace{0.7cm}
\centerline{ 
\scshape\arabic{section}.\ #1} \nopagebreak \vspace{0.2cm}}
\newcommand{\sect}[1]{
\vspace{0.4cm} \centerline{\large\scshape\rmfamily #1}
\vspace{0.2cm}}

\renewcommand{\subsection}{\secdef\subsct\sbsect}
\newcommand{\subsct}[2][default]{\refstepcounter{subsection}
\addcontentsline{toc}{subsection}
{{\tocsection{\!\!}{\hspace{1.2em}\thesubsection}{\!\!\!\!#1\dotfill}}{}}
\nopagebreak \vspace{0.45\baselineskip} {\flushleft\bf
\arabic{section}.\arabic{subsection}~\bf #1.~}
\\*[3mm]\noindent
\nopagebreak}
\newcommand{\sbsect}[1]{\vspace{0.1cm}\noindent
\textbf{#1.~}\vspace{0.1cm}}

\renewcommand{\subsubsection}{%
\secdef \subsubsect\sbsbsect}
\newcommand{\subsubsect}[2][default]{%
\refstepcounter{subsubsection}
\addcontentsline{toc}{subsubsection}{{\tocsection{\!\!}
{\hspace{3.05em}\thesubsubsection}{\!\!\!\!#1\dotfill}}{}}
\nopagebreak
\vspace{0.15\baselineskip} \nopagebreak {\flushleft\rmfamily
\itshape\arabic{section}.\arabic{subsection}.\arabic{subsubsection}
\ \rmfamily #1\/.}\ }
\newcommand{\sbsbsect}[1]{\vspace{0.1cm}\noindent
\rmfamily \itshape
\arabic{section}.\arabic{subsection}.\arabic{subsubsection} \
\sffamily #1\/.\ }

\newcommand{\dist}{\operatorname{dist}}
\newcommand{\supp}{\operatorname{supp}}
\newcommand{\diam}{\operatorname{diam}}

\newcommand{\Int}{\operatorname{Int}}
\newcommand{\Ext}{\operatorname{Ext}}
\newcommand{\per}{{\operatorname{per}}}

\newcommand{\textd}{\text{\rm d}}

\newcommand{\Arg}{\operatorname{Arg}}

\newcommand{\Vol}{{\text{\rm\Vol}}}

\newcommand{\CC}{\mathcal C}

\newcommand{\GG}{\mathcal G}

\newcommand{\KK}{\mathcal K}
\newcommand{\MM}{\mathcal M}

\newcommand{\QQ}{\mathcal Q}
\newcommand{\RR}{\mathcal R}
\newcommand{\CalS}{\mathcal S}

\newcommand{\ZZ}{\mathcal Z}

\newcommand{\C}{\mathbb C}
\newcommand{\D}{\mathbb D}

\newcommand{\N}{\mathbb N}

\newcommand{\R}{\mathbb R}

\newcommand{\T}{\mathbb T}

\newcommand{\V}{\mathbb V}

\newcommand{\Y}{\mathbb Y}
\newcommand{\Z}{\mathbb Z}

\newcommand{\twoeqref}[2]{(\ref{#1}--\ref{#2})}

\newcommand{\cc}{{\text{\rm c}}}

\newcommand{\abs}[1]{{\lvert #1\rvert}}
\newcommand{\RE}{{\Re \mathfrak e}}
\newcommand{\IM}{{\Im \mathfrak m}}

\newcommand{\ssK}{\cmss K}

\newcommand{\ssA}{\cmss A}

\newcommand{\ssX}{\cmss X}

\newcommand{\bi}{\boldsymbol i}

\newcommand{\bk}{\boldsymbol k}

\newcommand{\zz}{\fraktura z}
\newcommand{\zzT}{\zz^{\text{\rm T}}}
\newcommand{\frakg}{\fraktura g}

\newcommand{\aT}{a^{\text{\rm T}}}

\newcommand{\lambdac}{\lambda_{\text{\rm c}}}
\newcommand{\eusmN}{{\eusm N\mkern1.1mu}}

\newcommand{\scrG}{\mathscr{G}}
\newcommand{\scrU}{\mathscr{U}}
\newcommand{\scrO}{\mathscr{O}}
\newcommand{\scrL}{\mathscr{L}}
\newcommand{\scrS}{\mathscr{S}}

\begin{document}

\title[Partition function zeros at first-order phase transitions]
{\Large Partition function zeros at first-order\\phase
transitions: Pirogov-Sinai theory}

\renewcommand{\thefootnote}{\fnsymbol{footnote}}

\author[M.~Biskup, C.~Borgs, J.T.~Chayes, and R.~Koteck\'y]
{M.~Biskup,${}^1\,$ C.~Borgs,${}^2\,$
J.T.~Chayes${}^2\,$ and\, R.~Koteck\'y${}^3$}
\thanks{\hglue-4.5mm\fontsize{9.6}{9.6}\selectfont\footnotesize
\copyright\,2003 by the authors. Reproduction, by any means,
of the entire article for non-commercial purposes
is permitted without charge.\vspace{2mm}}
\maketitle

\thispagestyle{empty}

\vspace{-3mm}
\centerline{${}^1$%
\textit{Department of Mathematics, UCLA, Los Angeles, California, USA}}
\centerline{${}^2$%
\textit{Microsoft Research, One Microsoft Way, Redmond, Washington, USA}}
\centerline{${}^3$%
\textit{Center for Theoretical Study, Charles
University, Prague, Czech Republic}}

\vspace{3mm}
\begin{quote}
\footnotesize \textbf{Abstract:} This paper is a continuation of
our previous analysis~\cite{BBCKK2} of partition functions zeros
in models with first-order phase transitions and periodic boundary
conditions. Here it is shown that the assumptions under which the
results of~\cite{BBCKK2} were established are satisfied by a large
class of lattice models. These models are characterized by two
basic properties: The existence of only a finite number of ground
states and the availability of an appropriate contour
representation. This setting includes, for instance, the Ising,
Potts and Blume-Capel models at low temperatures. The combined
results of~\cite{BBCKK2} and the present paper provide complete
control of the zeros of the partition function with periodic
boundary~conditions for all models in the above class.
\end{quote}
\vspace{2mm}

\ifeleven
\vspace{7cm}
\else
\vspace{4cm}
\fi

\begin{quote}
This paper is dedicated to Elliott Lieb on the occasion of his
70$^{\text{th}}$ birthday. Elliott was thesis advisor to one of us
(JTC) and an inspiration to us all.
\end{quote}

\vfill

\begin{tabular}{lp{13cm}}
\multicolumn{2}{l}
{\footnotesize\it AMS Subject Classification: }
\footnotesize
82B05, 
82B26, 
26C10, 
82B20. 
\end{tabular}

\newpage
\section{Introduction}
\vspace{-6mm}
\subsection{Overview}
\label{sec1.1}\noindent
In the recent papers~\cite{BBCKK1,BBCKK2},
we presented a general theory of partition function zeros in
models  with periodic boundary conditions and interaction
depending on one complex parameter. The analysis was based on a
set of assumptions, called Assumptions~A and~B in~\cite{BBCKK2},
which are essentially statements concerning differentiability
properties of certain free energies supplemented by appropriate
non-degeneracy conditions. On the basis of these assumptions we
characterized the topology of the resulting phase diagram and
showed that the partition function zeros are in one-to-one
correspondence with the solutions to specific (and simple)
equations. In addition, the maximal degeneracy of the zeros was
proved to be bounded by the number of thermodynamically stable
phases, and the distance between the zeros and the corresponding
solutions was shown to be generically exponentially small in the
linear size of the system.

The reliance on Assumptions~A and~B in~\cite{BBCKK2} permitted us
to split the analysis of partition function zeros into two parts,
which are distinct in both mathematical and physical content: one
concerning the zeros of a complex (in fact, analytic)
function---namely the partition function with periodic boundary
conditions---subject to specific requirements, and the other
concerning the control of the partition function in a
statistical mechanical model depending on one complex parameter.
The former part of the analysis was carried out in~\cite{BBCKK2};
the latter is the subject of
this paper. Explicitly, the
principal goal of this paper can be summarized as follows: We will
define a large class of lattice spin models (which includes
several well-known systems, e.g., the Ising and Blume-Capel
models) and show that Assumptions~A and~B are satisfied for every
model in this class. On the basis of~\cite{BBCKK2}, for any model
in this class we then have complete control of the zeros of the
partition function with periodic boundary conditions.

The models we consider are characterized by two properties: the
existence of only a finite number of \emph{ground states} and the
availability of a \emph{contour} representation. In our setting,
the term ground state will simply mean a constant---or, after some
reinterpretations, a periodic---infinite volume spin
configuration. Roughly speaking, the contour representation will
be such that the contours correspond to finite, connected subsets
of the lattice where the spin configuration differs from any of
the possible ground states. A precise definition of these notions
is a bit technical; details will be provided in
Section~\ref{sec3}. Besides these properties, there will also be a
few quantitative requirements on the ground state energies and the
scaling of the excess contour energy with the size of the
contour---the Peierls condition---see Sections~\ref{sec2.1}
and~\ref{sec3.2}.

These two characteristic properties enable us to apply
Pirogov-Sinai theory---a general method for determining
low-temperature properties of a statistical mechanical model by
perturbing about zero-temperature. The first formulation of this
perturbation technique~\cite{PSa,PSb} applied to a class of models
with real, positive weights. The original ``Banach space''
approach of~\cite{PSa,PSb} was later replaced by inductive
methods~\cite{KP1}, which resulted in a complete classification of
translation-invariant Gibbs states~\cite{Z1}. The inductive
techniques also permitted a generalization of the characterization
of phase stability/coexis\-tence to models with complex
weights~\cite{BI}. However, most relevant for our purposes are the
results of~\cite{BK}, dealing with finite-size scaling in the
vicinity of first-order phase transitions. There Pirogov-Sinai
theory was used to derive detailed asymptotics of finite volume
partition functions. The present paper provides, among other
things, a variant of \cite{BK} that ensures appropriate
differentiability of the so-called metastable free energies as
required for the analysis of partition function zeros.

\smallskip
The remainder of this paper is organized as follows.
Section ~\ref{sec1.2} outlines the class of models of interest.
Section~\ref{sec1.3} defines the ground state and excitation
energies and introduces the torus partition function---the main
object of interest in this paper. Section~\ref{sec2.1} lists the
assumptions on the models and Section~\ref{sec2.2} gives the
statements of the main results of this paper. These immediately
imply Assumptions~A and~B of \cite{BBCKK2} for all models in the
class considered. Sections~\ref{sec3} and~\ref{sec4} introduce the
necessary tools from Pirogov-Sinai theory. These are applied in
Section~\ref{sec5} to prove the main results of
the paper.

\subsection{Models of interest}
\label{sec1.2}\noindent
Here we define the class of models
to be considered in this paper. Most of what is to
follow in this and the forthcoming sections is inspired by classic
texts on spin models, Gibbs states and Pirogov-Sinai theory,
e.g.,~\cite{Georgii,Ruelle,Sinai,Z1}.

\smallskip
We will consider finite-state spin models on the $d$-dimensional
hypercubic lattice~$\Z^d$ for $d\ge2$. At each site~$x\in\Z^d$ the
\emph{spin}, denoted by $\sigma_x$, will take values in a finite
set~$\CalS$. A \emph{spin configuration}
$\sigma=(\sigma_x)_{x\in\Z^d}$ is an assignment of a spin to each
site of the lattice. The interaction Hamiltonian will be described
using a collection of potentials~$(\Phi_\Lambda)$, where~$\Lambda$
runs over all finite subsets of~$\Z^d$. The~$\Phi_\Lambda$ are
functions on configurations from~$\CalS^{\Z^d}$ with the following
properties:
\begin{enumerate}
\item[(1)] The value $\Phi_\Lambda(\sigma)$ depends only
on~$\sigma_x$ with $x\in\Lambda$.
\item[(2)]
The potential is
translation invariant, i.e., if~$\sigma'$ is a translate
of~$\sigma$ and~$\Lambda'$ is the corresponding translate
of~$\Lambda$, then
$\Phi_{\Lambda'}(\sigma)=\Phi_\Lambda(\sigma')$.
\item[(3)]
There exists an $R\ge1$ such that $\Phi_\Lambda\equiv0$ for
all~$\Lambda$ with diameter exceeding~$R+1$.
\end{enumerate}
Here the \emph{diameter} of a cubic box with~$L\times\dots\times
L$ sites is defined to be~$L$ while for a general~$A\subset\Z^d$
it is the diameter of the smallest cubic box containing~$A$. The
constant~$R$ is called the \emph{range of the interaction}.

\begin{remark}
\label{rem1}
Condition~(2) has been included mostly for convenience of
exposition. In fact, all of the results of this paper hold under
the assumption that~$\Phi_\Lambda$ are periodic in the sense that
$\Phi_{\Lambda'}(\sigma)=\Phi_\Lambda(\sigma')$ holds
for~$\Lambda$ and~$\sigma$ related to~$\Lambda'$ and~$\sigma'$ by
a translation from~$(a\Z)^d$ for some fixed integer~$a$. This is
seen by noting that the periodic cases can always be converted to
translation-invariant ones by considering block-spin variables and
integrated potentials.
\end{remark}

As usual, the energy of a spin configuration is specified by the
Hamiltonian. Formally, the Hamiltonian is represented by a
collection of functions $(\beta H_\Lambda)$ indexed by finite
subsets of~$\Z^d$, where~$\beta H_\Lambda$ is defined by the
formula
\begin{equation}
\label{1.1}
\beta H_\Lambda(\sigma)
=\sum_{\Lambda'\colon\Lambda'\cap\Lambda\ne\emptyset}
\Phi_{\Lambda'}(\sigma).
\end{equation}
(The superfluous~$\beta$, playing the role of the inverse
temperature, appears only to maintain formal correspondence with
the fundamental formulas of statistical mechanics.) In light of
our restriction to finite-range interactions, the sum is always
finite.

\smallskip
We proceed by listing a few well known examples of models in the
above class. With the exception of the second example, the range
of each interaction is equal to $1$:

\smallskip\noindent
\textit{Ising model.\ } Here $\CalS=\{-1,+1\}$ and
$\Phi_\Lambda(\sigma)\not\equiv0$ only for~$\Lambda$ containing a
single site or a nearest-neighbor pair. In this case we have
\begin{equation}
\label{H-Ising}
\Phi_\Lambda(\sigma)=\begin{cases}
-h\sigma_x,\qquad&\text{if }\Lambda=\{x\},
\\
-J\sigma_x\sigma_y,\qquad&\text{if }\Lambda=\{x,y\}\text{ with }|x-y|=1.
\end{cases}
\end{equation}
Here~$J$ is the coupling constant,~$h$ is an external field
and~$|x-y|$ denotes the Euclidean distance between~$x$ and~$y$.

\smallskip\noindent
\textit{Perturbed Ising model.\ } Again $\CalS=\{-1,+1\}$, but now
we allow for arbitrary finite range perturbations. Explicitly,
\begin{equation}
\label{H-perturbed-Ising}
\Phi_\Lambda(\sigma)=\begin{cases}
-h\sigma_x,\qquad&\text{if }\Lambda=\{x\},
\\
-J_\Lambda\prod_{x\in\Lambda}\sigma_x&\text{if }|\Lambda|\geq 2
\text{ and }\diam\Lambda\leq R+1.
\end{cases}
\end{equation}
The coupling constants~$J_\Lambda$ are assumed to be translation
invariant (i.e., $J_\Lambda=J_{\Lambda'}$ if~$\Lambda$
and~$\Lambda'$ are translates of each other).  The constant~$h$ is
again the external field.

\smallskip\noindent
\textit{Blume-Capel model.\ } In this case $\CalS=\{-1,0,+1\}$ and
$\Phi_\Lambda(\sigma)\equiv0$ unless~$\Lambda$ is just a single
site or a nearest-neighbor pair. Explicitly, we have
\begin{equation}
\label{H-BC}
\Phi_\Lambda(\sigma)=
\begin{cases}
-\lambda\sigma_x^2-h\sigma_x,
\qquad&\text{if }\Lambda=\{x\},
\\
J(\sigma_x-\sigma_y)^2,
\qquad&\text{if }\Lambda=\{x,y\}\text{ with }|x-y|=1.
\end{cases}
\end{equation}
Here~$J$ is the coupling constant,~$\lambda$ is a parameter
favoring~$\pm1$ against~$0$-spins and~$h$ is an external field
splitting the symmetry between~$+1$ and~$-1$.

\smallskip\noindent
\textit{Potts model in an external field.\ } The state space
has~$q$ elements, $\CalS=\{1,\dots,q\}$ and~$\Phi_\Lambda$ is
again nontrivial only if~$\Lambda$ is a one-element set or a pair
of nearest-neighbor sites. Explicitly,
\begin{equation}
\label{H-Potts}
\Phi_\Lambda(\sigma)=\begin{cases}
-h\delta_{\sigma_x,1},\qquad&\text{if }\Lambda=\{x\},
\\
-J\delta_{\sigma_x,\sigma_y},\qquad&\text{if }
\Lambda=\{x,y\}\text{ with }|x-y|=1.
\end{cases}
\end{equation}
Here~$\delta_{\sigma,\sigma'}$ equals one if $\sigma=\sigma'$ and
zero otherwise,~$J$ is the coupling constant and~$h$ is an
external field favoring spin value~$1$. Actually, the results of
this paper will hold only for the low-temperature regime
(which in our parametrization corresponds to~$J\gg\log q$);
a more general argument covering \emph{all} temperatures (but under the
condition that~$q$ is sufficiently large) will be presented
elsewhere~\cite{BBCK1,BBCK2}.

\smallskip
Any of the constants appearing in the above Hamiltonian can in
principle be complex. However, not all complex values of, e.g.,
the coupling constant will be permitted by our additional
restrictions. See Section~\ref{sec2.3} for more discussion.

\subsection{Ground states, excitations and torus partition function}
\label{sec1.3}\noindent
The key idea underlying our formulation is
that \emph{constant} configurations represent the potential ground
states of the system. (A precise statement of this fact appears in
Assumption~C2 below.) This motivates us to define the
dimensionless \emph{ground state energy density}~$e_m$ associated
with spin $m\in\CalS$ by the formula
\begin{equation}
\label{1.5}
e_m=\sum_{\Lambda\colon\Lambda\ni0}
\frac1{|\Lambda|}\Phi_\Lambda(\sigma^m),
\end{equation}
where~$|\Lambda|$ denotes the cardinality of the set~$\Lambda$ and
where~$\sigma^m$ is the spin configuration that is equal to~$m$ at
every site. By our restriction to finite-range interactions, the
sum is effectively finite.

The constant configurations represent the states with minimal
energy; all other configurations are to be regarded as
excitations. Given a spin configuration~$\sigma$,
let~$B_R(\sigma)$ denote the union of all cubic boxes
$\Lambda\subset\Z^d$ of diameter $2R+1$ such
that ~$\sigma$ is not constant in~$\Lambda$.
We think of~$B_R(\sigma)$ as the set on which~$\sigma$ is ``bad''
in the sense that it is not a ground state at scale~$R$. The
set~$B_R(\sigma)$ will be referred to as the \emph{$R$-boundary}
of~$\sigma$. Then the \emph{excitation energy}~$E(\sigma)$ of
configuration~$\sigma$ is defined by
\begin{equation}
\label{1.6}
E(\sigma)=
\sum_{x\in B_R(\sigma)}\sum_{\Lambda\colon x\in\Lambda}
\frac{1}{|\Lambda|}\Phi_\Lambda(\sigma).
\end{equation}
To ensure that the sum is finite (and therefore meaningful) we
will only consider the configurations~$\sigma$ for
which~$B_R(\sigma)$ is a finite set.

The main quantity of interest in this paper is the partition
function with periodic boundary conditions which we now define.
Let $L\geq 2R+1$, and let~$\T_L$ denote the torus
of $L\times
L\times\dots\times L$ sites in~$\Z^d$, which can be thought of as
the factor of~$\Z^d$ with respect to the action of the
subgroup~$(L\Z)^d$.
Let us consider the Hamiltonian $\beta
H_L\colon\CalS^{\T_L}\to\C$ defined by
\begin{equation}
\label{1.7}
\beta H_L(\sigma)=\!\sum_{\Lambda\colon\Lambda\subset\T_L}
\Phi_\Lambda(\sigma),
\qquad \sigma\in\CalS^{\T_L},
\end{equation}
where~$\Phi_\Lambda$ are retractions of the corresponding
potentials
from~$\Z^d$ to~$\T_L$. (Here we use the translation invariance
of~$\Phi_\Lambda$.) Then the \emph{partition function with
periodic boundary conditions} in~$\T_L$ is defined~by
\begin{equation}
Z_L^\per=\sum_{\sigma\in\CalS^{\T_L}}e^{-\beta H_L(\sigma)}.
\end{equation}
In general,~$Z_L^\per$ is a complex quantity which depends on all
parameters of the Hamiltonian. We note that various other
partition functions will play an important role throughout this
paper. However, none of these will be needed for the statement of
our main results in Section~\ref{sec2}, so we postpone the
additional definitions and discussion to Section~\ref{sec4}.

\smallskip
We conclude this section with a remark concerning the
interchangeability of the various spin states. There are natural
examples (e.g., the Potts model) where several spin values are
virtually indistinguishable from each other. To express this
property mathematically, we will consider the situation where there exists a
subgroup~$\mathfrak G$ of the permutations of~$\CalS$ such that if
$\pi\in\mathfrak G$ then $e_{\pi(m)}=e_m$ and
$E(\pi(\sigma))=E(\sigma)$ for each $m\in\CalS$ and each
configuration~$\sigma$ with
finite~$B_R(\sigma)$, where~$\pi(\sigma)$ is the spin
configuration taking value~$\pi(\sigma_x)$ at each~$x$. (Note that
$B_R(\pi(\sigma))=B_R(\sigma)$ for any such permutation~$\pi$.)
Then we call two spin states~$m$ and~$n$ \emph{interchangeable}
if~$m$ and~$n$ belong to the same orbit of the group~$\mathfrak G$
on~$\CalS$.

While this extra symmetry has absolutely no effect on the contour
analysis of the torus partition sum, it turns out that
interchangeable spin states cannot be treated separately in our
analysis of partition function zeros. (The precise reason is that
interchangeable spin states would violate our non-degeneracy
conditions; see Assumption~C3-C4 and Theorem~\ref{thmA}3-4 below.)
To avoid this difficulty, we will use the factor set
$\RR=\CalS/\mathfrak G$ instead of the original index set~$\CalS$
when stating our assumptions and results.  In accordance with the
notation of~\cite{BBCKK2}, we will also use~$r$ to denote the
cardinality of the set~$\RR$, i.e., $\RR=\{1, 2,\dots,r\}$,
and~$q_m$ to denote the cardinality of the orbit corresponding to
$m\in\RR$.

\section{Assumptions and results}
\label{sec2}\noindent
In this section we list our
precise assumptions on the models of interest and state the
main results of this paper.

\subsection{Assumptions}
\label{sec2.1}\noindent
We will consider the setup outlined in
Sections~\ref{sec1.2}--\ref{sec1.3} with the additional assumption
that the parameters of the Hamiltonian depend on one complex
parameter~$z$ which varies in some open subset~$\tilde\scrO$ of
the complex plane.
Typically, we will take~$z=e^h$ or $z=e^{2h}$ where~$h$ is an
external field; see the examples at the end of
Section~\ref{sec1.2}.
Throughout this paper we will
assume that
the spin space~$\CalS$, the factor set~$\RR$, the integers~$q_m$
and the range of the interaction are independent of the parameter~$z$.
We will also assume that the spatial dimension~$d$ is no less than two.

\smallskip
The assumptions below will be expressed in
terms of complex derivatives with respect to $z$. For
brevity of exposition, let us use the standard notation
\begin{equation}
\partial_z=\tfrac12\bigl(\tfrac\partial{\partial x}
-i\tfrac\partial{\partial y}\bigr)
\quad\text{and}\quad
\partial_{\bar z}=\tfrac12\bigl(\tfrac\partial{\partial x}
+i\tfrac\partial{\partial y}\bigr)
\end{equation}
for the derivatives with respect to~$z$ and~$\bar z$, respectively.
Here $x=\RE z$ and $y=\IM z$.
Our assumptions will be formulated for the exponential weights
\begin{equation}
\varphi_\Lambda(\sigma,z)=e^{-\Phi_\Lambda(\sigma,z)},
\quad
\rho_z(\sigma)=e^{-E(\sigma,z)}
\quad\text{and}\quad\theta_m(z)=e^{-e_m(z)},
\end{equation}
where we have now made the dependence on~$z$
notationally explicit.
In terms of the
$\theta_m$'s and the quantity
\begin{equation}
\label{thetadef}
\theta(z)=\max_{m\in\RR}|\theta_m(z)|
\end{equation}
we define the set~$\scrL_\alpha(m)$ by
\begin{equation}
\scrL_\alpha(m)
=\bigl\{z\in\tilde\scrO\colon|\theta_m(z)|\ge \theta(z)e^{\alpha}\bigr\}.
\end{equation}
Informally,~$\scrL_\alpha(m)$ is the set of~$z$ for which~$m$ is
``almost'' a ground state of the Hamiltonian.

Since we want to refer back to Assumptions~A and~B of
\cite{BBCKK2}, we will call our new hypothesis Assumption~C.

\medskip\noindent
\textbf{Assumption~C.\ }
There exist a
domain $\tilde\scrO\subset\C$ and constants $\alpha,M,\tau\in(0,\infty)$
such that the following conditions are satisfied.
\settowidth{\leftmargini}{(11)}
\begin{enumerate}
\item[(0)]
For each $\sigma\in\CalS^{\Z^d}$ and each
finite~$\Lambda\subset\Z^d$, the function
$z\mapsto\varphi_\Lambda(\sigma,z)$
is holomorphic in~$\tilde\scrO$.
\item[(1)]
For all $m\in\CalS$, all
$z\in\tilde\scrO$ and all $\ell=0,1,2$, the
ground state weights obey the bounds
\begin{equation}
\bigl| {\partial_z^\ell}\theta_m(z)
\bigr|
\leq M^\ell\theta(z)
\end{equation}
In addition, the quantity $\theta(z)$ is uniformly bounded away from zero
in $\tilde\scrO$.
\item[(2)]
For every configuration~$\sigma$ with finite
$R$-boundary~$B_R(\sigma)$, the Peierls condition
\begin{equation}
\label{Peierls}
\bigl|\partial_z^\ell\rho_z(\sigma)\bigr|
\le\bigl(M|B_R(\sigma)|\bigr)^\ell
\bigl(e^{-\tau}\theta(z)\bigr)^{|B_R(\sigma)|}
\end{equation}
holds for all $z\in\tilde\scrO$ and $\ell=0,1,2$.
\item[(3)]
For all distinct
$m,n\in\RR$ and all $z\in\scrL_\alpha(m)\cap\scrL_\alpha(n)$, we
have
\begin{equation}
\Bigl|\frac{\partial_z\theta_m(z)}{\theta_m(z)}
-\frac{\partial_z\theta_n(z)}{\theta_n(z)}\Bigr|\ge\alpha.
\end{equation}
\item[(4)]
If~$\QQ\subset\RR$ is such that~$|\QQ|\ge3$, then for any
$z\in\bigcap_{m\in\QQ}\scrL_\alpha(m)$ we
assume that the complex
quantities $v_m(z)=\theta_m(z)^{-1}\,\partial_z\theta_m(z)$,
$m\in\QQ$, regarded as vectors
in~$\R^2$, are vertices of a strictly convex polygon. Explicitly,
we demand that the bound
\begin{equation}
\inf\biggl\{\,\Bigl|\,
v_m(z)
-\!\!\!\!\sum_{n\in\QQ\smallsetminus\{m\}}
\!\!\!\!\omega_n
v_n(z)
\,\Bigr|
\colon\omega_n\ge0,\!\!\sum_{n\in\QQ\smallsetminus\{m\}}
\!\!\!\!\omega_n=1\biggr\}\ge\alpha
\end{equation}
holds for every $m\in\QQ$ and every $z\in\bigcap_{n\in\QQ}\scrL_\alpha(n)$.
\end{enumerate}

\medskip
Assumptions~C0-2 are very natural; indeed, they are typically a
consequence of the fact that the
potentials~$\varphi_\Lambda(\sigma,z)$---and hence also~$\theta_m(z)$
and~$\rho_z(\sigma)$---arise by analytic continuation from
the positive real axis.
Assumptions~C3-4 replace the ``standard''
multidimensional non-degeneracy conditions which are typically
introduced to control the topological structure of the phase
diagram,
see e.g.~\cite{PSa,PSb,Sinai}. (However, unlike for the
``standard'' non-degeneracy conditions, here this control requires
a good deal of extra work, see~\cite{BBCKK2}.) Assumption~C4 is
only important in the vicinity of multiple coexistence points
(see Section~\ref{sec3.2});
otherwise, it can be omitted.

\begin{remark}
\label{rem2} For many models, including the first three of our
examples, the partition function has both zeros and poles, and
sometimes even involves non-integer powers of $z$.  In this
situation it is convenient to multiply the partition function by a
suitable power of~$z$ to obtain a function that is analytic in a
larger domain. Typically, this different normalization also leads
to a larger domain~$\tilde\scrO$ for which Assumption C holds.
Taking, e.g., the Ising model with $z=e^{2h}$, one easily verifies
that for low enough temperatures, Assumption C holds
everywhere in the complex plane---provided we replace
the term $-h\sigma_x$ by $-h(\sigma_x+1)$.  By contrast, in the
original representation (where $\varphi_{\{x\}}(\sigma,z)=(\sqrt
z)^{\sigma_x}$), one needs to take out a neighborhood of the
negative real axis (or any other ray from zero to infinity) to
achieve the analyticity required by Assumption C0.
\end{remark}

\begin{remark}
\label{rem3} If we replace the term $-h\sigma_x$ in
\twoeqref{H-Ising}{H-BC} by $-h(\sigma_x+1)$, Assumption~C (with
$z=e^{2h}$ for the Ising models, and $z=e^h$ for the Blume Capel
and Potts model) holds for all four examples listed in
Section~\ref{sec1.2}, provided that the nearest-neighbor couplings
are ferromagnetic and the temperature is low enough. (For the
perturbed Ising model, one also needs that the nearest-neighbor
coupling is sufficiently dominant.)
\end{remark}

\subsection{Main results}
\label{sec2.2}\noindent
Now we are in a position to state our main
results, which show that Assumptions~A and~B from~\cite{BBCKK2}
are satisfied and hence our conclusions concerning the partition
function zeros hold.
The structure of these theorems parallels the
structure of  Assumptions~A and~B. We caution the reader that the
precise
statement of these results is quite technical. For a discussion
of the implications of these theorems, see Section~\ref{sec2.3}.
The first theorem establishes the existence of metastable free
energies and their relation to the quantities $\theta_m$.

\renewcommand{\thetheorem}{A}
\begin{theorem}
\label{thmA}
%
%
Let $M\in(0,\infty)$ and $\alpha\in(0,\infty)$.
Then there is a
%
%
constant~$\tau_0$
depending
on~$M$,~$\alpha$,
the number of spin states~$|\CalS|$ and the dimension~$d$ such
that if Assumption~C holds for
the constants~$M$,~$\alpha$,
some open domain $\tilde\scrO\subset\C$
and
some~$\tau\geq\tau_0$,
then there are functions~$\zeta_m\colon\tilde\scrO\to\C$, $m\in\RR$,
for which the following holds:
\settowidth{\leftmargini}{(11)}
\begin{enumerate}
\item[(1)]
There are functions $s_m\colon\tilde\scrO\to\C$, $m\in\RR$,
such that $\zeta_m(z)$ can be expressed as
\begin{equation}
\label{zeta-theta}
\zeta_m(z)=\theta_m(z) e^{s_m(z)}
\quad
\text{and}
\quad
|s_m(z)|\leq e^{-\tau/2}.
\end{equation}
In particular, the quantity~$\zeta(z)=\max_{m\in\RR}\abs{\zeta_m(z)}$
is uniformly positive in~$\tilde\scrO$.
\item[(2)]
Each function~$\zeta_m$, viewed as a function of two
real variables~$x=\RE z$ and~$y=\IM z$, is twice
continuously differentiable on~$\tilde\scrO$ and  satisfies the
Cauchy-Riemann  equations~$\partial_{\bar z}\zeta_m(z)=0$ for
all~$z\in\scrS_m$, where
\begin{equation}
\scrS_m=\bigl\{z\in\tilde\scrO\colon |\zeta_m(z)|=\zeta(z)\bigr\}.
\end{equation}
In particular,~$\zeta_m$ is analytic in the interior of~$\scrS_m$.
\item[(3)]
For any pair of distinct indices~$m,n\in\RR$ and
any~$z\in\scrS_m\cap\scrS_n$ we have
\begin{equation}
\label{nondeg}
\biggr|\frac{\partial_z\zeta_m(z)}{\zeta_m(z)}-
\frac{\partial_z\zeta_n(z)}{\zeta_n(z)} \biggl|\ge\alpha-2e^{-\tau/2}.
\end{equation}
\item[(4)]
If~$\QQ\subset \RR$ is such that~$|\QQ|\ge 3$, then for
any~$z\in\bigcap_{m\in\QQ}\scrS_m$,
\begin{equation}
\label{tripledeg}
v_m(z)=\frac{\partial_z\zeta_m(z)}{\zeta_m(z)},\quad m\in\QQ,
\end{equation}
are the vertices of a strictly convex polygon in~$\C\simeq\R^2$.
\end{enumerate}
\end{theorem}

Theorem~\ref{thmA} ensures the validity of Assumption~A
in~\cite{BBCKK2} for any model satisfying Assumption~C with~$\tau$
sufficiently large. Assumption~A, in turn, allows us to establish
several properties of the topology of the phase diagram, see
Section~\ref{sec2.3} below for more details.

\smallskip
Following~\cite{BBCKK2},
we will refer to the indices in~$\RR$ as
\emph{phases},
and call a phase $m\in\RR$ \emph{stable at~$z$} if
$|\zeta_m(z)|=\zeta(z)$.
We will say that a point $z\in\tilde\scrO$ is a
point of \emph{phase coexistence} if there are at least two phases
$m\in\RR$ which are stable at $z$.
In~\cite{BBCKK2} we introduced these definitions without further
motivation, anticipating, however, the present work which provides
the technical justification of
these concepts.
Indeed, using the expansion techniques developed in
Sections~\ref{sec3} and~\ref{sec4}, one can show that,
for each $m\in\CalS$ that corresponds to a stable phase in~$\RR$,
the finite volume states with $m$-boundary conditions tend to
a unique infinite-volume limit~$\langle\cdot\rangle_m$ in the
sense of weak convergence on linear
functionals on local observables. (Here a local observable refers
to a function depending only on a finite number of spins).
The limit state is invariant under translations of~$\Z^d$,
exhibits exponential clustering,
and is a small perturbation of the ground state~$\sigma^m$
in the sense that
$\langle\delta_{\sigma_x,k}\rangle_m=\delta_{m,k}+O(e^{-\tau/2})$
for all~$x\in\Z^d$.

\begin{remark}
\label{rem4}
Note that two states $\langle\cdot\rangle_m$
and $\langle\cdot\rangle_{m'}$ are considered as
two different versions of the same phase
if $m$ and $m'$ are indistinguishable, in accordance
with our convention that $\RR$, and not $\CalS$, labels
phases.  Accordingly, the term phase coexistence refers
to the coexistence of \emph{distinguishable} phases,
and not to the coexistence of two states labelled by
different indices in the same orbit~$\RR$.  This interpretation of
a ``thermodynamic phase'' agrees with that used in physics,
but disagrees with that sometimes used in the mathematical
physics literature.
\end{remark}

\smallskip
While Theorem~\ref{thmA} is valid in the whole domain
$\tilde\scrO$, our next theorem will require that we restrict
ourselves to a subset $\scrO\subset\tilde\scrO$ with the property
that there exists some $\epsilon>0$ such that
for each point $z\in\scrO$, the disc $\D_\epsilon(z)$ of
radius $\epsilon$ centered at $z$ is contained in $\tilde\scrO$.
(Note that this condition requires $\scrO$ to be a strict subset
of $\tilde\scrO$, unless $\tilde\scrO$ consists of the whole
complex plane).
In order to state the next theorem,
we will need to recall some
notation from~\cite{BBCKK2}. Given any~$m\in\RR$ and $\delta>0$,
let~$\scrS_\delta(m)$ denote the region where the phase~$m$ is
``almost stable,''
\begin{equation}
\label{2.7a}
\scrS_\delta(m)= \bigl\{z\in\scrO\colon \abs{\zeta_m(z)}>
e^{-\delta} \zeta(z)\bigr\}.
\end{equation}
For any~$\QQ\subset\RR$, we also introduce the region where all
phases from~$\QQ$ are ``almost stable'' while the remaining ones
are not,
\begin{equation}
\scrU_\delta(\QQ)=\bigcap_{m\in\QQ} \scrS_\delta(m)\setminus
\bigcup_{n\in\QQ^\cc}\overline{\scrS_{\delta/2}(n)},
\label{2.7}
\end{equation}
with the bar denoting the set closure.

\renewcommand{\thetheorem}{B}
\begin{theorem}
\label{thmB}
Let $M,\alpha,\epsilon\in(0,\infty)$, and let
$\tau\geq\tau_0$, where $\tau_0$ is the constant from
Theorem~\ref{thmA}, and let $\kappa=\tau/4$.
Let $\tilde \scrO\subset\C$ and $\scrO\subset\tilde\scrO$ be open
domains such that that Assumption~C holds in  $\tilde \scrO$ and
$\D_\epsilon(z)\subset\tilde\scrO$ for all $z\in\scrO$.
Then there are constants $C_0$ (depending only on $M$), $M_0$
(depending on $M$ and $\epsilon$), and $L_0$ (depending on $d$,
$M$, $\tau$ and $\epsilon$) such that for each $m\in\RR$ and each $L\geq L_0$
there is a function~$\zeta_m^{(L)}\colon\scrS_{\kappa/L}(m)\to\C$
such that the following holds for all $L\geq L_0$:
\settowidth{\leftmargini}{(11)}
\begin{enumerate}
\item[(1)]
The function~$Z_L^\per$ is analytic
%
%
%
in~$\tilde\scrO$.
\item[(2)]
Each~$\zeta_m^{(L)}$ is non-vanishing and analytic
in~$\scrS_{\kappa/L}(m)$. Furthermore,
\begin{equation}
\label{zetamL}
\biggl|\log\frac{\zeta_m^{(L)}(z)}{\zeta_m(z)}\biggr| \le
e^{-\tau L/8}
\end{equation}
and
\begin{equation}
\label{der-zetamL}
\biggl|\partial_z\log\frac{\zeta_m^{(L)}(z)}{\zeta_m(z)}\biggr| +
\biggl|\partial_{\bar z}\log\frac{\zeta_m^{(L)}(z)}{\zeta_m(z)}\biggr|
\le e^{-\tau L/8}
\end{equation}
hold for all~$m\in\RR$ and all~$z\in\scrS_{\kappa/L}(m)$.
\item[(3)]
For each~$m\in\RR$,
all~$\ell\geq 1$,
and all $z\in\scrS_{\kappa/L}(m)$, we have
\begin{equation}
\label{uplogderL}
\biggl|\frac{\partial^\ell_z\zeta_m^{(L)}(z)}{\zeta_m^{(L)}(z)}
\biggr|\le
(\ell!)^2 M_0^\ell.
\end{equation}
Moreover, for all distinct~$m,n\in\RR$ and
all~$z\in\scrS_{\kappa/L}(m)\cap\scrS_{\kappa/L}(n)$,
\begin{equation}
\label{nondegL}
\biggr|\frac{\partial_z\zeta_m^{(L)}(z)}{\zeta_m^{(L)}(z)}-
\frac{\partial_z\zeta_n^{(L)}(z)}{\zeta_n^{(L)}(z)}
\biggl|\ge\alpha-2e^{-\tau/2}.
\end{equation}
\item[(4)]
For any~$\QQ\subset\RR$, the difference
\begin{equation}
\label{ZLper} \varXi_{\QQ,L}(z)=Z_L^\per(z)-\sum_{m\in\QQ} q_m
\bigl[\zeta_m^{(L)}(z)\bigr]^{L^d}
\end{equation}
satisfies the bound
\begin{equation}
\label{error}
\bigl|\partial_z^\ell \varXi_{\QQ,L}(z)\bigr|\le
\ell!(C_0L^d)^{\ell+1}
\zeta(z)^{L^d}\biggl(\,\sum_{m\in\RR}q_m\biggr)
e^{-\tau L/16}
\end{equation}
for all $\ell\geq 0$ and all
$z\in\scrU_{\kappa/L}(\QQ)$.
\end{enumerate}
\end{theorem}
\renewcommand{\thetheorem}{\arabic{section}.\arabic{theorem}}

Theorem~\ref{thmB} proves the validity of Assumption~B from
\cite{BBCKK2}. Together with Theorem~\ref{thmA}, this in turn
allows us to give a detailed description of the positions of the
partition function zeros for all models in our class,
see Section~\ref{sec2.3}.

The principal result of Theorem~\ref{thmB}
is stated in part~(4): The torus partition
function can be approximated by a finite sum of
terms---one for
each ``almost stable'' phase $m\in\RR$---which have well
controlled analyticity properties.
As a consequence, the zeros of the partition function arise
as a result of destructive interference between almost
stable phases, and all zeros are near to the set of
coexistence points ~$\scrG=\bigcup_{m\ne n}\scrS_m\cap\scrS_n$;
see Section~\ref{sec2.3} for further details.
Representations of the form \eqref{ZLper} were crucial for the
analysis of finite-size scaling near first-order phase
transitions~\cite{BK}. The original derivation goes back
to~\cite{BI}. In our case the situation is complicated by the
requirement of analyticity; hence the restriction
to~$z\in\scrU_{\kappa/L}(\QQ)$ in~(4).

\subsection{Discussion}
\label{sec2.3}\noindent
As mentioned previously,
Theorems~\ref{thmA} and~\ref{thmB} imply the validity of
Assumptions~A and~B of~\cite{BBCKK2}, which in turn imply the
principal conclusions of~\cite{BBCKK2} for any model
of the kind introduced in Section~\ref{sec1.2}
that satisfies
Assumption~C with~$\tau$ sufficiently large.
Instead of giving the
full statements of the results of~\cite{BBCKK2}, we will only
describe these theorems on a qualitative level. Readers interested
in more details are referred to Section~2 of~\cite{BBCKK2}.

\smallskip
Our first result concerns the set of coexistence
points,~$\scrG=\bigcup_{m\ne n}\scrS_m\cap\scrS_n$, giving rise to
the complex phase diagram. Here~Theorem~2.1 of~\cite{BBCKK2}
asserts that~$\scrG$ is the union of a set of simple, smooth (open
and closed) curves such that exactly two phases coexist at any
interior point of the curve, while at least three phases coexist
at the endpoints---these are the
\emph{multiple points}. Moreover, in
each compact set, any two such curves cannot get too
close without
intersecting and there are only a finite number of multiple
points. These properties are of course direct consequences of the
non-degeneracy conditions expressed in Theorem~\ref{thmA}3-4.

Having discussed the phase diagram, we can now turn our attention
to the zeros of~$Z_L^\per$. The combined results of
Theorems~2.2-2.4 of~\cite{BBCKK2} yield the following: First, all
zeros lie within~$O(L^{-d})$ of the set~$\scrG$. Second, along the
two-phase coexistence lines with stable phases~$m,n\in\RR$, the
zeros are within~$O(e^{-cL})$,
for some~$c>0$,
of the solutions to the
equations
\begin{align}
\label{Req}
{}&q_m^{1/L^d}|\zeta_m(z)|=q_n^{1/L^d}|\zeta_n(z)|,
\\
\label{Imq} {}&L^d\Arg \bigl(\zeta_m(z)/\zeta_n(z)\bigr)
=\pi\operatorname{ mod }2\pi.
\end{align}
Consecutive solutions to these equations are separated by
distances of order~$L^{-d}$,
i.e., there are of the other~$L^d$ zeros per unit length of the
coexistence line. Scaling by~$L^d$, this allows us to define
a \emph{density of zeros} along each two-phase coexistence line,
which in the limit~$L\to\infty$ turns out to be a smooth
function varying only over distances of order one.

Near the multiple points the zeros are
still in one-to-one correspondence with the solutions of a certain
equation. However, our control of the errors here is less precise
than in the two-phase coexistence region. In any case, all zeros
are at most~$(r-1)$-times degenerate. In addition, for models with
an Ising-like plus-minus symmetry, Theorem~2.5 of~\cite{BBCKK2}
gives conditions under which zeros will lie exactly on the unit
circle. This is the local Lee-Yang theorem.

\newcounter{obrazek}

\begin{figure}[t]
\refstepcounter{obrazek}
\label{fig1}
\vspace{.1in}
\ifpdf
\IfFileExists{Ising.pdf}
{\centerline{\includegraphics[width=2.6in]{Ising.pdf}}}{}
\else
\centerline{\includegraphics[width=2.6in]{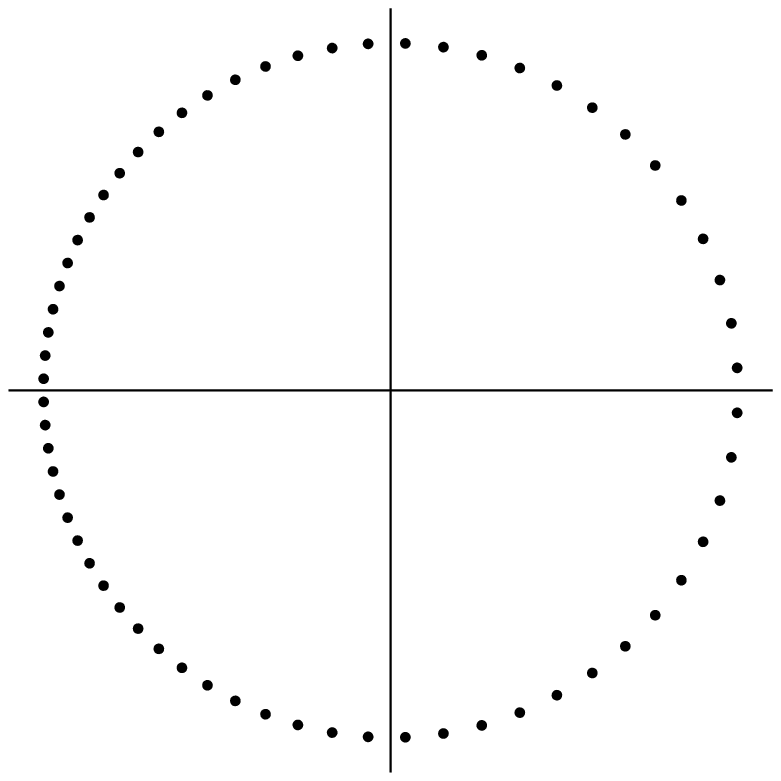}}
\fi
\vspace{.1in}
\bigskip
\begin{quote}
\fontsize{9.5}{7}\selectfont {\sc Figure~\theobrazek.\ } A
schematic figure of the solutions to \twoeqref{Req}{Imq} giving
the approximate locations of partition function zeros of the Ising
model in parameter~$z$ which is related to the external field~$h$
by~$z=e^{2h}$. The plot corresponds to dimension~$d=2$ and torus
side~$L=8$. The expansion used for calculating the
quantities~$\zeta_\pm$ is shown in \eqref{2.22}.
To make the non-uniformity of the spacing between zeros
more apparent,
the plot has been rendered for the choice~$e^{2J}=2.5$
even though this is beyond the region where we can
prove convergence of our expansions.
\normalsize
\end{quote}
\vskip-.1in
\end{figure}

\smallskip
Let us demonstrate these results in the
context of some of our examples from
Section~\ref{sec1.2}. We will begin with the standard Ising model
at low temperatures.
In this case there are two possible phases, labeled~$+$ and~$-$, with the
corresponding metastable free energies given as functions
of~$z=e^{2h}$ by
\begin{equation}
\label{2.22}
\zeta_\pm(z)=\exp\bigl\{\pm h+e^{-2dJ\mp2h}+O(e^{-(4d-2)J})\bigr\}.
\end{equation}
Symmetry considerations now imply that~$|\zeta_+(z)|=|\zeta_-(z)|$
if and only if~$\RE h=0$, i.e.,~$|z|=1$, and, as already known
from the celebrated Lee-Yang Circle Theorem~\cite{LY}, the same is
true
for the actual zeros of~$Z_L^\per$. However, our analysis
allows us to
go further and approximately calculate the solutions to the
system~\twoeqref{Req}{Imq}, which shows that the zeros
of~$Z_L^\per$ lie near the points~$z=e^{i\theta_k}$, where
$k=0,1,\dots,L^d-1$ and
\begin{equation}
\theta_k=\frac{2k+1}{L^d}\pi
+2e^{-2dJ}\sin\Bigl(\frac{2k+1}{L^d}\pi\Bigr)+O(e^{-(4d-2)J}).
\end{equation}
Of course, as~$L$ increases, higher and higher-order terms
in~$e^{-J}$ are needed to pinpoint the location of any particular
zero (given that the distance of close zeros is of the order
$L^{-d}$). Thus, rather than providing the precise location of any
given zero, the above formula should be used to calculate the
quantity~$\theta_{k+1}-\theta_k$, which is essentially the
distance between two consecutive zeros. The resulting derivation
of the \emph{density of zeros} is new even in the case of the
standard Ising model. A qualitative picture of how the zeros span
the unit circle is provided in~Fig.~\ref{fig1}.

A similar discussion applies to the ``perturbed'' Ising model,
provided the nearest-neighbor coupling is ferromagnetic and the
remaining terms in the Hamiltonian are small in some appropriate
norm. In the case of general multi-body couplings, the zeros will
lie on a closed curve which, generically, is not a circle. (For
instance, this is easily verified for the three-body interaction.)
However, if only even terms in~$(\sigma_x)$ appear in the
Hamiltonian, the models have the plus-minus symmetry required by
Theorem~2.5 of~\cite{BBCKK2} and all of the zeros will lie exactly
on the unit circle. This shows that the conclusions of the
Lee-Yang theorem hold well beyond the set of models to which the
classic proof applies.

Finally, in order to demonstrate the non-trivial topology of the
set of zeros, let us turn our attention to the Blume-Capel model.
In this case there are three possible stable phases, each corresponding to
a particular spin value. In terms of the complex parameter~$z=e^h$,
the corresponding metastable free
energies are computed from the formulas
\begin{equation}
\begin{aligned}
\zeta_+(z)&=z\,e^\lambda\exp\left\{z^{-1}e^{-2dJ-\lambda}
+dz^{-2}e^{-(4d-2)J-2\lambda}+O(e^{-4dJ})\right\},
\\
\zeta_-(z)&=z^{-1}e^\lambda\exp\left\{ze^{-2dJ-\lambda}
+dz^2e^{-(4d-2)J-2\lambda}+O(e^{-4dJ})\right\},
\\
\zeta_0(z)&=\exp\left\{(z+z^{-1})e^{-2dJ+\lambda}
+d(z^2+z^{-2})e^{-(4d-2)J+2\lambda}+O(e^{-4dJ})\right\}.
\end{aligned}
\end{equation}
Here it is essential that the energy of the plus-minus neighboring
pair exceeds that of zero-plus (or zero-minus) by a factor of four.


\begin{figure}[t]
\refstepcounter{obrazek}
\label{fig2}
\vspace{.2in}
\ifpdf
\IfFileExists{BC.pdf}
{\centerline{\includegraphics[width=5.4in]{BC.pdf}}}{}
\else
\centerline{\includegraphics[width=5.4in]{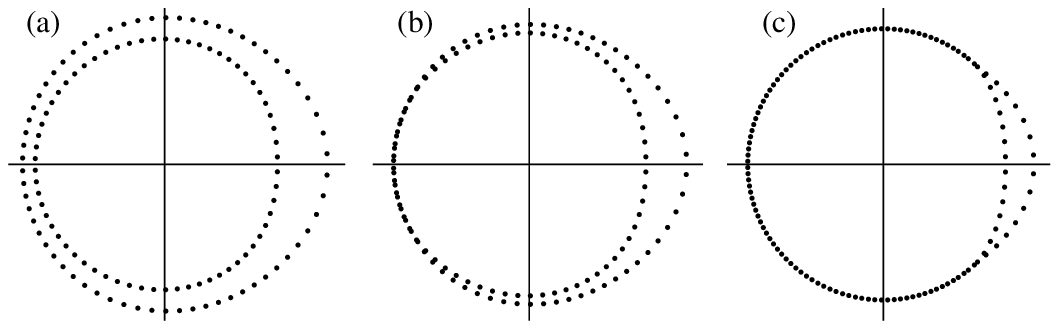}}
\fi
\vspace{.2in}
\bigskip
\begin{quote}
\fontsize{9}{5}\selectfont {\sc Figure~\theobrazek.\ }
A picture
demonstrating the location of partition function zeros of the
Blume-Capel model. Here the zeros concentrate on two curves,
related by the circle inversion, which may or may not coincide
along an arc of the unit circle. There are two critical values
of~$\lambda$, denoted by~$\lambdac^\pm$,
both of order~$e^{-2dJ}$, such that
for~$\lambda<\lambdac^- < 0$, the two curves do not
intersect; see~(a). Once~$\lambda$ increases through~$\lambdac^-$,
a common piece starts to develop which grows as~$\lambda$
increases through the interval~$[\lambdac^-,\lambdac^+]$, see~(b)
and~(c). Finally, both curves collapse on the unit circle
at~$\lambda=\lambdac^+ > 0$ and stay there for
all~$\lambda>\lambdac^+$. With the exception of the
``bifurcation'' points, the zeros lie \emph{exactly} on the unit
circle along the shared arc. The non-uniform spacing of the zeros
in~(b) comes from the influence of the ``unstable'' phase near
the
multiple points.
\normalsize
\end{quote}
\end{figure}

A calculation~\cite{BBCKK1} shows that the zeros lie on two
curves which are symmetrical with respect to circle inversion and
which may coincide along an arc of the unit circle, depending on
the value of~$\lambda$; see Fig.~\ref{fig2}. As~$\lambda$
increases, the shared portion of these curves grows and, for
positive~$\lambda$ exceeding
a constant of order~$e^{-2dJ}$,
all zeros will lie on
the unit circle.
Note that by the methods of ~\cite{Lieb-Sokal}, the last result can be
established~\cite{Lieb} for all temperatures provided~$\lambda$
is sufficiently large, while our results give the correct critical~$\lambda$
but only hold for low temperatures.

\section{Contour models and cluster expansion}
\label{sec3}\noindent
Let us turn to the proofs.  We begin by establishing
the necessary tools for applying Pirogov-Sinai theory.
Specifically, we will define contours and show that spin
configurations and collections
of matching contours are in one-to-one correspondence. This will
induce a corresponding relation between the contour and spin
partition functions. We will also summarize the
facts we will need from the theory of cluster expansions.

\subsection{Contours}
\label{sec3.1}\noindent
The goal of this section is to represent
spin configurations in terms of contours. Based on the
fact---following from Assumption C---that
the constant configurations are the only possible minima of
(the real part of) the energy, we will define contours as the
regions where the spin configuration is not constant.

Recalling our assumption $L\geq 2R+1$, let~$\sigma$ be a spin
configuration on~$\T_L$ and let~$B_R(\sigma)$ be the~$R$-boundary
of~$\sigma$. We equip~$B_R(\sigma)$ with a graph structure by
placing an edge between any two distinct sites~$x,y\in
B_R(\sigma)$ whenever $x$ and $y$ are contained in a cubic box
$\Lambda\subset\T_L$ of diameter $2R+1$ where~$\sigma$ is not
constant. We will denote the resulting graph by~$G_R(\sigma)$.
Some of our definitions will involve the connectivity induced by
the graph~$G_R(\sigma)$ but we will also use the usual concept of
connectivity on~$\T_L$ (or~$\Z^d$): We say that a set of
sites~$\Lambda\subset\T_L$ is connected if every two sites
from~$\Lambda$ can be connected by a nearest-neighbor path
on~$\Lambda$. Note that the connected components of~$B_R(\sigma)$
and the (vertex sets corresponding to the) components of the
graph~$G_R(\sigma)$ are often very different sets.

Now we are ready to define contours.  We start with contours
on~$\Z^d$, and then define contours on the torus in such a way
that they can be easily embedded into~$\Z^d$.

\begin{definition}
\label{def1}
A \emph{contour} on~$\Z^d$ is a pair $Y=(\supp Y,\sigma_Y)$
where~$\supp Y$ is a \emph{finite} connected subset of~$\Z^d$ and
where~$\sigma_Y$ is a spin configuration on~$\Z^d$ such that the
graph~$G_R(\sigma_Y)$ is connected and $B_R(\sigma_Y)=\supp Y$.

A \emph{contour} on~$\T_L$ is a pair $Y=(\supp
Y,\sigma_Y)$ where~$\supp Y$ is a non-empty, connected subset
of~$\T_L$ with diameter strictly less than~$L/2$ and
where~$\sigma_Y$ is a spin configuration on~$\T_L$ such that the
graph~$G_R(\sigma_Y)$ is connected and $B_R(\sigma_Y)=\supp Y$.

A \emph{contour network} on~$\T_L$ is a pair
$\eusmN=(\supp\eusmN,\sigma_{\eusmN})$, where~$\eusmN$ is a
(possibly empty or non-connected) subset
of~$\T_L$ and where~$\sigma_{\eusmN}$ is a spin configuration
on~$\T_L$ such that~$B_R(\sigma_{\eusmN})=\supp\eusmN$ and such
that the diameter of the vertex set of each component
of~$G_R(\sigma_{\eusmN})$ is at least~$L/2$.
\end{definition}

Note that each contour on~$\T_L$ has an embedding
into~$\Z^d$ which is unique up to translation by
multiples of~$L$.
(Informally, we just need to unwrap the torus without
cutting through the contour.) As long as we restrict attention
only to finite contours, the concept of a contour network has no
counterpart on~$\Z^d$, so there we will always assume
that~$\eusmN=\emptyset$.

Having defined contours and contour networks on
$\T_L$ abstractly, our
next task is to identify the contours~$Y_1,\dots,Y_n$ and the
contour network~$\eusmN$ from a general spin configuration
on~$\T_L$. Obviously, the supports of $Y_1,\dots,Y_n$ will be
defined as the vertex sets of the components of the
graph~$G_R(\sigma)$ with diameter less than~$L/2$,
while~$\supp\eusmN$ will be the remaining vertices
in~$B_R(\sigma)$. To define the corresponding spin configurations
we need to demonstrate that the restriction of~$\sigma$ to~$\supp
Y_i$ (resp.,~$\supp\eusmN$) can be extended to spin
configurations~$\sigma_{Y_i}$ (resp.,~$\sigma_{\eusmN}$) on~$\T_L$
such that $B_R(\sigma_{Y_i})=\supp Y_i$ (resp.,
$B_R(\sigma_{\eusmN})=\supp\eusmN$).
It will turn out to be
sufficient to show that~$\sigma$ is constant on the
\emph{boundary} of each connected component of~$\T_L\setminus
B_R(\sigma)$.

Given a set $\Lambda\subset\T_L$ (or $\Lambda\subset\Z^d$),
let~$\partial\Lambda$ denote the external boundary of~$\Lambda$,
i.e.,
$\partial\Lambda=\{x\in\T_L\colon\dist(x,\Lambda)=1\}$.
For the purposes of this section, we also need to define the
set~$\Lambda^\circ$ which is just~$\Lambda$ reduced by the
boundary of its complement,
$\Lambda^\circ=\Lambda\setminus\partial(\T_L\setminus\Lambda)$. An
immediate consequence of Definition~\ref{def1}
(and the restriction to~$2R+1\ge3$)
is the following
fact:

\begin{lemma}
\label{lemma3.0}
Let~$(\Lambda,\sigma)$ be either a contour or a
contour network on~$\T_L$,
and let ~$C$ be a connected component of~$\T_L\setminus\Lambda^\circ$.
Then~$\sigma$ is constant on $C$.
If~$(\Lambda,\sigma)$ is a contour on~$\Z^d$,
then~$\sigma$ is constant on each
connected component~$C$ of~$\Z^d\setminus\Lambda^\circ$,
with~$\Lambda^\circ$ now defined as
$\Lambda^\circ=\Lambda\setminus\partial(\Z^d\setminus\Lambda)$.
\end{lemma}

\begin{proofsect}{Proof}
Assume that~$\sigma$ is not constant on~$C$.
Then there must exist a pair
of nearest-neighbor sites~$x,y\in C$ such that
$\sigma_x\ne\sigma_y$. But then~$x$ and all of its nearest
neighbors lie in~$\Lambda=B_R(\sigma)$.
Since~$C\cap\Lambda^\circ=\emptyset$ and $x\in C$, we are forced
to conclude that $x\in\Lambda\setminus\Lambda^\circ$. But that
contradicts the fact that all of the neighbors of~$x$ also lie
in~$\Lambda$.
The same proof applies to contours on~$\Z^d$.
\end{proofsect}

\begin{definition}
\label{def2}
Let~$(\Lambda,\sigma)$ be either a contour or a contour network
on~$\T_L$
and let~$C$ be a connected component of~$\T_L\setminus\Lambda$.
The common value of the spin on this component in
configuration~$\sigma$ will be called the \emph{label} of~$C$.
The same definition applies to contours on~$\Z^d$, and to
connected components~$C$ of~$\Z^d\setminus\Lambda$.
\end{definition}

Let $\Lambda\subset\T_L$ be a connected set~with diameter less
than~$L/2$. Since the diameter was defined by enclosure into a
``cubic'' box (see Sect.~\ref{sec1.2}), it follows that each
such~$\Lambda$ has a well defined exterior and interior. Indeed,
any box of side less than~$L/2$ enclosing~$\Lambda$ contains less
than $(L/2)^d\leq L^d/2$ sites, so we can define the
\emph{exterior} of~$\Lambda$, denoted by~$\Ext\Lambda$, to be the
unique component of~$\T_L\setminus\Lambda$ that contains more
than~$L^d/2$ sites. The \emph{interior}~$\Int\Lambda$ is defined
simply by
putting~$\Int\Lambda=\T_L\setminus(\Lambda\cup\Ext\Lambda)$. On
the other hand, if~$\Lambda$ is the union of disjoint connected
sets each with diameter at least~$L/2$ we
define~$\Ext\Lambda=\emptyset$
and~$\Int\Lambda=\T_L\setminus\Lambda$.
These definitions for
connected sets imply the following definitions for contours on
$\T_L$:
\begin{definition}
\label{def3}
Let~$Y$ be a contour or a contour network on~$\T_L$.
We then define the
\emph{exterior} of~$Y$, denoted by~$\Ext Y$, as the set $\Ext\supp
Y$, and the~\emph{interior} of~$Y$, denoted by~$\Int Y$, as the
set~$\Int\supp Y$. For each $m\in\CalS$, we let $\Int_mY$ be the
union of all components of~$\Int Y$ with label~$m$. If
$Y$ is a contour on $T_L$, we say
that~$Y$ is a \emph{$m$-contour} if the label of~$\Ext Y$ is~$m$.

Analogous definitions apply to contours on~$\Z^d$, except that
the exterior of a contour $Y$ is now defined as the
infinite component of $\Z^d\setminus \supp Y$, while the interior
is defined as the union of all finite components of
$\Z^d\setminus \supp Y$.
\end{definition}

\smallskip
While most of the following statements can be easily modified to
hold for~$\Z^d$ as well as for the torus $\T_L$, for the sake of
brevity, we henceforth restrict ourselves to the torus.

\begin{lemma}
\label{lemma3.2b}
Let $R\geq 1$
and fix $L>2R+1$. Let~$\sigma$
be a spin configuration on~$\T_L$ and let~$\Lambda$ be either the
vertex set of a component of the graph~$G_R(\sigma)$ with diameter
less than~$L/2$ or the union of the vertex sets of all components
with diameter at least~$L/2$. Let~$\Lambda'$ be of the same form
with~$\Lambda'\ne\Lambda$. Then exactly one of the following is
true:
\settowidth{\leftmargini}{(11111)}
\begin{enumerate}
\item[(1)]
$\Lambda\cup\Int\Lambda\subset\Int\Lambda'$
and~$\Lambda'\cup\Ext\Lambda'\subset\Ext\Lambda$, or
\item[(2)]
$\Lambda'\cup\Int\Lambda'\subset\Int\Lambda$
and~$\Lambda\cup\Ext\Lambda\subset\Ext\Lambda'$, or
\item[(3)]
$\Lambda\cup\Int\Lambda\subset\Ext\Lambda'$
and~$\Lambda'\cup\Int\Lambda'\subset\Ext\Lambda$.
\end{enumerate}
\end{lemma}

\begin{proofsect}{Proof}
It is clearly enough to prove the first half of each of the
statements (1--3), since the second half follow from the first by
taking complements (for example in (3), we just use that
$\Lambda\cup\Int\Lambda\subset\Ext\Lambda'$ implies
$\T_L\setminus(\Lambda\cup\Int\Lambda)\supset\T_L\setminus\Ext\Lambda'$,
which is nothing but the statement
that~$\Lambda'\cup\Int\Lambda'\subset\Ext\Lambda$ by our
definition of interiors and exteriors).

In order to prove the first
halves of the statements (1--3),
we first assume that both~$\Lambda$ and~$\Lambda'$ are vertex sets
of components of the graph $G_R(\sigma)$ with diameter less than~$L/2$.
Clearly, since~$\Lambda$ and~$\Lambda'$
correspond to
different components of~$G_R(\sigma)$, we
have~$\Lambda\cap\Lambda'=\emptyset$. Moreover,~$\Lambda$
and~$\Lambda'$ are
both
connected (as subsets of~$\T_L$) so we have
either~$\Lambda\subset\Int\Lambda'$
or~$\Lambda\subset\Ext\Lambda'$ and \emph{vice versa}.
Hence,
exactly one of the following four statements is true:
\settowidth{\leftmargini}{(11111)}
\begin{enumerate}
\item[(a)]
$\Lambda\subset\Int\Lambda'$ and~$\Lambda'\subset\Int\Lambda$, or
\item[(b)]
$\Lambda\subset\Int\Lambda'$
and~$\Lambda'\subset\Ext\Lambda$, or
\item[(c)]
$\Lambda\subset\Ext\Lambda'$ and~$\Lambda'\subset\Int\Lambda$, or
\item[(d)]
$\Lambda\subset\Ext\Lambda'$ and~$\Lambda'\subset\Ext\Lambda$.
\end{enumerate}

We claim that
the case (a) cannot happen.
Indeed, suppose that~$\Lambda\subset\Int\Lambda'$ and observe that
if~$B$ is a box of size less than~$L^d/2$ such
that~$\Lambda'\subset B$, then~$\Ext\Lambda'\supset\T_L\setminus
B$. Hence~$\Int\Lambda'\subset B$. But then~$B$ also
encloses~$\Lambda$ and
thus~$\Ext\Lambda\cap\Ext\Lambda'\supset\T_L\setminus
B\ne\emptyset$. Now~$\Lambda'\cup\Ext\Lambda'$ is a connected set
intersecting~$\Ext\Lambda$ but not intersecting~$\Lambda$ (because
we assumed that~$\Lambda\subset\Int\Lambda'$). It follows
that~$\Lambda'\cup\Ext\Lambda'\subset\Ext\Lambda$, and hence
~$\Int\Lambda'\supset\Lambda\cup\Int\Lambda$. But then we cannot
have~$\Lambda'\subset\Int\Lambda$ as well. This excludes the case
(a) above, and also shows that (b) actually gives
$\Lambda\cup\Int\Lambda\subset\Int\Lambda'$,
which is  the first part of the claim (1), while (c) gives
$\Lambda'\cup\Int\Lambda'\subset\Int\Lambda$,
which is  the first part of the claim (2).

Turning to the remaining case~(d), let us observe that
$\Lambda'\subset\Ext\Lambda$ implies
$\Int\Lambda\cap\Lambda'\subset\Int\Lambda\cap\Ext\Lambda=\emptyset$.
Since $\Lambda\cap\Lambda'=\emptyset$ as well, this implies
$(\Lambda\cup\Int\Lambda)\cap\Lambda'=\emptyset$. But
$\Lambda\cup\Int\Lambda$ is a connected subset of $\T_L$, so
either $\Lambda\cup\Int\Lambda\subset\Int\Lambda'$ or
$\Lambda\cup\Int\Lambda\subset\Ext\Lambda'$.
Since
$\Lambda\subset\Ext\Lambda'$ excludes the first possibility, we
have shown that in case (d), we necessarily have
$\Lambda\cup\Int\Lambda\subset\Ext\Lambda'$, which is the first
part of statement (3). This concludes the proof of the lemma for
the case when both~$\Lambda$ and~$\Lambda'$ are vertex sets of
components of the graph $G_R(\sigma)$ with diameter less than
$L/2$.

Since it is not possible that both~$\Lambda$ and~$\Lambda'$ are
the union of the vertex sets of all components of diameter at
least $L/2$, it remains to show the statement of the lemma for the
case when~$\Lambda$ is the vertex set of a component of the graph
$G_R(\sigma)$ with diameter less than $L/2$, while~$\Lambda'$ is
the union of the vertex sets of all components of diameter at
least $L/2$.  By definition we now have $\Ext\Lambda'=\emptyset$,
so we will have to prove that $\Lambda\cup\Int\Lambda\subset
\Int\Lambda'$, or equivalently, $\Lambda'\subset\Ext\Lambda$. To
this end, let us first observe that
$\Lambda\cap\Lambda'=\emptyset$, since~$\Lambda$ has diameter less
than $L/2$ while all components of~$\Lambda'$ have diameter at
least $L/2$.  Consider the set $\Int\Lambda$.  Since~$\Lambda$ has
diameter less than $L/2$, we can find a box~$B$ of side length
smaller than $L/2$ that contains~$\Lambda$, and hence also
$\Int\Lambda$.  But this implies that none of the components of
$\Lambda'$ can lie in $\Int\Lambda$ (their diameter is too large).
Since all these components are connected subsets of
$\Int\Lambda\cup\Ext\Lambda$, we conclude that they must be part
of $\Ext\Lambda$. This gives the desired conclusion
$\Lambda'\subset \Ext\Lambda$.
\end{proofsect}

The previous lemma allows us to organize the components of
$G_R(\sigma)$ into a tree-like structure by regarding~$\Lambda'$
to be the ``ancestor'' of~$\Lambda$ (or, equivalently,~$\Lambda$
to be a ``descendant'' of~$\Lambda'$) if the first option in
Lemma~\ref{lemma3.2b} occurs. Explicitly, let $W_R(\sigma)$ be the
collection of all sets $\Lambda\subset \T_L$ that are either the
vertex set of a connected component of $G_R(\sigma)$ with diameter
less than $L/2$ or the union of the vertex sets of all connected
components of diameter at least $L/2$.
We use~$\Lambda_0$ to denote the latter.
If there is no component of diameter $L/2$ or larger,
we define $\Lambda_0=\emptyset$ and set $\Int\Lambda_0=\T_L$.

We now define a \emph{partial order} on~$W_R(\sigma)$ by setting
$\Lambda\prec\Lambda'$ whenever
$\Lambda\cup\Int\Lambda\subset\Int\Lambda'$. If
$\Lambda\prec\Lambda'$, but there is no $\Lambda''\in W_R(\sigma)$
such that $\Lambda\prec\Lambda''\prec\Lambda'$, we say that~$\Lambda$ is a
child of~$\Lambda'$ and~$\Lambda'$ is a parent of~$\Lambda$. Using
Lemma~\ref{lemma3.2b}, one easily shows that no child has more
than one parent, implying that the parent child relationship leads
to a tree structure on $W_R(\sigma)$, with root~$\Lambda_0$.
This opens the possibility for inductive arguments from the
innermost contours (the leaves in the above tree) to the outermost
contours (the children of the root). Our first use of such an
argument will be to prove that unique labels can be assigned to
the connected components of the complement of~$B_R(\sigma)$ .

\begin{lemma}
\label{lemma3.1} Let~$\sigma$ be a spin configuration on~$\T_L$
and let~$\Lambda$ be either the vertex set of a component of the
graph~$G_R(\sigma)$ with diameter less than~$L/2$ or the set of
sites in~$B_R(\sigma)$ that are not contained in any such
component. If~$C$ is a connected component
of~$\T_L\setminus\Lambda^\circ$, then~$\sigma$ is constant
on~$C\cap\Lambda$.
\end{lemma}
The proof is based on the following fact which is presumably well known:

\begin{lemma}
\label{lemma3.05} Let~$A\subset\Z^d$ be a finite connected set
with a connected complement. Then~$\partial A^\cc$ is
$*$-connected in the sense that any two sites~$x,y\in\partial
A^\cc$ are connected by a path on~$\partial A^\cc$ whose
individual steps connect only pairs of sites of~$\Z^d$ with
Euclidean distance not exceeding~$\sqrt2$.
\end{lemma}

\begin{proofsect}{Proof}
The proof will proceed in three steps.  In the first step, we will
prove that the \emph{edge} boundary of~$A$, henceforth denoted
by~$\delta A$, is a \emph{minimal cutset}. (Here we recall that a
set of edges~$E'$ in a graph $G=(V,E)$ is called a cutset if the
graph $G'=(V,E\setminus E')$ has at least two components, and a
cutset~$E'$ is called minimal if any proper subset of~$E'$ is not
a cutset.) In the second step, we will prove that the dual of the
edge boundary~$\delta A$ is a connected set of facets, and in the
third step we will use this fact to prove that~$\partial A^\cc$ is
$*$-connected.

Consider thus a set~$A$ which is connected and whose complement is
connected. Let~$\delta A$ be the edge boundary of~$A$ and
let~$E_d$ be the set of nearest-neighbor edges in~$\Z^d$. The
set~$\delta A$ is clearly a cutset since any nearest-neighbor path
joining~$A$ to~$A^\cc$ must pass through one of the edges
in~$\delta A$. To show that~$\delta A$ is also minimal, let~$E'$
be a proper subset of~$\delta A$, and let $e\in\delta A\setminus
E'$. Since both~$A$ and~$A^\cc$ are connected, an arbitrary pair
of sites $x,y\in\Z^d$ can be joined by a path that uses only edges
in $\{e\}\cup( E_d\setminus\delta A) \subset E_d\setminus E'$.
Hence such~$E'$ is not a cutset which implies that~$\delta A$ is
minimal as claimed.

To continue with the second part of the proof, we need to
introduce some notation.  As usual, we use the symbol ${\Z^*}^d$
to denote the set of all points in $\R^d$ with half-integer
coordinates.  We say that a set $c\subset{\Z^*}^d$ is a
\emph{$k$-cell} if the vertices in~$c$ are the ``corners'' of a
$k$-dimensional unit cube in~$\R^d$.  A $d$-cell
$c\subset{\Z^*}^d$ and a vertex $x\in Z^d$ are called dual to each
other if $x$ is the center of $c$ (considered as a subset of
$\R^d$).  Similarly, a facet~$f$ (i.e., a $(d-1)$-cell in
${\Z^*}^d$) and a nearest-neighbor edge $e\subset\Z^d$ are called
dual to each other if the midpoint of~$e$ (considered as a line
segment in~$\R^d$) is the center of~$f$. The boundary $\partial C$
of a set~$C$ of $d$-cells in ${\Z^*}^d$ is defined as the set of
facets that are contained in an odd number of cells in~$C$, and
the boundary $\partial F$ of a set $F$ of facets in ${\Z^*}^d$ is
defined as the set of $(d-2)$-cells that are contained in an odd
number of facets in $F$.  Finally, a set of facets $F$ is called
connected if any two facets $f,f'\in F$ can be joined by a path of
facets $f_1=f,\dots,f_n=f'$ in $F$ such that for all $i=1,\dots,
n-1$, the facets~$f_i$ and~$f_{i+1}$ share a $(d-2)$-cell in
${\Z^*}^d$.

Note that an arbitrary finite set of facets $F$ has empty boundary
if and only if there exists a finite set of cubes $C$ such that
$F=\partial C$, which follows immediately from the fact
$\R^d$ has trivial homology.  Using this fact, we now prove that
the set $F$ of facets dual to~$\delta A$ is connected.  Let $W$ be
the set of $d$-cells dual to~$A$, and let $F=\partial W$ be the
boundary of $W$.  We will now prove that $F$ is a connected set of
facets.  Indeed, since $F=\partial W$, we have that $F$ has empty
boundary, $\partial F=\emptyset$.  Assume that $F$ has more than
one component, and let $\tilde F\subset F$ be one of them. Then
$\tilde F$ and $F\setminus\tilde F$ are not connected to each
other, and hence share no $(d-2)$-cells.  But this implies that the
boundary of $\tilde F$ must be empty itself, so that $\tilde F$ is
the boundary of some set $\tilde W$.  This in turn implies that
the dual of $\tilde F$ is a cutset, contradicting the fact that
$\delta A$ is a minimal cutset.

Consider now two points $x,y\in \partial A^\cc\subset A$.  Then
there are points $\tilde x, \tilde y\in A^\cc$ such that $\{x,\tilde
x\}$ and $\{y,\tilde y\}$ are edges in~$\delta A$.  Taking into
account the connectedness of the dual of~$\delta A$, we can find a
sequence of edges $e_1=\{x,\tilde x\}$, $\dots$, $e_n=\{y,\tilde
y\}$ in~$\delta A$ such that for all $k=1,\dots, n-1$, the facets
dual to $e_k$
and $e_{k+1}$ share a $(d-2)$ cell in ${\Z^*}^d$.  As a consequence,
the edges $e_k$ and $e_{k+1}$ are either parallel, and the four
vertices in these two edges form an elementary plaquette
of the form $\{x,x+\mathbf{n}_1,x+\mathbf{n}_2,
x+\mathbf{n}_1+\mathbf{n}_2\}$ where $\mathbf{n}_1$ and
$\mathbf{n}_1$ are unit vectors in two different lattice directions,
or $e_k$ and $e_{k+1}$ are orthogonal and share exactly one endpoint.
Since both $e_k$ and $e_{k+1}$ are edges in~$\delta A$,
each of them must contains a point in $\partial A^\cc$, and by the above
case analysis, the two points are at most $\sqrt 2$ apart.  The sequence
$e_1,\dots,e_n$ thus gives rise to a sequence of (not necessarily
distinct) points $x_1,\dots,x_n\in\partial A^\cc$ such that
$x=x_1$, $y=x_n$ and $\dist(x_k,x_{k+1})\leq \sqrt 2$ for
all $k=1,\dots,n-1$.  This proves that $\partial A^\cc$ is
$*$-connected.
\end{proofsect}

\begin{proofsect}{Proof of Lemma~\ref{lemma3.1}}
Relying on Lemma~\ref{lemma3.2b}, we will prove the statement by
induction from innermost to outermost components of diameter less
than~$L/2$. Let~$\Lambda$ be the vertex set of a component of the
graph~$G_R(\sigma)$ with diameter less than~$L/2$ and
suppose~$B_R(\sigma)\cap\Int\Lambda=\emptyset$. (In other
words,~$\Lambda$ is an innermost component of~$B_R(\sigma)$.) Then
the same argument that was used in the proof of Lemma~\ref{lemma3.0}
shows that all connected components of~$\Int\Lambda$ clearly have
the desired property, so we only need to focus on~$\Ext\Lambda$.

Let us pick two sites
$x,y\in\partial\Ext\Lambda=\Lambda\cap\partial\Ext\Lambda$
and
let~$\Lambda'=\Lambda\cup\Int\Lambda$.
Then~$\Lambda'$ is connected with a connected complement and
since~$\Lambda$ has a diameter less than~$L/2$, we may as well
think of~$\Lambda'$ as a subset of~$\Z^d$. Now
Lemma~\ref{lemma3.05} guarantees
that~$\partial (\Lambda')^\cc=\partial\Ext\Lambda$ is $*$-connected and
hence~$x$ and~$y$ are connected by a $*$-connected path entirely
contained in~$\partial\Ext\Lambda$. But the spin configuration
must be constant on any box $(z+[-R,R]^d)\cap\Z^d$
with~$z\in\partial\Ext\Lambda$ and thus the spin is constant along
the path. It follows that~$\sigma_x=\sigma_y$.

The outcome of the previous argument is that now we can ``rewrite''
the configuration on~$\Lambda'$ without changing the rest
of~$B_R(\sigma)$. The resulting configuration will have fewer
connected components of diameter less than~$L/2$ and, proceeding
by induction, the proof is reduced to the cases when there are no
such components at all. But then we are down to the case
when~$\Lambda$ simply equals~$B_R(\sigma)$. Using again the
argument in the proof of Lemma~\ref{lemma3.0}, the spin must
be constant on each connected component~$C$ of~$\T_L\setminus
B_R(\sigma)^\circ$.
\end{proofsect}

The previous lemma shows that each component of the
graph~$G_R(\sigma)$ induces a unique label on every connected
component~$C$ of its complement. Consequently, if two contours
share such a component---which
includes the case when
their supports are adjacent to each other---they must induce the
same label on it. A precise statement of this ``matching''
condition is as follows. (Note, however, that not all collections
of contours will have this matching property.)

\begin{definition}
\label{def4}
We say that the pair $(\Y,\eusmN)$---where~$\Y$ is a
set of contours and~$\eusmN$ is a contour network on~$\T_L$---is a
\emph{collection of matching contours} if the following is true:
\settowidth{\leftmargini}{(11)}
\begin{enumerate}
\item[(1)]
$\supp Y \cap \supp Y'=\emptyset$ for any two distinct
$Y,Y'\in\Y$ and $\supp Y\cap\supp\eusmN=\emptyset$ for
any~$Y\in\Y$.
\item[(2)]
If~$C$ is a connected component of
$\T_L\setminus[(\supp\eusmN)^\circ\cup\bigcup_{Y\in\Y}(\supp
Y)^\circ]$, then the restrictions of the spin
configurations~$\sigma_Y$ (and~$\sigma_{\eusmN}$) to~$C$ are the
same for all contours~$Y\in\Y$ (and contour network~$\eusmN$)
with $\supp Y\cap C\neq\emptyset$ ($\supp\eusmN\cap C\neq\emptyset$).
 In other words, the contours/contour network
intersecting~$C$ induce the same label on~$C$.
\end{enumerate}
Here we use the convention that there are altogether~$|\CalS|$
distinct pairs~$(\Y,\eusmN)$ with both~$\Y=\emptyset$
and~$\eusmN=\emptyset$, each of which corresponds to
one~$m\in\CalS$.
\end{definition}

Definition~\ref{def4} has an obvious analogue for sets $\Y$
of contours on~$\Z^d$, where we require that
(1) $\supp Y \cap \supp Y'=\emptyset$ for any
two distinct $Y,Y'\in\Y$ and (2)
all contours
intersecting a connected component~$C$ of
$\Z^d\setminus[\bigcup_{Y\in\Y}(\supp Y)^\circ]$
induce the same label on~$C$.

It remains to check the intuitively obvious fact that spin
configurations and collections of matching contours are in
one-to-one correspondence:

\begin{lemma}
\label{lemma3.2} For each spin
configuration~$\sigma\in\CalS^{\T_L}$, there exists a unique
collection $(\Y,\eusmN)$ of matching contours
on $\T_L$ and for any
collection $(\Y,\eusmN)$ of matching contours on~$\T_L$, there
exists a unique spin configuration~$\sigma\in\CalS^{\T_L}$ such
that the following is true: \settowidth{\leftmargini}{(11)}
\begin{enumerate}
\item[(1)] The supports of the contours in~$\Y$ (of the contour
network~$\eusmN$) are the vertex sets (the union of the vertex
sets) of the connected components of the graph $G_R(\sigma)$ with
diameter strictly  less than (at least)~$L/2$.
\item[(2)] The spin configuration corresponding to a collection
$(\Y,\eusmN)$ of matching contours arise by restricting~$\sigma_Y$
for each $Y\in\Y$ as well as $\sigma_{\eusmN}$ to the support of
the corresponding contour (contour network) and then extending the
resulting configuration by the common label of the adjacent
connected components.
\end{enumerate}
\end{lemma}

\begin{proofsect}{Proof}
Let~$\sigma$ be a spin configuration and let~$\Lambda$ be a
component of the graph~$G_R(\sigma)$
 with diameter less
than~$L/2$. Then Lemma~\ref{lemma3.1} ensures that~$\sigma$ is
constant on the
boundary $\partial C$ of each component $C$
of~$\Lambda^\cc$.
Restricting $\sigma$ to~$\Lambda$ and extending
the resulting configuration in such a way that
the new configuration, $\tilde\sigma$, restricted to a component
component $C$ of~$\Lambda^\cc$, is equal
to the old configuration on $\partial C$,
the pair $(\Lambda,\tilde\sigma)$ thus defines
a contour.
Similarly, if~$\Lambda$ is the union of all components of the
graph~$G_R(\sigma)$ with diameter at least~$L/2$ and~$C$ is a
connected component of~$\T_L\setminus\Lambda^\circ$, then~$\sigma$
is, after removal of all contours, constant on~$C$. The
contours/contour network $(\Y,\eusmN)$ then arise from~$\sigma$ in
the way described. The supports of these objects are all disjoint,
so the last property to check is that the labels induced on the
adjacent connected components indeed match. But this is a direct
consequence of the construction.

To prove the converse, let $(\Y,\eusmN)$ denote a set of matching
contours and let~$\sigma$ be defined by the corresponding contour
configuration on the support of the contours (or contour network)
and by the common value of the spin in contour configurations for
contours adjacent to a connected component of
$\T_L\setminus[(\supp\eusmN)^\circ\cup\bigcup_{Y\in\Y}(\supp
Y)^\circ]$. (If at least one of~$\Y,\eusmN$ is nonempty, then this
value is uniquely specified because of the matching condition;
otherwise, it follows by our convention that empty~$(\Y,\eusmN)$
carries an extra label.)

It remains to show that~$\Y$ are the contours and~$\eusmN$ is the
contour network of~$\sigma$. Let~$A$ be a component of the
graph~$G_R(\sigma)$. We have to show that it coincides with $\supp
Y$ for some $Y\in\Y$ or with a component of $\supp\eusmN$ (viewed
as a graph). We start with the observation that
$A\subset\supp\eusmN\cup(\bigcup_{Y\in\Y}\supp Y)$. Next we note
that for each $Y\in\Y$, the graph $G_R(\sigma_Y)$ is connected.
Since the restriction of $\sigma_Y$ to $\supp Y$ is equal to the
corresponding restriction of $\sigma$, we conclude that $\supp
Y\cap A\neq \emptyset$ implies $\supp Y\subset A$, and similarly
for the components of $\supp\eusmN$. To complete the proof, we
therefore only have to exclude that $\supp Y\subset A$ for more
than one contour $Y\in\Y$, or that $\Lambda\subset A$ for more
than one component~$\Lambda$ of $\supp\eusmN$, and similarly for
the combination of contours in $Y$ and components of
$\supp\eusmN$.

Let us  assume that $\supp Y\subset A$ for more
than one contour $Y\in\Y$. Since~$A$ is a connected component of
the graph $G_R(\sigma)$, this implies that there exists a box
$B_z=(z+[-R,R]^d)\cap\Z^d$ and two contours $Y_1,Y_2\in Y$ such
that $\sigma$ is not constant on~$B_z$, $\supp Y_1\cup \supp
Y_2\subset A$ and~$B_z$ is intersecting both~$\supp Y_1$
and~$\supp Y_2$. But this is in contradiction with the fact that
$\Y$ is a collection of matching contours (and a configuration on
any such box not contained in the support of one of the contours
in $\Y$ or in a  component of $\supp\eusmN$ must be
constant). In the same way one excludes the
case combining  $\supp Y$ with a
component of $\supp\eusmN$ or combining two components of
$\supp\eusmN$. Having excluded  everything else, we thus have
shown that~$A$ is either the support of one of the contours in
$Y$, or one of the components of $\supp\eusmN$.
\end{proofsect}

\subsection{Partition functions and Peierls' condition}
\label{sec3.2}\noindent
A crucial part of our forthcoming
derivations concerns various contour partition functions, so our
next task will be to define these quantities.
We need some notation:
Let $(\Y,\eusmN)$ be a collection of matching contours on $\T_L$.
A contour $Y\in\Y$ is called an \emph{external contour in $\Y$} if
$\supp Y\subset\Ext Y'$ for all $Y'\in\Y$ different from~$Y$, and
we will call two contours $Y,Y'\in\Y$  \emph{mutually external} if
$\supp Y\subset\Ext Y'$ and $\supp Y'\subset\Ext Y$. Completely
analogous definitions apply to a set of matching contours
$\Y$ on~$\Z^d$
(recall that  on ~$\Z^d$, we always set $\eusmN=\emptyset$).
Note that, by
Lemma~\ref{lemma3.2b},
two contours of a configuration
$\sigma$ on $\T_L$ are either
mutually external or one is contained in the interior of the
other.
Inspecting the proof of this Lemma~\ref{lemma3.2b},
the reader may easily verify that this remains true for
configurations  on~$\Z^d$,
provided the set~$B_R(\sigma)$ is finite.

Given a contour~$Y=(\supp Y,\sigma_Y)$ or a contour
network~$\eusmN=(\supp\eusmN,\sigma_{\eusmN})$
let~$E(Y,z)$ and~$E(\eusmN,z)$ denote
the
corresponding excitation energies~$E(\sigma_Y,z)$ and
$E(\sigma_{\eusmN},z)$ from \eqref{1.6}.
We then introduce exponential
weights~$\rho_z(Y)$ and $\rho_z(\eusmN)$, which are related to the
quantities~$E(Y,z)$ and~$E(\eusmN,z)$ according to
\begin{equation}
\rho_z(Y)=e^{-E(Y,z)}
\quad\text{and}\quad
\rho_z(\eusmN)=e^{-E(\eusmN,z)}.
\end{equation}
The next lemma states that the exponential weights
$\theta_m(z)$, $\rho_z(Y)$ and $\rho_z(\eusmN)$ are analytic functions
of~$z$.

\begin{lemma}
\label{lemAnaly}
Suppose that Assumption~C0 holds, let $q\in\CalS$, let $Y$ be a $q$-contour
and let $\eusmN$ be a contour network.  Then
$\theta_q(z)$, $\rho_z(Y)$ and $\rho_z(\eusmN)$ are analytic
functions of $z$ in $\tilde\scrO$.
\end{lemma}

\begin{proofsect}{Proof}
By assumption~C0, the functions
$z\mapsto\varphi_\Lambda(\sigma,z)=\exp\{-\Phi_\Lambda(\sigma,z)\}$
are holomorphic in~$\tilde\scrO$.  To prove the lemma, we will
show that $\theta_q(z)$, $\rho_z(Y)$ and $\rho_z(\eusmN)$ can be
written as products over the exponential potentials
$\varphi_\Lambda(\sigma,z)$, with $\sigma=\sigma^q$,
$\sigma=\sigma_Y$ and $\sigma=\sigma_{\eusmN}$, respectively.

Let us start with $\theta_q(z)$.  Showing that
$\theta_q$ is
the product of
exponential potentials
$\varphi_\Lambda(\sigma^q,z)$ is clearly equivalent to showing that $e_q$ can
be rewritten in the form
\begin{equation}
\label{e_q.alt}
e_q=\sum_{\Lambda\in \V_e}\Phi_\Lambda(\sigma^q),
\end{equation}
where $\V_e$ is a collection of subsets
$\Lambda\subset\T_L$.  But this is obvious from the definition
\eqref{1.5} of $e_q$: just choose $\V_e$ in such a way that it
contains exactly one representative from each equivalence class
under translations.

Consider now a contour $Y=(\supp Y,\sigma_Y)$ and the corresponding
excitation energy~$E(Y,z)$.  We will want to show that~$E(Y,z)$
can be written in the form
\begin{equation}
\label{EYz.alt}
E(Y,z)=\sum_{\Lambda\in \V_Y}\Phi_\Lambda(\sigma_Y),
\end{equation}
where $\V_Y$ is again a collection of subsets $\Lambda\subset\T_L$.
Let $\Lambda_q=\Ext Y\cup\Int_q Y$, and $\Lambda_m=\Int_m Y$
for $m\neq q$.  Consider a point $x\in\Lambda_m$.  Since
$x\notin\supp Y=B_R(\sigma_Y)$, the configuration $\sigma_Y$
must be constant on any subset $\Lambda\subset\T_L$ that
has diameter $2R+1$ or less and contains the point $x$, implying
that
\begin{equation}
\sum_{\Lambda\colon x\in\Lambda}\frac 1{|\Lambda|}\Phi_\Lambda(\sigma_Y)
=
\sum_{\Lambda\colon x\in\Lambda}\frac 1{|\Lambda|}\Phi_\Lambda(\sigma^m)
=e_m
\end{equation}
whenever $x\in\Lambda_m$.  Using these facts, we now rewrite
$E(Y,z)$ as
\begin{equation}
\begin{aligned}
\label{1.7alt}
E(Y,z)
&=
\beta H_L(\sigma_Y)-\sum_{x\in\T_L\smallsetminus\supp Y}
\sum_{\Lambda:x\in\Lambda}
\frac1{|\Lambda|}\Phi_\Lambda(\sigma_Y)
\\
&=\sum_{\Lambda\subset\T_L}\Phi_\Lambda(\sigma_Y)
- \sum_{m\in\CalS}|\Lambda_m|e_m
\\
&=\sum_{\Lambda\subset\supp Y}\Phi_\Lambda(\sigma_Y)
+\sum_{m\in\CalS}\Biggl\{\biggl(
\sum_{\begin{subarray}{c}
\Lambda\subset\T_L\\\Lambda\cap\Lambda_m\neq\emptyset
\end{subarray}}\Phi_\Lambda(\sigma^m)\biggr)-
|\Lambda_m|e_m\Biggr\}.
\end{aligned}
\end{equation}
To complete the proof, we note that
the sum over all~$\Lambda$ with $\Lambda\cap\Lambda_m\neq\emptyset$
contains at least $|\Lambda_m|$ translates of each $\Lambda\subset\T_L$
contributing to the right hand side of \eqref{e_q.alt}.  As a consequence,
the difference on the right hand side of \eqref{1.7alt} can
be written in the form \eqref{EYz.alt}, proving that~$E(Y,z)$
is of the form \eqref{EYz.alt}. The proof that
$\rho_z(\eusmN)$ is an analytic function of~$z$ is virtually identical.
\end{proofsect}

Next we define partition functions in finite subsets of~$\Z^d$.
Fix an index $q\in\CalS$. Let $\Lambda\subset\Z^d$
be a finite set
and let~$\MM(\Lambda,q)$ be the set of all collections~$\Y$ of
matching contours in~$\Z^d$
with the following properties:
\settowidth{\leftmargini}{(11111)}
\begin{enumerate}
\item[(1)]
For each $Y\in\Y$, we have $\supp Y\cup\Int Y\subset\Lambda$.
\item[(2)]
The external contours in~$\Y$ are~$q$-contours.
\end{enumerate}
Note that $\supp Y\cup\Int Y\subset\Lambda$ is implied by the
simpler condition that $\supp Y\subset\Lambda$ if
$\Z^d\setminus\Lambda$ is connected, while in the case where
$\Z^d\setminus\Lambda$ is not connected, the condition $\supp
Y\cup\Int Y\subset\Lambda$ is stronger, since it implies that none
of the contours $Y\in\Y$ contain any hole of~$\Lambda$ in its
interior. (Here a hole is defined as a finite component
of~$\Z^d\setminus\Lambda$.) In the sequel, we will say that
\emph{$Y$ is a contour in~$\Lambda$} whenever~$Y$ obeys the
condition~$\supp Y\cup\Int Y\subset\Lambda$.

The \emph{contour partition function} in~$\Lambda$ with boundary
condition~$q$ is then defined by
\begin{equation}
\label{Z_q}
Z_q(\Lambda,z)=\sum_{\Y\in\MM(\Lambda,q)}\Bigl[ \prod_{m\in\CalS}
\theta_m(z)^{|\Lambda_m(\Y)|}\Bigr] \prod_{Y\in\Y}\rho_z(Y),
\end{equation}
where~$\Lambda_m(\Y)$ denotes the union of all components of
$\Lambda\setminus\bigcup_{Y\in\Y}\supp Y$ with label~$m$, and
$|\Lambda_m(\Y)|$ stands for the cardinality of~$\Lambda_m(\Y)$.

\smallskip
If we add the condition that the contour network $\eusmN$ is
empty, the definitions of the set~$\MM(\Lambda,q)$ and the
partition function $Z_q(\Lambda,z)$
clearly extends to any
subset~$\Lambda\subset\T_L$, because on~$\T_L$ every contour has a
well defined exterior and interior. However, our goal is to have a
contour representation for the full torus partition function.
Let~$\MM_L$ denote the set of all collections $(\Y,\eusmN)$ of
matching contours in~$\T_L$
which, according to our convention, include an extra label~$m\in\CalS$
when both~$\Y$ and~$\eusmN$ are empty.
If~$(\Y,\eusmN)\in\MM_L$ is such a collection,
let~$\Lambda_m(\Y,\eusmN)$ denote the union of the components of
$\T_L\setminus(\supp\eusmN\cup\bigcup_{Y\in\Y}\supp Y)$ with
label~$m$. Then we have:

\begin{proposition}[Contour representation]
\label{prop3.6}
The partition function on the torus~$\T_L$ is given by
\begin{equation}
\label{ZL1}
Z_L^\per(z)=\sum_{(\Y,\eusmN)\in\MM_L}\Bigl[ \prod_{m\in\CalS}
\theta_m(z)^{|\Lambda_m(\Y,\eusmN)|}\Bigr]\,
\rho_z(\eusmN)\prod_{Y\in\Y}\rho_z(Y).
\end{equation}
In particular, we have
\begin{equation}
\label{ZL2}
Z_L^\per(z)=\sum_{(\emptyset,\eusmN)\in\MM_L}
\rho_z(\eusmN)\,
\prod_{m\in\CalS}Z_m\bigl(\Lambda_m(\emptyset,\eusmN),z\bigr).
\end{equation}
\end{proposition}

\begin{proofsect}{Proof}
By Lemma~\ref{lemma3.2}, the spin configurations~$\sigma$ are in
one-to-one correspondence with the pairs $(\Y,\eusmN)\in\MM_L$.
Let $(\Y,\eusmN)$ be the pair corresponding to~$\sigma$. Rewriting
\eqref{1.7} as
\begin{equation}
\label{1.7a}
\beta H_L(\sigma)=\!\sum_{x\in\T_L}\,\,\sum_{\begin{subarray}{c}
\Lambda\colon\Lambda\subset\T_L\\\Lambda\ni x
\end{subarray}}
\frac1{|\Lambda|}\Phi_\Lambda(\sigma),
\end{equation}
we can now split the first sum into several parts: one for each
$m\in\CalS$ corresponding to $x\in\Lambda_m(\Y,\eusmN)$, one for
each $Y\in\Y$ corresponding to $x\in\supp Y$, and finally,
one for
the part of the sum corresponding to $x\in\supp\eusmN$. Invoking the
definitions of the energies~$e_m(z)$,~$E(Y,z)$ and~$E(\eusmN,z)$,
this gives
\begin{equation}
\label{1.7b}
\beta
H_L(\sigma)=
\sum_{m\in\CalS}e_m(z)\bigl|\Lambda_m(\Y,\eusmN)\bigr|
+\sum_{Y\in\Y}E(Y,z)+E(\eusmN,z).
\end{equation}
Strictly speaking, the fact that the excitation energy
factors (technically, sums) over contours and contour networks requires a
proof.
Since this is straightforward using
induction as in the proof of Lemma~\ref{lemma3.1},
starting again with
the innermost contours, we leave the formal proof to the reader.
Using the definitions
of~$\theta_m(z)$,~$\rho_z(Y)$ and~$\rho_z(\eusmN)$ and noting
that, by Lemma~\ref{lemma3.2}, the sum over~$\sigma$ can be
rewritten as the sum over~$(\Y,\eusmN)\in\MM_L$, formula
\eqref{ZL1} directly follows.

The second formula, \eqref{ZL2}, formally arises by a resummation
of all contours that can contribute together with a given contour
network~$\eusmN$. It only remains to check that if
$\Y_m\subset\Y$ is the set of~$Y\in\Y$ with $\supp
Y\subset\Lambda_m=\Lambda_m(\emptyset,\eusmN)$, then~$\Y_m$ can
take any value in~$\MM(\Lambda_m,m)$. But this follows directly
from Definition~\ref{def4} and the definition
of~$\MM(\Lambda_m,m)$.
\end{proofsect}

In order to be useful, the representations \eqref{ZL1} and
\eqref{ZL2} require that contours and contour networks are
sufficiently suppressed with respect to the maximal ground state
weight $\theta$.  This is ensured by Assumption~C2, which
guarantees
that $|\rho_z(Y)|\leq \theta(z)^{|Y|}e^{-\tau|Y|}$ and
$|\rho_z(\eusmN)|\leq \theta(z)^{|\eusmN|}e^{-\tau|\eusmN|}$,
where we used the symbols $|Y|$ and $|\eusmN|$ to denote the
cardinality of~$\supp Y$ and $\supp\eusmN$, respectively.

\subsection{Cluster expansion}
\label{sec3.3}\noindent The last ingredient that we  will need is
the \emph{cluster expansion}, which will serve as our principal
tool for evaluating and estimating logarithms of various partition
functions. The cluster expansion is conveniently formulated in the
context of so-called abstract polymer models~\cite{Sei,KP2,D,MS}.
Let~$\ssK$ be a countable set---the set of all
\emph{polymers}---and let~$\not\sim$ be the \emph{relation of
incompatibility} which is a reflexive and symmetric binary
relation on~$\ssK$. For each~$\ssA\subset\ssK$, let~$\MM(\ssA)$ be
the set of multi-indices $\ssX\colon\ssK\to\{0\}\cup\N$ that are
finite, $\sum_{\gamma\in\ssK}\ssX(\gamma)<\infty$, and that
satisfy~$\ssX(\gamma)=0$ whenever~$\gamma\not\in\ssA$. Further,
let $\CC(\ssA)$ be the set of all multi-indices $\ssX\in\MM(\ssA)$
with values in $\{0,1\}$ that satisfy
$\ssX(\gamma)\ssX(\gamma')=0$ whenever $\gamma\not\sim\gamma'$
and~$\gamma\ne\gamma'$.

Let $\zz\colon\ssK\to\C$ be a polymer
functional. For each finite subset $\ssA\subset\ssK$, let us
define the polymer partition function~$\ZZ(\ssA)$ by the formula
\begin{equation}
\label{3.10}
\ZZ(\ssA)=\sum_{\ssX\in\CC(\ssA)}
\prod_{\gamma\in\ssK}\zz(\gamma)^{\ssX(\gamma)}.
\end{equation}
In the most recent formulation~\cite{D,MS}, the cluster expansion
corresponds to a multidimensional Taylor series for the
quantity~$\log\ZZ(\ssA)$, where the complex variables are
the~$\zz(\gamma)$.
Here \emph{clusters} are simply  multi-indices~$\ssX\in \MM(\ssK)$
for which any
nontrivial decomposition of~$\ssX$ leads to incompatible
multi-indices. Explicitly, if~$\ssX$ can be written
as~$\ssX_1+\ssX_2$ with~$\ssX_1,\ssX_2\not\equiv0$, then there
exist two (not necessary distinct)
polymers~$\gamma_1,\gamma_2\in\ssK$,
$\gamma_1\not\sim\gamma_2$,
such
that~$\ssX_1(\gamma_1)\ssX_2(\gamma_2)\neq 0$.

Given a finite sequence
$\Gamma=(\gamma_1,\dots,\gamma_n)$
of polymers in $\ssK$, let $n(\Gamma)=n$  be the length
of the sequence $\Gamma$, let
$\GG(\Gamma)$ be the set of all connected graphs
on $\{1,\dots,n\}$ that have no edge between
the vertices
$i$ and $j$ if $\gamma_i\sim\gamma_j$, and let
${\ssX}_\Gamma$ be the multi-index for which
${\ssX}_\Gamma(\gamma)$ is equal to the number
of times that~$\gamma$ appears
in~$\Gamma$.
For a finite multi-index $\ssX$, we then define
\begin{equation}
\label{aT}
\aT(\ssX)=
\sum_{\Gamma:{\ssX}_\Gamma=\ssX}\frac 1{n(\Gamma)!}
\sum_{\frakg\in\GG(\Gamma)}(-1)^{|\frakg|},
\end{equation}
with~$|\frakg|$ denoting the  number of edges
in~$\frakg$, and
\begin{equation}
\label{zT}
\zzT(\ssX)=\aT(\ssX)\prod_{\gamma\in\ssK}\zz(\gamma)^{\ssX(\gamma)}.
\end{equation}
Note that $\GG(\Gamma)=\emptyset$ if $\ssX_\Gamma$ is not a
cluster, implying, in particular, that~$\zzT(\ssX)=0$
whenever~$\ssX$ is not a cluster.
We also use the notation  $\ssX\not\sim \gamma$ whenever
$\ssX$ is a cluster such that
$\ssX(\gamma')>0$ for at least one
$\gamma'\not\sim\gamma$.

\smallskip
The main result of~\cite{MS} (building upon~\cite{D}) is then as
follows:

\begin{theorem}[Cluster expansion]
\label{thmKP} Let~$a\colon\ssK\to[0,\infty)$ be a
function and let $\zz_0\colon\ssK\to[0,\infty)$ be polymer weights
satisfying the bound
\begin{equation}
\label{3.11}
\sum_{\begin{subarray}{c}
\gamma'\in\ssK\\\gamma'\not\sim\gamma
\end{subarray}}
\zz_0(\gamma')\,e^{a(\gamma')}\le a(\gamma),
\qquad\gamma\in\ssK.
\end{equation}
Then~$\ZZ(\ssA)\ne0$ for any finite set~$\ssA\subset\ssK$ and
any collection of polymer weights~$\zz\colon\ssK\to\C$
in the multidisc $\D_{\ssA}=\{(\zz(\gamma))\colon
|\zz(\gamma)|\le\zz_0(\gamma),\,\gamma\in\ssA\}$.
Moreover, if we define~$\log\ZZ(\ssA)$ as
the unique continuous branch of the complex logarithm of~$\ZZ(\ssA)$
on~$\D_{\ssA}$ normalized so that~$\log\ZZ(\ssA)=0$ when~$\zz(\gamma)=0$ for
all~$\gamma\in\ssA$, then
\begin{equation}
\label{3.12}
\log\ZZ(\ssA)=\sum_{\ssX\in\MM(\ssA)}\zzT(\ssX)
\end{equation}
holds for each finite set $\ssA\subset\ssK$. Here the power series
on the right hand side converges absolutely on the
multidisc~$\D_{\ssA}$. Furthermore, the bounds
\begin{equation}
\label{3.13}
\sum_{\begin{subarray}{c}
\ssX\in\MM(\ssK)\\\ssX(\gamma)\ge1
\end{subarray}}
\bigl|\zzT(\ssX)\bigr|\le
\sum_{\ssX\in\MM(\ssK)}\ssX(\gamma)
\bigl|\zzT(\ssX)\bigr|\le\bigl|\zz(\gamma)\bigr|e^{a(\gamma)}
\end{equation}
and
\begin{equation}
\label{3.13a}
\sum_{\begin{subarray}{c}
\ssX\in\MM(\ssK)\\\ssX\not\sim\gamma
\end{subarray}}
\bigl|\zzT(\ssX)\bigr|\le a(\gamma)
\end{equation}
hold for each~$\gamma\in\ssK$.
\end{theorem}

\begin{proofsect}{Proof}
This is essentially the main result of~\cite{MS} stated under the
(stronger) condition \eqref{3.11}, which is originally due
to~\cite{N,KP2}. To make the correspondence with \cite{MS} explicit,
let
\begin{equation}
\mu(\gamma)=\log\bigl(1+|\zz(\gamma)|e^{a(\gamma)}\bigr)
\end{equation}
and note
that
$\mu(\gamma)\le|\zz(\gamma)|e^{a(\gamma)}\le\zz_0(\gamma)e^{a(\gamma)}$.
The condition \eqref{3.11} then guarantees
that we have
$\sum_{\gamma'\not\sim\gamma}\mu(\gamma')\le a(\gamma)$ and
hence
\begin{equation}
|\zz(\gamma)|=(e^{\mu(\gamma)}-1)e^{-a(\gamma)}
\le(e^{\mu(\gamma)}-1)
\exp\Bigl\{-\sum_{\gamma'\not\sim\gamma}\mu(\gamma')\Bigr\}.
\end{equation}
This implies that any collection of weights~$\zz\colon\ssK\to\C$
such that~$|\zz(\gamma)|\le\zz_0(\gamma)$ for all~$\gamma\in\ssK$
will fulfill the principal condition of the main theorem
of~\cite{MS}. Hence, we can conclude that $\ZZ(\ssA)\ne0$
in~$\D_{\ssA}$ and that \eqref{3.12} holds. Moreover, as shown
in~\cite{MS}, both quantities on the left-hand side of
\eqref{3.13} are bounded by $e^{\mu(\gamma)}-1$ which simply
equals~$|\zz(\gamma)|e^{a(\gamma)}$.  The bound \eqref{3.13}
together with the condition~\eqref{3.11} immediately give
\eqref{3.13a}.
\end{proofsect}

To facilitate the future use of this result, we will extract the
relevant conclusions into two lemmas. Given a spin state
$q\in\CalS$, let~$\ssK_q$ denote the set of all~$q$-contours
in~$\Z^d$. If $Y,Y'\in\ssK_q$, let us call~$Y$ and~$Y'$
\emph{incompatible} if $\supp Y\cap\supp Y'\ne\emptyset$.
If~$\ssA$ is a finite set of $q$-contours, we will let~$\ZZ(\ssA)$
be the polymer sum \eqref{3.10} defined using this incompatibility
relation. Then we have:

\begin{lemma}
\label{lemma3.7}
There exists a constant $c_0=c_0(d,|\CalS|)\in(0,\infty)$ such
that, for all~$q\in\CalS$ and all contour functionals
$\zz\colon\ssK_q\to\C$ satisfying
the condition
\begin{equation}
\label{z<z_0}
|\zz(Y)|\le\zz_0(Y)=e^{-(c_0+\eta)|Y|} \quad \text{for all}\quad Y\in\ssK_q,
\end{equation}
for some~$\eta\ge0$, the following holds for all $k\ge 1$:

\smallskip\noindent
(1)~$\ZZ(\ssA)\ne0$
for all
finite~$\ssA\subset\KK_q$ with $\log \ZZ(\ssA)$ given by
\eqref{3.12}, and
\begin{equation}
\label{3.14}
\sum_{\begin{subarray}{c}
\ssX\in\MM(\ssK_q)
\\
V(\ssX)\ni0,\, \Vert\ssX \Vert\ge k\end{subarray}}
\bigl|\zzT(\ssX)\bigr|\le e^{-\eta k}.
\end{equation}
Here $V(\ssX)=\bigcup_{Y\colon\ssX(Y)>0} V(Y)$
with $V(Y)=\supp Y\cup\Int Y$  and  $\Vert\ssX\Vert=\sum_{Y\in\ssK_q}\ssX(Y)|Y|$.

\smallskip\noindent
(2)~Furthermore, if the activities
$\zz(Y)$ are twice continuously differentiable
(but not necessarily analytic) functions of a complex
parameter~$z$ such that the bounds
\begin{equation}
\label{3.21}
\bigl|\partial_w\zz(Y)\bigr|\le
\zz_0(Y)
\quad\text{and}\quad
\bigl|\partial_w\partial_{w'}\zz(Y)\bigr|
\le
\zz_0(Y)
\end{equation}
hold for any~$w,w'\in\{z,\bar z\}$ and any~$Y\in\ssK_q$, then
\begin{equation}
\label{der3.14}
\sum_{\begin{subarray}{c}\ssX
\in
\MM(\ssK_q)
\\
V(\ssX)\ni0,\, \Vert\ssX \Vert\ge k
\end{subarray}}
\bigl|\partial_w\zzT(\ssX)\bigr|
\le e^{-\eta k}\quad
\text{and}\quad
\sum_{\begin{subarray}{c}\ssX
\in
\MM(\ssK_q)
\\
V(\ssX)\ni0,\, \Vert\ssX \Vert\ge k
\end{subarray}}
\bigl|\partial_w\partial_{w'}\zzT(\ssX)\bigr|
\le e^{-\eta k}.
\end{equation}
for any~$w,w'\in\{z,\bar z\}$.
\end{lemma}

Using, for any finite~$\Lambda\subset\Z^d$,
the notation~$\ssK_{q,\Lambda}=\{Y\in\ssK_q\colon\supp Y\cup\Int
Y\subset\Lambda\}$ and $\partial\Lambda$ for the set of sites
in~$\Z^d\setminus\Lambda$ that have a nearest neighbor in~$\Lambda$,
we get
the following lemma
as an
easy
corollary:

\begin{lemma}
\label{lemma3.7-cor}
Suppose that the weights~$\zz$ satisfy the bound \eqref{z<z_0} and are
invariant under the translations of~$\Z^d$. Then
the \emph{polymer pressure}~$s_q=
\lim_{\Lambda\uparrow\Z^d}|\Lambda|^{-1}\log\ZZ(\ssK_{q,\Lambda})$
exists and is given~by
\begin{equation}
\label{3.19}
s_q=\sum_{\ssX\in
\MM(\ssK_q)
\colon V(\ssX)\ni0}
\frac1{|V(\ssX)|}\,\zzT(\ssX).
\end{equation}
Moreover, the bounds
\begin{equation}
\label{sqeta}
|s_q|\le e^{-\eta}
\end{equation}
and
\begin{equation}
\label{3.20}
\bigl|\log\ZZ(\ssK_{q,\Lambda})-s_q|\Lambda|\bigr|
\le e^{-\eta}|\partial\Lambda|
\end{equation}
hold.
Finally, if the conditions \eqref{3.21} on derivatives of
the weights $\zz(Y)$
are also met, the polymer pressure $s_q$
is twice continuously differentiable in~$z$ with the bounds
\begin{equation}
\label{3.22}
\bigl|\partial_w s_q \bigr|\le e^{-\eta}
\quad\text{and}\quad
\bigl| \partial_w\partial_{w'}s_q \bigr|\le e^{-\eta},
\end{equation}
valid for any~$w,w'\in\{z,\bar z\}$.
\end{lemma}

\begin{proofsect}{Proof of Lemma \ref{lemma3.7}}
Let us consider a polymer model where polymers are either a single
site of~$\Z^d$ or a~$q$-contour from~$\ssK_q$.
(The reason for including single
sites as polymers will become apparent below.)
Let the
compatibility
between
contours be defined by
disjointness of their
supports while that between a contour~$Y$ and a site~$x$
by disjointness
of~$\{x\}$ and $\supp Y\cup\Int Y$.
If we let~$\zz(\gamma)=0$ whenever~$\gamma$ is just a single site,
this polymer model is indistinguishable from the one considered in
the statement of the lemma.
Let us choose $c_0$ so that
\begin{equation}
\label{c0def}
\sum_{Y\in\ssK_q\colon V(Y)\ni 0}
e^{(2-c_0)|Y|}\le 1.
\end{equation}
To see that this is possible with a constant $c_0$
depending only on the dimension and the cardinality of~$\CalS$, we
note that each polymer is a connected subset of~$\Z^d$. As is well
known, the number of such sets of size~$n$ containing the origin
grows only exponentially with~$n$. Since there are only finitely
many spin states, this shows that it is possible to choose $c_0$
as claimed.

Defining $a(\gamma)=1$ if~$\gamma$ is a single site and
$a(Y)=|Y|$ if $Y$ is a $q$-contour in~$\ssK_q$, the
assumption~\eqref{3.11} of Theorem~\ref{thmKP} is then satisfied.
(Note that assumption~\eqref{3.11} requires slightly less than
\eqref{c0def}, namely the analogue of \eqref{c0def}
with the exponent of $(1-c_0)|Y|$ instead of $(2-c_0)|Y|$;
the reason why we chose $c_0$ such that \eqref{c0def} holds will become
clear momentarily.)
Theorem~\ref{thmKP} guarantees that $\ZZ(\ssA)\ne0$ and
\eqref{3.12} holds for the corresponding cluster weights~$\zzT$.
Actually,   assumption~\eqref{3.11} is, for all $\eta\geq 0$,
 also satisfied  when  $\zz(Y)$ is replaced by $\zz(Y)e^{b(Y)}$
with $b(Y)=\eta |Y|$, yielding
\begin{equation}
\label{3.13aa}
\sum_{\begin{subarray}{c}
\ssX\in\MM(\ssK)\\\ssX\not\sim\gamma
\end{subarray}}
e^{b(\ssX)}\bigl|\zzT(\ssX)\bigr|\le a(\gamma)
\end{equation}
with $b(\ssX)=\eta \Vert\ssX\Vert$ instead of \eqref{3.13a}.
Using \eqref{3.13aa} with~$\gamma$ chosen to be the polymer represented by
the site at the origin and  observing that the
quantity $b(\ssX)$  exceeds~$\eta k$ for any cluster contributing to the sum
in \eqref{3.14}, we get the bound
\begin{equation}
e^{\eta k}\sum_{\begin{subarray}{c}
\ssX\in
\MM(\ssK_q)
\\
V(\ssX)\ni0,\, \Vert\ssX \Vert\ge k\end{subarray}}
\bigl|\zzT(\ssX)\bigr|
\le \sum_{\begin{subarray}{c}
\ssX\in
\MM(\ssK_q)
\\
V(\ssX)\ni0\end{subarray}}
\bigl|\zzT(\ssX)\bigr|e^{b(\ssX)}\le1,
\end{equation}
i.e., the bound \eqref{3.14}.

In order to prove the bounds \eqref{der3.14}, we
first notice that, in view of \eqref{zT} and \eqref{3.21} we have
\begin{equation}
\bigl|\partial_w\zzT(\ssX)\bigr|\le
\Vert\ssX\Vert\bigl|\zzT_0(\ssX)\bigr|\le
e^{\Vert\ssX\Vert}\bigl|\zzT_0(\ssX)\bigr|
\end{equation}
and
\begin{equation}
\bigl|\partial_w\partial_{w'}\zzT(\ssX)\bigr|
\le
\Vert\ssX\Vert^2\bigl|\zzT_0(\ssX)\bigr|\le
e^{\Vert\ssX\Vert}\bigl|\zzT_0(\ssX)\bigr|.
\end{equation}
Using \eqref{3.13aa} with
$b(Y)=(\eta+1)|Y|$
(which is also possible
since we choose $c_0$ such that \eqref{c0def}
holds as stated, instead of the weaker condition
where $(2-c_0)|Y|$ is replaced by $(1-c_0)|Y|$)
we get \eqref{der3.14} in the same way as \eqref{3.14}.
\end{proofsect}

\begin{proofsect}{Proof of Lemma~\ref{lemma3.7-cor}}
The bound \eqref{3.14} for $k=1$ immediately implies that the sum in
\eqref{3.19} converges with~$|s_q|\le e^{-\eta}$.
Using \eqref{3.12} and standard resummation techniques,
we rewrite the left hand side of \eqref{3.20} as
\begin{equation}
\label{3.20a}
\bigl|\log\ZZ(\ssK_{q,\Lambda})-s_q|\Lambda|\bigr|
=\Bigl|
\sum_{\begin{subarray}{c}
\ssX\in\MM(\ssK_q)\\ V(\ssX)\not\subset\Lambda
\end{subarray}}
\frac{|V(\ssX)\cap\Lambda|}{|V(\ssX)|}
\zzT(\ssX)\Bigr|.
\end{equation}
Next we note that for any cluster $\ssX\in\MM(\ssK_q)$,
the set $V(\ssX)$ is a connected subset
of~$\Z^d$,
which
follows immediately from
the observations that  $\supp Y\cup\Int Y$ is connected for
all contours $Y$, and
that
incompatibility of two contours
$Y,Y'$ implies that $\supp Y\cap\supp Y'\neq\emptyset$.
 Since only clusters
with $V(\ssX)\cap\Lambda\neq\emptyset$ and
$V(\ssX)\cap\Lambda^\cc\neq\emptyset$ contribute
to the right hand side of \eqref{3.20a}, we conclude
that the right hand side of \eqref{3.20a}
can be bounded by a sum over clusters $\ssX\in\ssK_q$
with $V(\ssX)\cap\partial\Lambda\neq\emptyset$.
Using this fact and the bound \eqref{3.14} with $k=1$,
\eqref{3.20} is proved.

Similarly, using the bounds \eqref{der3.14} in combination with
explicit expression \eqref{3.19}
in terms of absolutely converging cluster expansions, the claims
\eqref{3.22} immediately follow.
\end{proofsect}

\begin{remark}
The proof of Lemma~\ref{lemma3.7} holds without changes if
we replace the set of all $q$-contours in~$\Z^d$ by the set of
all $q$-contours on the torus $\T_L$.  This is not true, however,
for the proof of the bound \eqref{3.20} from Lemma~\ref{lemma3.7-cor}
since one also has to take into account the difference between
clusters wrapped around the torus and clusters in~$\Z^d$.
The corresponding modifications will be discussed in Section~\ref{sec4.4}.
\end{remark}

\section{Pirogov-Sinai analysis}
\label{sec4}\noindent The main goal of this section is to develop
the techniques needed to control the torus partition function.
Along the way we will establish some basic properties of the
metastable free energies which will be used to prove the
statements concerning the quantities~$\zeta_m$. Most of this
section concerns the contour model on~$\Z^d$. We will return to
the torus in Sections~\ref{sec4.4} and~\ref{sec5}.

All of the derivations in this section are based on assumptions
that are slightly more general than Assumption~C. Specifically, we
only make statements concerning a contour model satisfying the
following three conditions
(which depend on two parameters,~$\tau$ and~$M$):
\settowidth{\leftmargini}{(11)}
\begin{enumerate}
\item[(1)] The partition functions~$Z_q(\Lambda,z)$
and~$Z_L^\per(z)$ are expressed in terms of the energy
variables~$\theta_m(z)$ and contour weights~$\rho_z$ as stated in
\eqref{Z_q} and \eqref{ZL1}, respectively.
\item[(2)] The weights~$\rho_z$ of contours and contour networks
are translation invariant and are
twice continuously differentiable functions on~$\tilde\scrO$.
They obey the bounds
\begin{equation}
\label{Pei4-Y}
\bigl|\partial_z^\ell\partial_{\bar z}^{\bar\ell}
\rho_z(Y)\bigr|
\le (M|Y|)^{\ell+\bar\ell} e^{-\tau |Y|}\theta(z)^{|Y|},
\end{equation}
and
\begin{equation}
\label{Pei4-N}
\bigl|\partial_z^\ell\partial_{\bar z}^{\bar\ell}
\rho_z(\eusmN)\bigr|
\le (M|\eusmN|)^{\ell+\bar\ell}
e^{-\tau|\eusmN|}\theta(z)^{|\eusmN|}
\end{equation}
as long as $\ell,\bar\ell\geq 0$ and $\ell+\bar\ell\leq 2$.
\item[(3)] The energy variables~$\theta_m$ are twice continuously
differentiable functions on $\tilde\scrO$ and obey the bounds
\begin{equation}
\label{Pei4-Theta}
\bigl|\partial_z^\ell\partial_{\bar z}^{\bar\ell}
\theta_m(z)\bigr|
\le (M)^{\ell+\bar\ell}
\theta(z)
\end{equation}
as long as $\ell,\bar\ell\geq 0$ and $\ell+\bar\ell\leq 2$. We
will assume that~$\theta(z)$ is bounded uniformly from below
throughout~$\tilde\scrO$. However, we allow that some of
the~$\theta_m$ vanish at some~$z\in\tilde\scrO$.
\end{enumerate}
In particular, throughout this section we will not require that any of the
quantities~$\theta_m$,~$\rho_z(Y)$ or~$\rho_z(\eusmN)$ is analytic in~$z$.

\subsection{Truncated contour weights}
The key idea of contour expansions is that, for phases that are
thermodynamically stable, contours appear as heavily suppressed
perturbations of the corresponding ground states. At the points of
the phase diagram where all ground states lead to stable phases,
cluster expansion should then allow us to calculate all important
physical quantities. However, even in these special circumstances,
the direct use of the cluster expansion on \eqref{Z_q} is
impeded by the presence of the energy
terms~$\theta_m(z)^{|\Lambda_m(\Y)|}$ and, more seriously, by the
requirement that the contour labels match.

To solve these problems, we will express the partition function
in a form which does not involve any matching condition.
First we will rewrite
the sum in \eqref{Z_q} as
a sum
over mutually external contours $\Y^{\text{ext}}$
times a sum over collections of contours which are contained
in the interior of one of the contours in $\Y^{\text{ext}}$.
For a fixed contour $Y\in \Y^{\text{ext}}$, the sum
over all contours inside $\Int_mY$ then contributes
%
%
%
%
%
the factor $Z_m(\Int_mY,z)$, while the exterior
of the contours in $\Y^{\text{ext}}$ contributes
the factor
$\theta_m(z)^{|\Ext|}$,
where
$\Ext=\Ext_\Lambda(\Y^{\text{ext}})=
\bigcap_{Y\in\Y^{\text{ext}}}(\Ext
Y\cap\Lambda)$.
As a consequence, we can rewrite
the partition function \eqref{Z_q} as
\begin{equation}
\label{extcontrep}
Z_q(\Lambda,z)
=\sum_{\Y^{\text{ext}}}
\theta_q(z)^{|\Ext|}
\prod_{Y\in\Y^{\text{ext}}}\Bigl\{\,\rho_z(Y)\prod_m Z_m(\Int_mY,z)\Bigr\},
\end{equation}
where the sum goes over all collections of compatible external
$q$-contours in~$\Lambda$.

At this point, we use
an idea which originally goes back to \cite{KP1}.
Let us multiply
each term in the above sum by~$1$ in the form
\begin{equation}
1=\prod_{Y\in\Y^{\text{ext}}}\prod_m
\frac{Z_q(\Int_mY,z)}{Z_q(\Int_mY,z)}.
\end{equation}
Associating the partition functions in the denominator with the
corresponding contour, we get
\begin{equation}
Z_q(\Lambda,z)
=
\sum_{\Y^{\text{ext}}}
\theta_q(z)^{|\Ext|}
\prod_{Y\in\Y^{\text{ext}}}
\Big(\theta_q(z)^{| Y|}K_q(Y,z) Z_q(\Int Y,z)\Big),
\end{equation}
where~$K_q(Y,z)$ is given by
\begin{equation}
\label{Kq}
K_q(Y,z)=\rho_{z}(Y)\,\theta_q(z)^{-|Y|}
\prod_{m\in\CalS}\frac{Z_m(\Int_mY,z)}{Z_q(\Int_mY,z)}.
\end{equation}
Proceeding by induction, this leads to the representation
\begin{equation}
\label{zeestar}
Z_q(\Lambda,z)=\theta_q(z)^{|\Lambda|}
\sum_{\Y\in\CC(\Lambda,q)}\prod_{Y\in\Y}K_q(Y,z),
\end{equation}
where~$\CC(\Lambda,q)$ denotes the set of all collections of
non-overlapping $q$-contours in~$\Lambda$. Clearly, the sum on the
right hand side is exactly of the form needed to apply cluster
expansion, provided the contour weights satisfy the necessary
convergence assumptions.

Notwithstanding the appeal of the previous
construction, a bit of caution is necessary. Indeed, in order for
the weights~$K_q(Y,z)$ to be well defined, we are forced to
assume---or prove by cluster expansion, provided we somehow know
that the weights~$K_q$ have the required decay---that
$Z_q(\Int_mY,z)\ne0$. In the ``physical'' cases when the contour
weights are real and positive (and the ground-state energies are
real-valued), this condition usually
follows automatically.
However, here we are considering contour models with general complex
weights and, in fact, our ultimate goal is actually to look at
situations where a partition function vanishes.

Matters get
even more complicated whenever there is a ground
state which fails to yield a stable state (which is what happens
at a generic point of the phase diagram). Indeed, for such ground
states, the occurrence of a large contour provides a
mechanism for flipping from an unstable to a stable phase---which is
the favored situation once the volume gain of free energy exceeds
the energy penalty at the boundary. Consequently, the relative
weights of (large) contours in unstable phases are generally
large, which
precludes the use of the cluster expansion
altogether. A classic solution to this difficulty is to
modify the contour functionals for unstable
phases~\cite{Z1,BI,BK}. We will follow the strategy of \cite{BK},
where contour weights are truncated with the aid of a smooth
mollifier.

\smallskip
To introduce the truncated contours weights, let us consider a
$C^2(\R)$-function $x\mapsto\chi(x)$, such that
$0\le\chi(\cdot)\le1$, $\chi(x)=0$ for $x\le-2$ and $\chi(x)=1$
for $x\ge-1$. Let~$c_0$ be the constant from Lemma~\ref{lemma3.7}.
Using~$\chi$ as a regularized truncation
factor, we shall inductively define new contour weights
$\widetilde K_q'(\cdot,z)$ so that $|\widetilde K_q'(Y,z)|\le
e^{-(c_0+\tau/2)|Y|}$ for all $q$-contours $Y$. By
Lemma~\ref{lemma3.7},
 the associated partition
functions~$Z'_q(\cdot,z)$ defined by
\begin{equation}
\label{Zqprime}
Z_q'(\Lambda,z)=\theta_q(z)^{|\Lambda|}
\sum_{\Y\in\CC(\Lambda,q)}\prod_{Y\in\Y}\widetilde K_q'(Y,z)
\end{equation}
can then be controlled by cluster expansion.  (Of course, later we
will show that $\widetilde K_q'(\cdot,z)=K_q(\cdot,z)$ and
$Z_q'(\Lambda,z)=Z_q(\Lambda,z)$ whenever the ground state~$q$
gives rise to a stable phase.)

Let $\theta_q(z)\ne0$,
let~$Y$ be a~$q$-contour in~$\Lambda$, and
suppose that~$Z_m'(\Lambda',z)$ has been
defined by \eqref{Zqprime}
for all
$m\in\CalS$  and all $\Lambda'\subsetneqq\Lambda$.
Let us further assume by induction that
$Z_q'(\Lambda',z)\neq 0$
for all
$m\in\CalS$  and all $\Lambda'\subsetneqq\Lambda$.
We then define a smoothed cutoff function
$\phi_q(Y,z)$ by
\begin{equation}
\label{phi}
\phi_q(Y,z)=\prod_{m\in\CalS}\chi_{q;m}(Y,z),
\end{equation}
where
\begin{equation}
\label{chiqm}
\chi_{q;m}(Y,z)=
\chi
\left(\frac \tau 4+
\frac 1{|Y|}
\log\left|{\textstyle\frac{Z_q'(\Int Y,z)\theta_q(z)^{|Y|}}
{Z_m'(\Int Y,z)\theta_m(z)^{|Y|}}}
\right|
\right).
\end{equation}
Here $\chi_{q;m}(Y,z)$ is interpreted as $1$ if $\theta_m(z)$
or $Z_m'(\Int Y,z)$
is zero.

As a consequence of the above definitions and the fact that
$\Int_mY\subsetneqq\Lambda$ for all~$m\in\CalS$, the expressions
\begin{equation}
\label{Kprime}
K_q'(Y,z)=\rho_z(Y)\,\theta_q(z)^{-|Y|}\phi_q(Y,z)
\prod_{m\in\CalS}\frac{Z_m(\Int_mY,z)}{Z_q'(\Int_mY,z)}
\end{equation}
and
\begin{equation}
\label{tildeKprime} \widetilde  K_q'(Y,z)=\begin{cases}
K_q'(Y,z),\quad&\text{if }|K_q'(Y,z)|\le e^{-(c_0+\tau/2)|Y|},
\\
0, &\text{otherwise},
\end{cases}
\end{equation}
are meaningful for all~$z$ with $\theta_q(z)\ne0$.  By
Lemma~\ref{lemma3.7} we now know that $Z_q'(\Lambda,z)\ne0$ and
the inductive definition can proceed.

In the exceptional case $\theta_q(z)=0$, we let
$\widetilde K_q'(\cdot,z)=K_q'(\cdot,z)\equiv0$ and
$Z_q'(\cdot,z)\equiv0$. Note that this is consistent
with~$\phi_q(Y,z)\equiv0$.

\begin{remark}
Theorem~\ref{thm4.2} stated and proved below will ensure that
$|K_q'(Y,z)|<e^{-(c_0+\tau/2)|Y|}$ for all~$q$-contours~$Y$ and
all $q\in\CalS$,
provided $\tau\geq 4c_0 +16$.
Hence, as it turns out \emph{a posteriori}, the
second alternative in \eqref{tildeKprime} never occurs and, once
we are done with the proof of Theorem~\ref{thm4.2}, we can safely
replace~$\widetilde K'_q$ everywhere by~$K'_q$.
The additional truncation allows us to define and use the relevant
metastable free energies before stating and proving the (rather involved)
Theorem~\ref{thm4.2}. An alternative strategy would be to define
scale dependent free energies as was done e.g.~in~\cite{BK}.
\end{remark}

\subsection{Metastable free energies}
Let us rewrite
$Z_q'(\Lambda,z)$ as
\begin{equation}
\label{Z-ZZ}
Z_q'(\Lambda,z)=\theta_q(z)^{|\Lambda|}\ZZ_q'(\Lambda,z)
\end{equation}
where
\begin{equation}
\label{ZZ-def}
\ZZ_q'(\Lambda,z)=
\sum_{\Y\in\CC(\Lambda,q)}\prod_{Y\in\Y}
\widetilde K_q'(Y,z).
\end{equation}
We then define
\begin{equation}
\label{zetadef}
\zeta_q(z)
=\theta_q(z)e^{s_q(z)},
\end{equation}
where
\begin{equation}
\label{sdef}
s_q(z)
=\lim_{|\Lambda|\to\infty,\,\frac{|\partial\Lambda|}{|\Lambda|}\to 0}
\frac 1{|\Lambda|}\log
\ZZ'_q(\Lambda,z)
\end{equation}
By Lemma~\ref{lemma3.7-cor}, the partition functions
$\ZZ_q'(\Lambda,z)$ and the polymer pressure $s_q(z)$
can be analyzed by a convergent cluster
expansion, leading to the following lemma.

\begin{lemma}
\label{lemma4.1}
For each $q\in\CalS$ and each $z\in\tilde\scrO$,
the van~Hove limit
\eqref{sdef} exists and obeys the  bound
\begin{equation}
\label{thetazeta}
|s_q(z)|
\le e^{-\tau/2}.
\end{equation}
If~$\Lambda$ is a finite subset
of~$\Z^d$ and $\theta_q(z)\neq 0$,
we further have that
$Z'_q(\Lambda,z)\neq 0$ and
\begin{equation}
\label{logZ-f}
\bigl| \log \bigl(\zeta_q(z)^{-|\Lambda|}
Z'_q(\Lambda,z)\bigr)\bigr|\le e^{-\tau/2}
|\partial\Lambda|,
\end{equation}
while $\zeta_q(z)=0$ and $Z'_q(\Lambda,z)=0$ if
$\theta_q(z)=0$.
\end{lemma}

\begin{proofsect}{Proof}
Recalling the definition of compatibility between $q$-contours
from the paragraph before Lemma~\ref{lemma3.7}, $\CC(\Lambda,q)$
is exactly the set of all compatible collections of $q$-contours
in~$\Lambda$. Using the bound  \eqref{tildeKprime},
the statements of the lemma
are now direct consequences of
Lemma~\ref{lemma3.7-cor},
the definition \eqref{zetadef},
the representation \eqref{Z-ZZ} for $Z'_q(\Lambda,z)$
and the fact that we set $\widetilde K'_q(Y,z)=0$
if $\theta_q(z)=0$.
\end{proofsect}

The logarithm of~$\zeta_q(z)$---or at least its real part---has a
natural interpretation as the \emph{metastable free energy} of the
ground state~$q$. To state our next theorem, we actually need to
define these (and some other) quantities explicitly:
For each $z\in\tilde\scrO$ and each $q\in\CalS$ with
$\theta_q(z)\neq 0$, let
\begin{equation}
\label{fdef}
\begin{aligned}
f_q(z)&=-\log|\zeta_q(z)|,\\
f(z)&=\min_{m\in\CalS} f_m(z),\\
a_q(z)&=f_q(z)-f(z).
\end{aligned}
\end{equation}
If $\theta_q(z)=0$, we set $f_q(z)=\infty$ and $a_q=\infty$. (Note
that $\sup_{z\in\tilde\scrO}f(z)<\infty$ by \eqref{zetadef}, the
bound \eqref{thetazeta} and our assumption that
$\theta(z)=\max_q|\theta_q(z)|$ is bounded away from zero.)

In accord with our previous definition, a phase~$q$
is
stable at~$z$ if $a_q(z)=0$.
We will also
say
that a $q$-contour~$Y$ is \emph{stable at~$z$} if
$K'_q(Y,z)=K_q(Y,z)$. As we will see, stability of the
phase $q$ implies that all $q$-contours are stable.
Now we can formulate an analogue of
Theorem~3.1 of~\cite{BI} and Theorem~1.7 of \cite{Z1}.

\begin{theorem}
\label{thm4.2}
Suppose that $\tau\geq 4c_0+16$ where $c_0$ is the constant from
Lemma~\ref{lemma3.7}, and let
$\tilde\epsilon=e^{-\tau/2}$. Then  the following holds for all
$z\in\tilde\scrO$:
\settowidth{\leftmargini}{(iiiii)}
\begin{enumerate}
\item[(i)\,\,\,]
For all $q\in\CalS$ and all~$q$-contours\/~$Y$, we have $|K_q'(Y,z)|<
e^{-(c_0+\tau/2)|Y|}$ and, in particular, $\widetilde
K_q'(Y,z)=K_q'(Y,z)$.
%
%
\item[(ii)\;]
\label{cor4.4}
If~$Y$ is a $q$-contour
with~$a_q(z)\diam Y\le\frac \tau 4$, then
$K_q'(Y,z)=K_q(Y,z)$.
\item[(iii)]
If $a_q(z)\diam\Lambda\le \frac \tau 4$, then
$Z_q(\Lambda,z)=Z'_q(\Lambda,z)\ne0$ and
\begin{equation}
\label{thm4.2.ii}
\left|{Z_q(\Lambda,z)}\right|\ge e^{-f_q(z)|\Lambda|-
\tilde\epsilon|\partial\Lambda|}.
\end{equation}
\item[(iv)]
If $m\in\CalS$,
then
\begin{equation}
\label{thm4.2.iv}
\left|
Z_m(\Lambda,z) \right|
\le
e^{-f(z)|\Lambda|}
e^{2\tilde\epsilon|\partial\Lambda|}.
\end{equation}
\end{enumerate}
\end{theorem}

Before proving Theorem~\ref{thm4.2}, we state and prove the
following simple lemma which will be used both in the proof
of Theorem~\ref{thm4.2} and in the proof of Proposition~\ref{prop4.3}
in the next subsection.

\begin{lemma}
\label{cor4.3}
Let~$m,q\in\CalS$, let~$z\in\tilde\scrO$ and let~$Y$ be a~$q$-contour.
\settowidth{\leftmargini}{(iiiii)}
\begin{enumerate}
\item[(i)\,\,\,]
If $\phi_q(Y,z)>0$, then
\begin{equation}
\label{eq2a}
a_q(|\Int Y|+|Y|)
\leq (\tau/4+2+4e^{-\tau/2})|Y|.
\end{equation}
\item[(ii)\,] If
$\phi_q(Y,z)>0$ and $\chi_{q;m}(Y,z)<1$, then
\begin{equation}
\label{eq2a-ii}
a_m(|\Int Y|+|Y|)
\leq (1+8e^{-\tau/2})|Y|.
\end{equation}
\end{enumerate}
\end{lemma}

\begin{proofsect}{Proof of Lemma~\ref{cor4.3}}
By the definitions \eqref{phi} and \eqref{chiqm},
the
condition $\phi_q(Y,z)>0$ implies that
\begin{equation}
\label{eq2}
\max_{n\in\CalS}\log
\left|\frac{Z'_n(\Int Y,z)\theta_n(z)^{|Y|}}
{Z'_q(\Int Y,z)\theta_q(z)^{|Y|}}\right|
\le( 2+\tau/4)|Y|.
\end{equation}
Next we observe that $\phi_q(Y,z)>0$ implies $\theta_q(z)\neq 0$.
Since the maximum in \eqref{eq2} is clearly
attained
for some $n$ with $\theta_n(z)\neq 0$, we may
use the bound \eqref{logZ-f} to estimate the partition functions
on the left hand side of \eqref{eq2}.  Combined with
\eqref{zetadef},
\eqref{thetazeta}, \eqref{fdef} and the estimate
$|\partial\Int Y|\leq |Y|$,
this immediately gives the bound \eqref{eq2a}.

Next we use that the condition $\chi_{q;m}(Y,z)<1$ implies that
\begin{equation}
\label{eq2aa}
\log\left|\frac{Z_m'(\Int Y,z)\theta_m(z)^{|Y|}}
{Z_q'(\Int Y,z)\theta_q(z)^{|Y|}}\right|
\geq
(1+\tau/4)|Y|.
\end{equation}
Since \eqref{eq2aa} is not consistent with $\theta_m(z)=0$,
we may again use
\eqref{logZ-f}, \eqref{zetadef}, \eqref{thetazeta} and \eqref{fdef}
to estimate the left hand side, leading to the bound
\begin{equation}
\label{eq2aaa}
(f_q-f_m)(|\Int Y|+|Y|)
\geq
(\tau/4+1-4e^{-\tau/2}){|Y|}.
\end{equation}
Combining \eqref{eq2aaa} with \eqref{eq2a} and expressing
$a_m$ as $a_q-(f_q-f_m)$, one easily
obtains the bound \eqref{eq2a-ii}.
\end{proofsect}

As in \cite{BI}, Theorem~\ref{thm4.2} is proved using induction on
the diameter of~$\Lambda$ and $Y$. The initial step for the
induction, namely,~(i-ii) for $\diam Y=1$---which is trivially
valid since no such contours exist---and (iii-iv) for
$\diam\Lambda=1$, is established by an argument along the same
lines as that which drives the induction, so we just need
to
prove the
induction step. Let $N\ge1$ and suppose that the claims (i-iv)
have been established (or hold automatically) for all
$Y',\Lambda'$ with $\diam Y', \diam\Lambda'<N$.
Throughout the proof we will omit the argument~$z$ in~$f_m(z)$
and~$a_m(z)$.

\smallskip
The proof of the induction step is given in four parts:

\begin{proofsect}{Proof of (i)}
Let~$Y$ be such that $\diam Y=N$. First we will show that the second
alternative in \eqref{tildeKprime} does not apply.
By the bounds \eqref{Pei4-Y} and
\eqref{thetazeta}, we have that
\begin{equation}
\label{eq1b} \left|\rho_z(Y)\theta_q(z)^{-|Y|}\right| \leq
e^{-\tau|Y|}\left(\frac{\theta(z)}{|\theta_q(z)|}\right)^{|Y|}
\leq e^{-(\tau-2\tilde\epsilon)|Y|}e^{a_q|Y|},
\end{equation}
while the inductive assumption (iv), the bound  \eqref{logZ-f}
and the fact
that
$\sum_m|\Int_mY|=|\Int Y|$ and
$\sum_m|\partial\Int_mY|=|\partial\Int Y|\leq |Y|$,
imply that
\begin{equation}\label{eq1c}
\left|\prod_{m\in\CalS}\frac{Z_m(\Int_mY,z)}{Z_q'(\Int_mY,z)}\right|
\leq e^{a_q|\Int Y|}e^{3\tilde\epsilon|Y|}.
\end{equation}
Assuming without loss of generality that $\phi_q(Y,z)>0$
(otherwise there is nothing to prove), we now combine the bounds
\eqref{eq1b} and \eqref{eq1c} with \eqref{eq2a} and the fact that
$\tilde\epsilon=e^{-\tau/2}\leq 2/\tau\leq  1/8$, to conclude
that
$|K_q'(Y,z)| \leq e^{-(\frac 34\tau-\frac 52-5\tilde\epsilon)|Y|}
< e^{-(\frac 34\tau-4)|Y|}$.
By the assumption
$\tau\geq 4c_0+16$, this
is bounded by $e^{-(c_0+\tau/2)|Y|}$, as desired.
\end{proofsect}

\begin{proofsect}{Proof of (ii)}
Let~$\diam Y=N$ and
suppose that~$Y$ is a~$q$-contour satisfying
$a_q\diam Y\leq \tau/4$.
Using
the bounds \eqref{thetazeta} and \eqref{logZ-f},
the definitions \eqref{zetadef} and \eqref{fdef}, and the fact
that
$|\partial\Int Y|\leq|Y|$
we can
conclude that
\begin{equation}
\begin{aligned}
\label{phione}
\max_{m\in\CalS}
\frac{1}{|Y|}
&\log\left|
\frac{Z_m'(\Int Y,z)\theta_m(z)^{|Y|}}{Z_q'(\Int Y,z)\theta_q(z)^{|Y|}}
\right|
\le
a_q{\frac{|\supp Y\cup\Int Y|}{| Y|}}
+4\tilde\epsilon
\le
\frac \tau 4+1.
\end{aligned}
\end{equation}
In the last inequality, we used the bound $|\supp Y\cup\Int Y|\le
|Y|\diam Y$, the assumption that $a_q\diam Y\leq \tau/4$ and the fact that
$4\tilde\epsilon\leq 1$.
We also used that $a_q<\infty$ implies $\theta_q\neq 0$,
which justifies the use of the bound \eqref{logZ-f}.
 By the definitions \eqref{phi}  and
\eqref{chiqm}, the bound \eqref{phione}
implies that $\phi_q(\Int Y,z)=1$. On the other
hand, $Z_q(\Int_mY,z)=Z'_q(\Int_mY,z)$ for all $m\in\CalS$ by the
inductive assumption (iii) and the fact that $\diam\Int_mY<\diam
Y = N$. Combined with the inductive assumption (i),  we infer that
$\widetilde K'_q(Y,z)=K'_q(Y,z)=K_q(Y,z)$.
\end{proofsect}

\begin{proofsect}{Proof of (iii)}
Let
$\Lambda\subset\Z^d$ be such that
$\diam\Lambda=N$
and~$a_q\diam\Lambda\le\tau/4$.
By
the fact that~(ii) is known to hold for all contours~$Y$ with~$\diam Y\le N$,
we have that $K'_q(Y,z)=K_q(Y,z)$ for all~$Y$ in~$\Lambda$, implying that
$Z_q(\Lambda,z)=Z_q'(\Lambda,z)$. Invoking  \eqref{logZ-f}
and \eqref{fdef}, the bound \eqref{thm4.2.ii}
follows directly.
\end{proofsect}

\begin{proofsect}{Proof of (iv)}
Let~$\Lambda$ be a subset of~$\Z^d$ with $\diam\Lambda=N$.
 Following~\cite{Z1,BI}, we will apply the cluster expansion only to
contours that are sufficiently suppressed and handle the other
contours by a crude upper bound. Given a compatible collection of
contours~$\Y$, recall that \emph{internal} contours are those
contained in~$\Int Y$ of some other $Y\in\Y$ while the others are
\emph{external}. Let us call an~$m$-contour~$Y$ \emph{small} if
$a_m \diam Y\le \tau/4$; otherwise we will call it \emph{large}. The
reason for this distinction is that if~$Y$ is small then it is
automatically stable.

Bearing in mind the above definitions, let us partition any
collection of contours
$\Y\in\MM(\Lambda,m)$
into three sets
$\Y^{\text{int}}\cup\Y_{\text{small}}^{\text{ext}}
\cup\Y_{\text{large}}^{\text{ext}}$ of internal, small-external
and large-external contours, respectively.
Fixing~$\Y_{\text{large}}^{\text{ext}}$ and resumming the remaining
two families of contours, the partition function $Z_m(\Lambda,z)$
can be recast in the form
\begin{equation}
\label{largecontdec}
Z_m(\Lambda,z)=\sum_{\widetilde\Y}Z^{\text{small}}_m(\Ext,z)
\prod_{Y\in\widetilde\Y}
\Bigl\{\rho_z(Y)\prod_{n\in\CalS}Z_n(\Int_nY)\Bigr\}.
\end{equation}
Here the sum runs over
all sets~$\widetilde\Y$ of
mutually external large~$m$-contours in~$\Lambda$, the symbol
$\Ext=\Ext_\Lambda(\widetilde{\Y})$ denotes the set
$\bigcap_{Y\in\widetilde\Y}(\Ext Y\cap\Lambda)$ and
$Z^{\text{small}}_m(\Ext,z)$ is the partition sum
in~$\Ext$ induced by~$\widetilde\Y$.
Explicitly,
$Z^{\text{small}}_m(\Lambda,z)$ is the quantity from~\eqref{Z_q}
with the sum restricted to the collections $\Y\in\MM(\Lambda,m)$
for which all external contours are small according to the above
definition.

In the special case where $\theta_m(z)=0$, all contours
are large by definition (recall that
$a_m=\infty$
if~$\theta_m$ vanishes) and the partition
function $Z^{\text{small}}_m(\Lambda,z)$ is defined to be zero unless
$\Lambda=\emptyset$, in which case we set it to one.
We will not pay special attention to the case $\theta_m=0$
in the sequel of this
proof, but as the reader may easily verify, all our estimates
remain true in this case, and can be formally derived by considering
the limit $a_m\to\infty$.

Using the inductive assumption (iv) to estimate the partition
functions $Z_n(\Int_n Y)$, the
Peierls condition
\eqref{Pei4-Y}
to bound the activities $\rho_z(Y)$, and the bound
\eqref{thetazeta} to estimate $\theta(z)$ by
$e^{-f}e^{\tilde\epsilon}$,
we~get
\begin{equation}
\begin{aligned}
\prod_{Y\in\widetilde\Y}
\Bigl\{\rho_z(Y)\prod_{n\in\CalS}Z_n(\Int_nY)\Bigr\}
&\leq
\prod_{Y\in\widetilde\Y}
\Bigl\{e^{-\tau|Y|}
e^{-f(|\Int Y|+|Y|)+3\tilde\epsilon |Y|}\Bigr\}
\\
&=
e^{-f|\Lambda\setminus\Ext|}
\prod_{Y\in\widetilde\Y}e^{-(\tau-3\tilde\epsilon)|Y|}.
\end{aligned}
\end{equation}
Next
we will estimate
the partition function $Z^{\text{small}}_m(\Ext,z)$.
Since all small~$m$-contours are stable by the inductive
hypothesis, this partition function can be analyzed by a
convergent cluster expansion.  Let us consider the ratio of
$Z^{\text{small}}_m(\Ext,z)$ and $Z_m'(\Ext,z)$.  Expressing the
logarithm of this ratio as a sum over clusters we obtain a sum
over clusters that contain at least one contour of size
$|Y|\ge\diam Y>\tau/a_m\geq2/a_m$.
Using the bound \eqref{3.14}
with~$\eta=\tau/2$
we conclude that
\begin{equation}
\Bigl|\frac{Z^{\text{small}}_m(\Ext,z)}{
Z_m'(\Ext,z)}\Bigr| \le e^{|\Ext|e^{-\tau/a_m}}.
\end{equation}
Combined with Lemma~\ref{lemma4.1}
and the definitions \eqref{fdef},
this gives
\begin{equation}
\label{ratio}
\left|Z^{\text{small}}_m(\Ext,z)
\right| \le
e^{-(f_m-e^{-\tau/a_m})|\Ext|}\,
e^{{\tilde\epsilon}|\partial\Lambda|}
\prod_{Y\in\widetilde\Y}e^{\tilde\epsilon|Y|}.
\end{equation}
We thus conclude that the left hand side of \eqref{largecontdec} is
bounded by
\begin{equation}
\begin{aligned}
\label{bound2}
\left| {Z_m(\Lambda,z)}\right|
&
\le
\max_{\widetilde\Y}
\biggl(e^{-(a_m/2)|\Ext|}
\prod_{Y\in\widetilde\Y}e^{-(\tau/4)|Y|}
\biggr)
\\
&\qquad\qquad\times
e^{-f|\Lambda|}e^{{\tilde\epsilon}|\partial\Lambda|}
\sum_{\widetilde\Y}e^{-b|\Ext|}
\prod_{Y\in\widetilde\Y}e^{-(3\tau/4-4\tilde\epsilon)|Y|},
\end{aligned}
\end{equation}
where $b=a_m/2-e^{-\tau/a_m}$.
Note that~$b\ge e^{-\tau/a_m}$
which is implied by
the fact that $4e^{-\tau/a_m}\le 4a_m/\tau\leq a_m$.

For the purposes of this proof, it suffices to
bound the first factor in \eqref{bound2} by~$1$.
In a later proof, however, we will use a more subtle bound.
To bound the second factor,
we will invoke Zahradn\'{\i}k's
method (see \cite[Main Lemma]{Z1}
or \cite[Lemma~3.2]{BI}): Consider the contour model with weights
$\widehat K(Y)=e^{-(3\tau/4-4)|Y|}$ if~$Y$ is a large~$m$-contour
and $\widehat K(Y)=0$ otherwise.  Let $\widehat Z(\Lambda)$ be
the corresponding polymer partition function
in~$\Lambda$---see \eqref{3.10}---and let~$\varphi$ be the
corresponding free energy.
Clearly $\widehat Z(\Lambda)\geq 1$ so that $-\varphi\geq 0$.
Since $3\tau/4-4\ge c_0+\tau/2$, we can use
Lemmas~\ref{lemma3.7} and~\ref{lemma3.7-cor} to obtain further bounds.
For the free energy, this gives
$0\le-\varphi\le \min\{\tilde\epsilon,e^{-\tau/a_m}\}$
because the weights of contours
smaller than~$2/a_m$ identically vanish.
Since $b\geq e^{-\tau/a_m}$, this allows us to bound
the sum on the right hand side of \eqref{bound2} by
\begin{equation}
\label{bound2-a}
\sum_{\widetilde\Y}e^{\varphi|\Ext|}
\prod_{Y\in\widetilde\Y}e^{-(3\tau/4-4\tilde\epsilon)|Y|}
\leq\sum_{\widetilde\Y}
e^{\varphi|\Ext|}
\prod_{Y\in\widetilde\Y}\Bigl\{
e^{\varphi|Y|}e^{-(3\tau/4-5\tilde\epsilon)|Y|}\Bigr\}.
\end{equation}
Using Lemma~\ref{lemma3.7-cor} once more, we have that
$\widehat Z(\Int Y)e^{\varphi|\Int Y|} e^{\tilde\epsilon|Y|}\geq 1$.
Inserting into \eqref{bound2-a}, we obtain
\begin{equation}
\begin{aligned}
\label{bound3} \sum_{\widetilde\Y}e^{-b|\Ext|}
\prod_{Y\in\widetilde\Y}e^{-(3\tau/4-4\tilde\epsilon)|Y|}
&\le
\sum_{\widetilde\Y}
e^{\varphi\left(|\Ext|+\sum_{Y\in\widetilde\Y}(|\Int Y|+|Y|)\right)}
\prod_{Y\in\widetilde\Y}
\Bigl\{\widehat Z(\Int Y)
\widehat K(Y)\Bigr\}
\\
&=e^{\varphi|\Lambda|}\sum_{\widetilde\Y}
\prod_{Y\in\widetilde\Y}
\Bigl\{\widehat Z(\Int Y)
\widehat K(Y)\Bigr\}.
\end{aligned}
\end{equation}
Consider, on the other hand, the polymer partition function
$\widehat Z(\Lambda)$ in the representation
\eqref{3.10}.  Resuming all contours but the external ones,
we obtain precisely the right hand side of \eqref{bound3},
except for the factor $e^{\varphi|\Lambda|}$.
This shows that the right hands side of \eqref{bound3}  is
equal to $\widehat Z(\Lambda)
e^{\varphi|\Lambda|}$ which---again by Lemma~\ref{lemma3.7-cor}---is
bounded by~$e^{\tilde\epsilon|\partial\Lambda|}$. Putting this
and \eqref{bound2} together we obtain the proof of the claim (iv).
\end{proofsect}

\subsection{Differentiability of free energies}
\label{sec4.3}\noindent
Our next item of concern will be the
existence of two continuous and bounded derivatives of the
metastable free energies.
To this end, we first prove the following proposition,
which establishes a bound of the form \eqref{thm4.2.iv} for the
derivatives of the partition functions $Z_m(\Lambda,z) $.

\begin{proposition}
\label{lem4.2v}
Let~$\tau$ and~$M$ be the constants from
\eqref{Pei4-Y} and \eqref{Pei4-Theta},
let~$\tilde\epsilon=e^{-\tau/2}$, and suppose that
$\tau\geq 4c_0+16$ where $c_0$ is the constant from
Lemma~\ref{lemma3.7}. Then
\begin{equation}
\label{thm4.2.v}
\left|
\partial_z^\ell \partial_{\bar z}^{\bar\ell}
Z_m(\Lambda,z) \right|
\le
e^{-f(z)|\Lambda|}
\bigl(2M|\Lambda|\bigr)^{\ell+\bar\ell}
e^{2\tilde\epsilon|\partial\Lambda|},
\end{equation}
holds for all
$z\in\tilde\scrO$, all $m\in\CalS$,
and all $\ell,\bar\ell\geq 0$ with
$\ell+\bar\ell\le 2$.
\end{proposition}

\begin{proofsect}{Proof}
Again, we proceed by induction on the diameter of~$\Lambda$.
We start from the representation \eqref{extcontrep} which we rewrite
as
\begin{equation}
\label{extcontrep-a}
Z_m(\Lambda,z)=\sum_{\Y^{\text{ext}}}\prod_{x\in\Ext}\theta_m(z)
\prod_{Y\in\Y^{\text{ext}}}Z(Y,z),
\end{equation}
where we abbreviated
$Z(Y,z)=\rho_z(Y)\prod_n Z_n(\Int_nY,z)$. Let
$1\le \ell<\infty$ be fixed (later, we will use that actually, $\ell\leq 2$)
and let us consider the impact of
applying~$\partial_z^\ell$ on~$Z_m(\Lambda,z)$. Clearly, each of
the derivatives acts either on some of~$\theta_m$'s, or on
some
of the~$Z(Y,z)$'s.
Let~$k_x$ be the number of times the term $\theta_m(z)$ is
differentiated ``at~$x$,'' and let~$i_Y$
be the number
of times the
factor $Z(Y,z)$ is
differentiated.
Let $\bk=(k_x)$ and $\bi=(i_Y)$
 be the corresponding multiindices. The resummation of all
contours~$Y$ for which $i_Y=0$ and~$k_x=0$ for all~$x\in\supp
Y\cup\Int Y$ then contributes a factor
$Z_m(\Ext_\Lambda(\overline\Y^{\text{ext}})\setminus\Lambda',z)$,
where we used $\overline\Y^{\text{ext}}$ to denote the set of all
those $Y\in\Y^{\text{ext}}$ for which $i_Y>0$,
$\Ext_\Lambda(\overline\Y^{\text{ext}})=
\Lambda\setminus\bigcup_{Y\in\overline\Y^{\text{ext}}}(\supp
Y\cup\Int Y)$, and  $\Lambda'= \{x\colon\, k_x>0\}$. (Remember the
requirement that no contour
in~$\Ext_\Lambda(\overline\Y^{\text{ext}})\setminus\Lambda'$
surrounds any of the ``holes.'') Using this notation, the result
of differentiating can be concisely written as
\begin{multline}
\label{1der}
\quad
\partial_z^\ell Z_m(\Lambda,z)=
\sum_{\overline\Y^{\text{ext}}}
\sum_{\Lambda'\subset\Ext_\Lambda(\overline\Y^{\text{ext}})}
Z_m(\Ext_\Lambda(\overline\Y^{\text{ext}})\setminus\Lambda',z)
\\
\times\sum_{\begin{substack}{ \bk,\bi\\
\bk+\bi=\ell}\end{substack}}
\frac{\ell!}{\bk!\,\bi!}
\,\prod_{x\in \Lambda'}\partial_z^{k_x}\theta_m(z)
\prod_{Y\in\overline\Y^{\text{ext}}}
\partial_z^{i_Y}Z(Y,z).
\quad
\end{multline}
Here the first sum goes over all collections (including the empty one)
$\overline\Y^{\text{ext}}$ of mutually external contours in
$\Lambda$ and the third sum goes over all pairs of multiindices
$(\bk,\bi)$, $k_x=1,2,\dots$, $x\in\Lambda'$, $i_Y=1,2,\dots$,
$Y\in\overline\Y^{\text{ext}}$. (The terms with $|\Lambda'| +
|\overline\Y^{\text{ext}}|>\ell$ vanish.) We write $\bk+\bi=\ell$
to abbreviate $\sum_x k_x+\sum_Y i_Y=\ell$ and use the
symbols~$\bk!$ and~$\bi!$  to denote the multi-index
factorials~$\prod_x k_x!$ and~$\prod_Y i_Y!$, respectively.

We now  use
\eqref{Pei4-Theta} and \eqref{thetazeta}
to bound
$|\partial_z^{k_x}\theta_m(z)|$ by
$(M)^{k_x}e^{\tilde\epsilon}e^{-f(z)}$. Employing
\eqref{Pei4-Y} and \eqref{thetazeta}
to bound
the  derivatives of $\rho_z(Y)$, and the inductive hypothesis to
bound the derivatives of $Z_m(\Int_m Y,z)$, we estimate
$|\partial_z^{i_Y}Z(Y,z)|$ by
$[2M|V(Y)|]^{i_Y}e^{-(\tau-3\tilde\epsilon)|Y|}
e^{-f(z)|V(Y)|}$
(recall that $V(Y)$ was defined as $\supp Y\cup \Int Y$).
Finally, we may use
the bound \eqref{thm4.2.iv}
to estimate
\begin{equation}
|Z_m(\Ext_\Lambda(\overline\Y^{\text{ext}})\setminus\Lambda',z)|\le
e^{2\tilde\epsilon|\partial(\Ext_\Lambda(\overline\Y^{\text{ext}})
\setminus\Lambda')}
e^{-f(z)|\Ext_\Lambda(\overline\Y^{\text{ext}})\setminus\Lambda'|}.
\end{equation}
Combining these estimates and invoking the
inequality
\begin{equation}
|\partial(\Ext_\Lambda(\overline\Y^{\text{ext}})\setminus\Lambda')|
\leq |\partial\Lambda|+
|\Lambda'|+
\sum_{Y\in\overline\Y^{\text{ext}}}|Y|,
\end{equation}
we get
\begin{multline}
\label{2der}
\left|{\partial_z^\ell Z_m(\Lambda,z)}\right| \le
e^{2\tilde\epsilon|\partial\Lambda|}e^{-f(z)|\Lambda|}
\sum_{\overline\Y^{\text{ext}}}
\sum_{\Lambda'\subset\Ext_\Lambda(\overline\Y^{\text{ext}})}
\sum_{\begin{substack}{ \bk,\bi\\
 \bk+\bi=\ell}\end{substack}}
\frac{\ell!}{\bk!\,\bi!}
\\
\times \prod_{x\in\Lambda'}
(M e^{3\tilde\epsilon})^{k_x}
\prod_{Y\in\overline\Y^{\text{ext}}}
\bigl(2M|V( Y)|\bigr)^{i_Y}e^{-(\tau-5\tilde\epsilon)|Y|}.
\qquad
\end{multline}

Let us now consider the case $\ell=1$ and $\ell=2$.  For $\ell=1$,
the sum on
the right
hand side of \eqref{2der} can be rewritten as
\begin{equation}
\label{ell=1}
\sum_{x\in\Lambda}
\Bigl(M e^{3\tilde\epsilon}+\sum_{Y:x\in V(Y)\subset\Lambda}
2Me^{-(\tau-5\tilde\epsilon)|Y|}\Bigr),
\end{equation}
while for $\ell=2$, it becomes
\begin{equation}
\label{ell=2}
\sum_{x,y\in\Lambda}
\Bigl(\bigl(M e^{3\tilde\epsilon})^2+
2M e^{3\tilde\epsilon}2M
\sum_{\begin{substack}{ Y:x\in\Lambda\setminus V(Y)\\
y\in V(Y)\subset \Lambda}\end{substack}}
e^{-(\tau-5\tilde\epsilon)|Y|}
+(2M)^2
\sum_{\overline\Y^{\text{ext}}}
\prod_{Y\in\overline\Y^{\text{ext}}}
e^{-(\tau-5\tilde\epsilon)|Y|}
\Bigr),
\end{equation}
where the last sum goes over sets of mutually external contours
$\overline\Y^{\text{ext}}$ in~$\Lambda$ such that
$\{x,y\}\subset \bigcup_{Y\in \overline\Y^{\text{ext}}}V(Y)$
and $\{x,y\}\cap V(Y)\neq\emptyset$ for each $Y\in \overline\Y^{\text{ext}}$.
Note that the last condition can only
be satisfied if $\overline\Y^{\text{ext}}$ contains either one or
two contours. Introducing the shorthand
\begin{equation}
S=\sum_{Y:0\in V(Y)\subset\Z^d}e^{-(\tau-5\tilde\epsilon)|Y|}
\end{equation}
we bound the expression \eqref{ell=1} by
$(e^{3\tilde\epsilon}+2S)M|\Lambda|$, and the
expression \eqref{ell=2} by
$(e^{6\tilde\epsilon}+4 e^{3\tilde\epsilon}S
+ 4(S+S^2))M^2|\Lambda|^2$.  Recalling that
$c_0$ was defined in such a way that the bound
\eqref{c0def} holds, we may now use the
fact that $\tau-5\tilde\epsilon-c_0\geq\frac 12\tau$
to bound $S$ by $e^{-2}\tilde\epsilon $.  Since
$\tilde\epsilon\leq 1/8$, this implies that
the above two terms can be estimated by
$(e^{3/8}+\frac 14 e^{-2})M|\Lambda|\leq
2M|\Lambda|$
and $(e^{6/8}+\frac 12 e^{3/8-2}
+ \frac 12(e^{-2}+\frac 18 e^{-4}))M^2|\Lambda|^2
\leq 4M^2|\Lambda|^2$, as desired.

This completes the proof for the derivatives
with respect to $z$.  The proof for the derivatives with
respect to $\bar z$ and the mixed derivatives is completely
analogous and is left to the reader.
\end{proofsect}

Next we will establish a bound on the
first two derivatives of the contour weights~$K_q'$.
Before formulating the next proposition, we recall the definitions
of the polymer partition function $\ZZ_q'(\Lambda,z)$
and the polymer pressure $s_q$ in
\eqref{sdef} and \eqref{ZZ-def} .

\begin{proposition}
\label{prop4.3}
Let~$\tau$ and~$M$ be the constants from
\eqref{Pei4-Y} and \eqref{Pei4-Theta},
let $c_0$ be the constant from Lemma~\ref{lemma3.7}, and
let~$\tilde\epsilon=e^{-\tau/2}$. Then there exists
%
%
a finite constant~$\tau_1\geq 4c_0+16$
depending only on $M$, $d$ and $|\CalS|$ such that
if~$\tau\ge\tau_1$,
the contour weights $K_q'(Y,\cdot)$ are twice
continuously differentiable in $\tilde\scrO$.
Furthermore,
the bounds
\begin{equation}
\label{derK}
\bigl|
\partial_z^\ell\partial_{\bar z}^{\bar\ell}
K_q'(Y,z)\bigr|\le e^{-(c_0+\tau/2)|Y|}
\end{equation}
and
\begin{equation}
\label{derZZ}
\bigl|
\partial_z^\ell\partial_{\bar z}^{\bar\ell}
\ZZ_q'(\Lambda,z)\bigr|\le |\Lambda|^{\ell+\bar\ell}
e^{s_q(z)|\Lambda|+\tilde\epsilon|\partial\Lambda|}
\end{equation}
hold for all $q\in\CalS$, all~$z\in\tilde\scrO$, all~$q$-contours~$Y$,
all finite $\Lambda\subset\Z^d$ and all $\ell,\bar\ell\ge 0$ with
$\ell+\bar\ell\leq 2$.
\end{proposition}

Proposition~\ref{prop4.3} immediately implies that the
polymer pressures $s_q$ are twice continuously differentiable
and obey the bounds of Lemma~\ref{lemma3.7-cor}.  For future
reference, we state this in the following corollary.

\begin{corollary}
\label{cor4.6}
Let $\tau_1$ be as in Proposition~\ref{prop4.3}.
If ~$\tau\geq \tau_1$
and $q\in\CalS$, then
$s_q$ is a twice continuously differentiable function in
$\tilde\scrO$ and obeys the bounds
\begin{equation}
\label{s-bounds}
\bigl|\partial_w s_q \bigr|\le e^{-\tau/2}
\quad\text{and}\quad
\bigl| \partial_w\partial_{w'}s_q \bigr|\le e^{-\tau/2},
\quad
\end{equation}
valid for any~$w,w'\in\{z,\bar z\}$ and any $z\in\tilde\scrO$.
\end{corollary}

\begin{proofsect}{Proof of Proposition~\ref{prop4.3}}
Let $\tau\geq\tau_1\geq 4c_0+16$.  Then
Theorem~\ref{thm4.2} is at our
disposal.
It will be convenient to cover the set $\tilde \scrO$
by the open sets
\begin{equation}
\tilde \scrO^{(q)}_1=\{z\in \tilde \scrO
\colon |\theta_q(z)|<e^{-(\tau/4+2+6\tilde\epsilon)}
\theta(z)\}
\end{equation}
and
\begin{equation}
\tilde \scrO^{(q)}_2=\{z\in \tilde \scrO
\colon |\theta_q(z)|>e^{-(\tau/4+2+8\tilde\epsilon)}
\theta(z)\}.
\end{equation}
We first note that
$K_q'(Y,z)=0$ if $z\in\tilde \scrO^{(q)}_1$.
Indeed, assuming $K_q'(Y,z)\neq 0$ we necessarily have
$\phi_q(Y,z)>0$, which, by \eqref{eq2a}, implies
that $a_q\leq \tau/4+2+4\tilde\epsilon$ and thus
$\log \theta(z)-\log|\theta_q(z)|\leq \tau/4+2+6\tilde\epsilon$,
which is incompatible with $z\in\tilde \scrO^{(q)}_1$.
Hence, the claims trivially hold in $\tilde\scrO^{(q)}_1$
and
it remains to prove
that
$K_q'(Y,\cdot)$ is twice continuously differentiable
in $\scrO^{(q)}_2$, and that \eqref{derK}
and \eqref{derZZ}
hold for all  $z\in\tilde\scrO^{(q)}_2$. As in the proof of
Theorem~\ref{thm4.2} we will proceed by induction on the diameter
of~$Y$ and~$\Lambda$. Let~$N\ge1$ and suppose that
$K_q'(Y,\cdot)\in C^2(\tilde\scrO^{(q)}_2)$ and obeys
the bounds
\eqref{derK}
 for all~$q\in\CalS$ and all~$q$-contours~$Y$ with~$\diam
Y<N$, and that \eqref{derZZ} holds for all~$q\in\CalS$ and all
$\Lambda\subset\Z^d$ with $\diam\Lambda<N-1$.

We start by proving
that $K_q'(Y,\cdot)\in C^2(\tilde\scrO^{(q)}_2)$
whenever $Y$ is a $q$-contour $Y$ of diameter $N$.  To this end,
we first observe that in $\tilde\scrO^{(q)}_2$,  we have that
$\theta_q(z)\neq 0$ and hence also $Z'_q(\Int Y,z)\neq 0$.
Using the inductive assumption,
this implies that the
quotient
\begin{equation}
Q_{m,Y}(z)=\frac{Z_m'(\Int Y,z)\theta_m(z)^{|Y|}}
{Z_q'(\Int Y,z)\theta_q(z)^{|Y|}}
\end{equation}
is twice continuously differentiable in $\tilde \scrO^{(q)}_2$, which in
turn implies that $\chi_{q;m}(Y,z)$ is twice continuously
differentiable.
Combined with the corresponding
continuous differentiability
of~$\rho_z(Y)$, $\theta_q(z)$,
$Z_m(\Int_mY,z)$, and $Z_q'(\Int_m Y,z)$,
this proves the existence of two continuous
derivatives of $z\mapsto K_q'(Y,z)$ with respect to both~$z$
and~$\bar z$.

Next we prove the bound \eqref{derZZ} for
$\diam\Lambda=N-1$.
As we will see, these bounds follow immediately from the inductive
assumptions \eqref{derK} and Lemma~\ref{lemma3.7-cor}.  Indeed,
let $\zz_q(Y)=K_q'(Y,z)$ if $\diam Y\leq N-1$, and
$\zz_q(Y)=0$ if $\diam Y> N-1$.  The inductive assumptions
\eqref{derK} then
guarantee the conditions \eqref{3.21} of
Lemma~\ref{lemma3.7-cor}.  Combining the representation
 \eqref{3.12} for $\log \ZZ_q'(\Lambda,z)$ with
the estimate  \eqref{der3.14} from Lemma~\ref{lemma3.7-cor}
we thus conclude that
\begin{equation}
\bigl|
\partial_z^\ell\partial_{\bar z}^{\bar\ell}
\log\ZZ_q'(\Lambda,z)\bigr|\le |\Lambda|\tilde\epsilon,
\end{equation}
while \eqref{3.20} gives the bound
\begin{equation}
\bigl|\ZZ_q'(\Lambda,z)\bigr|\le e^{s_q|\Lambda|
+\tilde\epsilon|\partial\Lambda|}.
\end{equation}
Combining these bounds with the estimates
$\tilde\epsilon|\Lambda|\leq|\Lambda|$
and $\tilde\epsilon^2|\Lambda|^2+\tilde\epsilon|\Lambda|\leq|\Lambda|^2$,
we obtain the desired bounds \eqref{derZZ}.

Before turning to the proof of \eqref{derK} we will show that
for $z\in\tilde\scrO^{(q)}_2$,
the bound \eqref{derZZ} implies
\begin{equation}
\label{derZ}
\bigl|
\partial_z^\ell\partial_{\bar z}^{\bar\ell}
 Z_q'(\Lambda,z)\bigr|
\le \Bigl(M_1e^{\tau/4+3}|\Lambda|\Bigr)^{\ell+\bar\ell}
e^{-f_q(z)|\Lambda|+\tilde\epsilon|\partial\Lambda|}
\end{equation}
with $M_1=1+M$.
Indeed, invoking the assumption
\eqref{Pei4-Theta},
the definition of $\tilde\scrO^{(q)}_2$, and the fact that
$\tilde\epsilon\leq 1/8$, we may estimate
the first and second derivative of
$\theta_q(z)^{|\Lambda|}$ by
\begin{equation}
\label{d1-thetaL}
\left|\partial_z^\ell\partial_{\bar z}^{\bar\ell}
\theta_q(z)^{|\Lambda|}\right|
\leq
\left(M|\Lambda|
\frac{\theta(z)}{|\theta_q(z)|}\right)^{\ell+\bar\ell}
\,|\theta_q(z)|^{|\Lambda|}
\leq
\left(M |\Lambda|e^{\tau/4+3}
\right)^{\ell+\bar\ell}
\,|\theta_q(z)|^{|\Lambda|}.
\end{equation}
Combined with \eqref{Z-ZZ} and
\eqref{derZZ}
this gives \eqref{derZ}.

Let~$Y$ be a $q$-contour
with~$\diam Y=N$, and let us consider the
derivatives with respect to~$z$; the other derivatives are handled
analogously.
By the assumption \eqref{Pei4-Y}
and the bound
\eqref{thetazeta}, we have
\begin{equation}
\bigl|\partial_z^\ell\rho_z(Y)\bigr|
\le |Y|^\ell M^\ell e^{-(\tau-2\tilde\epsilon)|Y|}
e^{a_q|Y|}|\theta_q(z)|^{|Y|},
\end{equation}
while
\eqref{Pei4-Theta}
and the assumption that $z\in\tilde\scrO_2^{(q)}$
(cf~\eqref{d1-thetaL})
yields
\begin{equation}
\bigl|\partial_z^\ell\theta_q(z)^{-|Y|}\bigr|
\le
(|Y|+1)^\ell(Me^{\tau/4+3})^\ell
|\theta_q(z)\bigr|^{-|Y|}.
\end{equation}
Further,
combining the bound~\eqref{derZ}
with
Theorem~\ref{thm4.2}
and Proposition~\ref{lem4.2v}
we
have
\begin{equation}
\biggl|\partial_z^\ell\prod_{m\in\CalS}
\frac{Z_m(\Int_mY,z)}{Z_q'(\Int_mY,z)}\biggr|
\le
|\Int Y|^\ell
\bigl(2M+2M_1e^{2\tilde\epsilon|Y|}e^{3+\tau/4}\bigr)^\ell
e^{3\tilde\epsilon|Y|}e^{a_q|\Int Y|}.
\end{equation}
Finally, let us consider one of the factors $\chi_{q;m}(Y,z)$.
To bound its
derivative, we may assume
that~$z$ is an accumulation point of~$z'$ with $\chi_{q;m}(Y,z')<1$
(otherwise its derivative is zero), so by
Lemma~\ref{cor4.3}(ii) we have that
$a_m\leq 1+8\tilde\epsilon$
and thus
$\log \theta(z)-\log|\theta_m(z)|\leq
1+10\tilde\epsilon<\tau/4+2+8\tilde\epsilon$,
implying that $z\in\tilde\scrO^{(m)}_2$.
We may therefore use the bounds \eqref{d1-thetaL} and~\eqref{derZ}
to estimate the derivatives of
$\chi_{q;m}(Y,z)$, yielding the bound
\begin{equation}
\begin{aligned}
\bigl|\partial_z^\ell\chi_{q;m}(Y,z)\bigr|
&\le C(|\Int Y|+|Y|)^\ell
\left(4M_1e^{3+\tau/4}e^{2\tilde\epsilon|Y|}\right)^\ell
\end{aligned}
\end{equation}
where~$C$ is a constant bounding both the first and the second
derivative of the mollifier function~$\chi$. Combining all these
estimates, we obtain
a bound of the form
\begin{equation}
\bigl|\partial_z^\ell K_q'(Y,z)\bigr|
\le
\tilde C(|\Int Y|+|Y|)^\ell e^{\ell\tau/4}
e^{-(\tau-\tilde c\tilde\epsilon)|Y|}e^{a_q(|\Int Y|+|Y|)}
\end{equation}
with a constant~$\tilde C$ that
depends on $M$ and the number
of spin states $|\CalS|$, and a constant $\tilde c$
that depends only on $|\CalS|$.
Using the bound \eqref{eq2a}
and the fact that
$e^{\ell\tau/4}\leq e^{(\tau/8)|Y|}$
(note that $|Y|\geq (2R+1)^d> 4$
by our definition of contours),
we conclude that
\begin{equation}
\bigl|\partial_z^\ell K_q'(Y,z)\bigr|\leq
\tilde C(|\Int Y|+|Y|)^\ell
e^{-(5\tau/8-3-\tilde c\tilde\epsilon)|Y|}.
\end{equation}
Increasing~$\tau_1$ if necessary to absorb all of
the prefactors, the bound \eqref{derK}
follows.
\end{proofsect}

We close the subsection with
a lemma concerning the Lipschitz continuity of real-valued
functions~$z\mapsto f(z)$ and~$z\mapsto e^{-a_q(z)}$ on~$\tilde\scrO$:

\begin{lemma}
\label{lemLipf}
Let $\tau_1$ be as in Proposition~\ref{prop4.3} and let
$\widetilde M_1=4M+1$.
If~$\tau\geq \tau_1$, $q\in\CalS$, and
if $z,z_0\in\tilde\scrO$ are such that
$[z_0,z]=\{sz+(1-s)z_0\colon 0\le s\le1\}\subset\tilde\scrO$,
then
\begin{equation}
\label{fbound}
|f(z_0)-f(z)|\leq \widetilde M_1|z-z_0|
\end{equation}
and
\begin{equation}
\label{abound}
\bigl|e^{-a_q(z)}-e^{-a_q(z_0)}\bigr|\leq
2\widetilde M_1|z-z_0|\,e^{\widetilde M_1|z-z_0|}.
\end{equation}
\end{lemma}

\begin{proofsect}{Proof}
Let $\zeta_q(z)$ be the quantity defined in \eqref{zetadef},
and let $\tilde\epsilon=e^{-\tau/2}$.
Combining the assumption \eqref{Pei4-Theta} with the bounds
\eqref{s-bounds} and \eqref{thetazeta}, we get the estimate
\begin{equation}
\bigl|\partial_w \zeta_q(z)\bigr|\le
(Me^{2\tilde\epsilon} +\tilde\epsilon)
e^{-f(z)},
\quad w,w'\in\{z,\bar z\}.
\end{equation}
With the help of the bound $Me^{2\tilde\epsilon} +\tilde\epsilon
\leq 2M +1/2=\widetilde M_1/2$, we conclude that
\begin{equation}
\label{e^fq-bound}
|e^{-f_q(z_1)}-e^{-f_q(z_2)}|\leq
\widetilde M_1\int_{[z_1,z_2]} e^{-f(z')}\,|\textd z'|,
\qquad z_1,z_2\in[z_0,z],
\end{equation}
where $|\textd z'|$ denotes the Lebesgue
measure on the interval~$[z_0,z]$.
Using that~$f=\max_qf_q$, this implies
\begin{equation}
\label{e^f-bound}
|e^{-f(z_1)}-e^{-f(z_2)}|\leq
\widetilde M_1\int_{[z_1,z_2]} e^{-f(z')}\,|\textd z'|,
\qquad z_1,z_2\in[z_0,z].
\end{equation}
Now if \eqref{fbound} is violated, i.e.,
when~$|f(z)-f(z_0)|\ge(\widetilde M
_1
+\epsilon)|z-z_0|$,
then the same is true either about
the first or the second half of the segment~$[z_0,z]$ .
This shows
that
there is a sequence of intervals~$[z_{1,n},z_{2,n}]$
of length~$2^{-n}|z_0-z|$ where~$|f(z_{1,n})-f(z_{2,n})|
\ge(\widetilde M
_1
+\epsilon)|z_{1,n}-z_{2,n}|$.
But that would be in contradiction with \eqref{e^f-bound}
which implies that
\begin{equation}
\lim_{n\to\infty}\frac{|f(z_{1,n})-f(z_{2,n})|}{|z_{1,n}-z_{2,n}|}=
\lim_{n\to\infty}\frac{|e^{-f(z_{1,n})}-e^{-f(z_{2,n})}|}
{\int_{[z_{1,n},z_{2,n}]} e^{-f(z')}\,|\textd z'|}
\le\widetilde M_1,
\end{equation}
where we use the mean-value Theorem and a compactness argument to
infer the first equality. Hence, \eqref{fbound} must be true after
all.

To prove \eqref{abound}, we combine the triangle
inequality and the bound $f_q(z_0)\geq f(z_0)$ with
\eqref{e^fq-bound} and \eqref{e^f-bound}
to conclude that
\begin{equation}
\begin{aligned}
|e^{-a_q(z)}-e^{-a_q(z_0)}|
&=\bigl|e^{f(z)}e^{-f_q(z)}-e^{f(z_0)}e^{-f_q(z_0)}\bigr|
\\
&\leq
e^{f(z)}|e^{-f_q(z)}-e^{-f_q(z_0)}|
+\frac{e^{-f_q(z_0)}}{e^{-f(z)}e^{-f(z_0)}}|e^{-f(z_0)}-e^{-f(z)}|
\\
&\leq 2\widetilde M_1 \int_{z_0}^z e^{f(z)-f(z')} |\textd z'|.
\end{aligned}
\end{equation}
Bounding $f(z)-f(z')$ by $\widetilde M_1|z-z_0|$, we obtain the
bound \eqref{abound}.
\end{proofsect}

\subsection{Torus partition functions}
\label{sec4.4}\noindent In this subsection we consider the
partition functions $Z_q(\Lambda,z)$, defined for
$\Lambda\subset\T_L$ in \eqref{Z_q}. Since all contours
contributing to $Z_q(\Lambda,z)$ have diameter strictly less than
$L/2$, the partition function $Z_q(\Lambda,z)$ can be represented
in the form \eqref{zeestar}, with $K_q(Y,z)$ defined by embedding
the contour $Y$ into~$\Z^d$. Let $Z_q'(\Lambda,z)$ be the
corresponding truncated partition function, defined with weights
$K_q'(Y,z)$
given by \eqref{Kprime}.
Notice, however, that even though every contour $Y\subset\Lambda$
can be individually embedded  into~$\Z^d$, the relation of incompatibility
is formulated
on torus.
The polymer partition function $\ZZ_q'(\Lambda,z)$
and $Z_q'(\Lambda,z)$
can then again be analyzed by a convergent cluster expansion,
bearing in mind, however,  the torus
incompatibility relation.
The torus analogue of Lemma~\ref{lemma4.1} is then as follows:

\begin{lemma}
\label{lemma4.5}
Assume that $\tau\geq\tau_1$, where $\tau_1$ is the constant from
Proposition~\ref{prop4.3}
and let $q\in\CalS$ and $z\in\tilde\scrO$
be such that $\theta_q(z)\neq 0$.
Then
\begin{equation}
\label{logZT-f}
\left|\partial_w^\ell
\log\left(
\zeta_q(z)^{-|\Lambda|}
Z_q'(\Lambda,z)\right)\right|
\leq
e^{-\tau/2}|\partial\Lambda|+2|\Lambda| e^{-\tau L/4}
\end{equation}
for any~$\Lambda\subset\T_L$,
any~$z\in\tilde\scrO$,
$\ell=0,1$, and~$w\in\{z,\bar z\}$.
\end{lemma}

\begin{proofsect}{Proof}
Let us write $Z_q'(\Lambda,z)$ in the form \eqref{Z-ZZ}.
Taking into account the torus compatibility relation
when comparing the cluster expansion for $\log \ZZ_q'(\Lambda,z)$
with the corresponding terms contributing to $s_q |\Lambda|$,
we see that the  difference stems not only from clusters passing
through the boundary $\partial\Lambda$,
but also from the clusters that are wrapped around the torus in the former
as well as the clusters that cannot be placed on the torus in the latter.
For such clusters, however,
we necessarily have $\sum_Y \ssX(Y)|Y|\geq L/2$.
Since the functional~$\zz(Y)=K_q'(Y,z)$ satisfies the bound
\eqref{z<z_0} with $\eta=\tau/2$, we may use the bound
\eqref{3.14} to estimate the contribution of these
clusters. This yields
\begin{equation}
\label{sqL1}
\Bigl|\log \ZZ_q'(\Lambda,z)-s_q|\Lambda|\Bigr|
\leq e^{-\tau/2}|\partial\Lambda|+2|\Lambda| e^{-\tau L/4},
\end{equation}
which is \eqref{logZT-f} for~$\ell=0$.
To handle the
case~$\ell=1$, we just need to recall that,
by
Proposition~\ref{prop4.3}, the functional~$\zz(Y)=K_q'(Y,z)$
satisfies the
bounds \eqref{3.21} with~$\eta=\tau/2$.
Then
the desired estimate for~$\ell=1$ follows with help of
\eqref{der3.14} by a straightforward generalization of the
above proof of~\eqref{sqL1}.
\end{proofsect}

Next we provide the corresponding  extension of
Theorem~\ref{thm4.2} to the torus:

\begin{theorem}
\label{thm4.6}
Let $\tau\geq 4c_0+16$ where~$c_0$ is the constant from
Lemma~\ref{lemma3.7},
and let
us abbreviate $\tilde\epsilon=e^{-\tau/2}$.
For all~$z\in\tilde\scrO$, the following holds
for all subsets~$\Lambda$ of the torus~$\T_L$:
\settowidth{\leftmargini}{(iiiii)}
\begin{enumerate}
\item[(i)\;]
If $a_q(z)\diam\Lambda\le \frac \tau 4$, then
$Z_q(\Lambda,z)=Z'_q(\Lambda,z)\ne0$ and
\begin{equation}
\label{thm4.6.i}
\left|{Z_q(\Lambda,z)}\right|\ge
e^{-f_q(z)|\Lambda|}
e^{-
\tilde\epsilon|\partial\Lambda|-2|\Lambda|e^{-\tau L/4}
}.
\end{equation}
\item[(ii)]
If $m\in\CalS$, then
\begin{equation}
\label{thm4.6.ii}
\left|Z_m(\Lambda,z) \right|\le
e^{-f(z)|\Lambda|+2\tilde\epsilon|\partial\Lambda|+
4|\Lambda|e^{-\tau L/4}}.
\end{equation}
\item[(iii)]
If $m\in\CalS$, then
\begin{equation}
\label{thm4.6.iii}
\left|Z_m(\T_L,z) \right|\le
e^{-f(z)L^d}
\max\bigl\{e^{-a_m(z)L^d/2},e^{-\tau L^{d-1}/4}\bigr\}
e^{4L^de^{-\tau L/4}}.
\end{equation}
\end{enumerate}
\end{theorem}

\begin{remark}
The bounds \eqref{thm4.6.i} and \eqref{thm4.6.ii} are obvious
generalizations of the corresponding bounds in
Theorem~\ref{thm4.2} to the torus. But unlike in
Proposition~\ref{prop4.3}, we will not need to prove the bounds
for the derivatives with respect to~$z$. When such bounds will be
needed in the next section, we will invoke analyticity in~$z$ and
estimate the derivatives using Cauchy's Theorem.
\end{remark}

\begin{proofsect}{Proof of (i)}
Since all contours can by
definition be embedded into~$\Z^d$, Theorem~\ref{thm4.2}(ii) guarantees
that~$K_q'(Y,z)=K_q(Y,z)$ for all $q$-contours in~$\Lambda$ and hence
$Z_q(\Lambda,z)=Z'_q(\Lambda,z)$. Then \eqref{thm4.6.i} follows by
Lemma~\ref{lemma4.5} and the definition of~$f_q$.
\end{proofsect}

\begin{proofsect}{Proof of (ii)}
We will only indicate the changes relative to the proof of
part~(iv) of~Theorem~\ref{thm4.2}. First, since all contours can
be embedded into~$\Z^d$, we have that a corresponding bound---
namely, \eqref{thm4.2.iv}---holds for the interiors of all
contours in~$\Lambda$. This means that all of the derivation
\twoeqref{largecontdec}{bound2} carries over, with the exception
of the factor~$e^{\tilde\epsilon|\partial\Lambda|}$ in
\eqref{ratio} and \eqref{bound2} which by Lemma~\ref{lemma4.5}
should now be replaced
by~$e^{\tilde\epsilon|\partial\Lambda|+2|\Lambda|e^{-\tau L/4}}$.
In order to estimate the last sum in \eqref{bound2}, we will again
invoke the trick described in \twoeqref{bound2-a}{bound3}. This
brings in yet another
factor~$e^{\tilde\epsilon|\partial\Lambda|+2|\Lambda|e^{-\tau
L/4}}$. From here \eqref{thm4.6.ii} follows.
\end{proofsect}

\begin{proofsect}{Proof of (iii)}
The estimate is analogous to that in~(ii); the only
difference is that now we have to make use of the extra decay from
the maximum in \eqref{bound2}. (Note that for~$\Lambda=\T_L$ we
have~$|\partial\Lambda|=0$ and~$|\Lambda|=L^d$.)
Following \cite{BI}, this is done as follows: If~$Y$ is a contour, a
standard isoperimetric inequality yields
\begin{equation}
|Y|\ge
\frac 1{2d}|\partial(\supp Y\cup\Int Y)|\ge |\supp
Y\cup\Int Y|^{\frac{d-1}d}.
\end{equation}
Hence, if~$\widetilde\Y$ is a collection of external contours
in~$\T_L$ and~$\Ext$ is the corresponding exterior set, we have
\begin{equation}
\sum_{Y\in\widetilde\Y}|Y| \ge
\sum_{Y\in\widetilde\Y}|
\supp Y\cup\Int Y|^{\frac{d-1}d}\ge
\biggl(\sum_{Y\in\widetilde\Y}|
\supp Y\cup\Int Y|\biggr)^{\frac{d-1}d}\!\!\!\!
=\bigl(L^d-|\Ext|\bigr)^{\frac{d-1}d}.
\end{equation}
Writing~$|\Ext|=(1-x)L^d$ where~$x\in[0,1]$, the
maximum in \eqref{bound2} is
bounded by
\begin{equation}
\sup_{x\in[0,1]}\exp\Bigl\{-\frac{a_m}2L^d(1-x)-
\frac \tau4
L^{d-1}x^{\frac{d-1}d}\Bigr\}.
\end{equation}
The function in the exponent is convex
and the supremum is thus clearly dominated by the bigger of the values
at~$x=0$ and~$x=1$. This gives the maximum in \eqref{thm4.6.iii}.
\end{proofsect}

Apart from the partition functions~$Z_m(\T_L,z)$, we will also
need to deal with the situations where there is a non-trivial
contour network.
To this end, we need a suitable estimate on the difference
\begin{equation}
Z_L^{\text{\rm big}}(z)
=Z_L^\per(z)-
\sum_{m\in\CalS}Z_m(\T_L,z).
\end{equation}
This is the content of
the last lemma of this section.

\begin{lemma}
\label{lemma5.1}
There exists a constant $\tilde c_0$ depending only on~$d$ and
$|\CalS|$ such that
for $\tau\geq 4\tilde c_0+16$ and all~$z\in\tilde\scrO$, we have
\begin{equation}
\label{Zbigbd}
|Z_L^{\text{\rm big}}(z)|\le L^d e^{-\tau L/4}
e^{5L^d e^{-\tau L/4}}\zeta(z)^{L^d}.
\end{equation}
\end{lemma}

\begin{proofsect}{Proof}
Let $c_0$ be the constant from Lemma~\ref{lemma3.7},
and let $\tilde c_0=\tilde c_0(d,|\CalS|)\geq c_0$ be such that
\begin{equation}
\label{tildec_0def}
\sum_{\Lambda\subset\T_L}(|\CalS|e^{-c_0})^{|\Lambda|}
\leq L^d,
\end{equation}
where the sum goes over all connected subsets~$\Lambda$ of the
torus $\T_L$ (the existence of such a constant follows immediately
from the fact that the number of connected subsets $\Lambda\subset\Z^d$
that contain a given point $x$ and have size $k$ is bounded by
a $d$-dependent constant raised to the power $k$).

The proof of the lemma is
now a straightforward corollary of Theorem~\ref{thm4.6}.
Indeed, invoking
the representation \eqref{ZL2} we have
\begin{equation}
Z_L^{\text{\rm
big}}(z)=\sum_{\begin{subarray}{c}
(\emptyset,\eusmN)\in\MM_L\\\eusmN\ne\emptyset
\end{subarray}}
\rho_z(\eusmN)\,
\prod_{m\in\CalS}Z_m\bigl(\Lambda_m(\emptyset,\eusmN),z\bigr),
\end{equation}
where~$\Lambda_m(\emptyset,\eusmN)$ is defined before
Proposition~\ref{prop3.6}.
Using
\eqref{Pei4-N} and
\eqref{thm4.6.ii} in conjunction with the
bounds
$\theta(z)\le \zeta(z)e^{2\tilde\epsilon}$
and $\sum_{m\in\CalS}|\partial\Lambda_m(\emptyset,\eusmN)|\le|\eusmN|$,
we get
\begin{equation}
\label{4.74}
|Z_L^{\text{\rm big}}(z)|\le
\zeta(z)^{L^d}e^{4L^de^{-\tau L/4}}
\sum_{\begin{subarray}{c}
(\emptyset,\eusmN)\in\MM_L\\\eusmN\ne\emptyset
\end{subarray}}
e^{-(\tau-4\tilde\epsilon)|\eusmN|}.
\end{equation}
Taking into account that each connected component of
$\supp\eusmN$ has size at least $L/2$, the last sum can
be bounded by
%
\begin{equation}
\label{4.75}
\sum_{\begin{subarray}{c}
(\emptyset,\eusmN)\in\MM_L\\\eusmN\ne\emptyset
\end{subarray}}
e^{-(\tau-4\tilde\epsilon)|\eusmN|}
\leq
\sum_{n=1}^\infty \frac 1{n!}
S^n\leq Se^S
\end{equation}
where
\begin{equation}
\label{S-sum}
S=\sum_{\begin{subarray}{c}
\Lambda\subset\T_L\\|\Lambda|\geq L/2
\end{subarray}}
\Bigl(|\CalS|e^{-(\tau-4\tilde\epsilon)}\Bigr)^{|\Lambda|}
\end{equation}
is a sum over connected sets $\Lambda\subset\T_L$ of size at
least $L/2$.  Extracting a factor $e^{-\tau L/4}$ from
the right hand side of \eqref{S-sum}, observing that
$\tau/2-4\tilde\epsilon\geq\tilde c_0$, and recalling
that $\tilde c_0$ was defined in such a way that \eqref{tildec_0def}
holds, we get the estimate $S\leq L^de^{-\tau L/4}$.
Combined with \eqref{4.74} and \eqref{4.75} this gives
the desired bound \eqref{Zbigbd}.
\end{proofsect}

\section{Proofs of main results}
\label{sec5}\noindent
We are finally in a position to prove our main results.
Unlike in Section~\ref{sec4}, all of the derivations will assume
the validity of Assumption~C.
Note that the assumptions
\twoeqref{Pei4-Y}{Pei4-Theta} follow from Assumptions C0-C2,
so all results from Section~\ref{sec4} are at our disposal.
Note also that $\rho_z(Y)$, $\rho_z(\eusmN)$ and $\theta_m(z)$
are analytic functions of $z$ by Lemma~\ref{lemAnaly},
implying that the partition functions $Z_m(\Lambda,\cdot)$ and
$Z_L^\per$ are analytic functions of $z$.

%
%
We will prove Theorems~\ref{thmA} and~\ref{thmB} for
\begin{equation}
\tau_0=\max\{\tau_1,4\tilde c_0+16,2\log(2/\alpha)\}
\end{equation}
where $\tau_1$ is the constant from Proposition~\ref{prop4.3},
$\tilde c_0$ is the constant from Lemma~\ref{lemma5.1}
and $\alpha$ is the constant from Assumption C.
Recall that
$\tau_1\geq 4c_0+16$, so for $\tau\geq\tau_0$
we can use all results of
Section~\ref{sec4}.

First, we will attend to the proof of
Theorem~\ref{thmA}:

\begin{proofsect}{Proof of Theorem~\ref{thmA}}
Most of the required properties have already been established.
Indeed,
let~$\zeta_q$ be as defined in
\eqref{zetadef}. Then \eqref{zeta-theta} is exactly
\eqref{thetazeta}
which proves part~(1) of the Theorem~\ref{thmA}.

In order to prove that~$\partial_{\bar z}\zeta_q(z)=0$
whenever~$z\in\scrS_q$,
we
recall that $\zeta_q(z)=\theta_q(z)e^{s_q(z)}$ where~$\theta_q(z)$
is holomorphic in~$\tilde\scrO$ and~$s_q(z)$ is given in terms of
its Taylor expansion in the contour activities $K'_q(Y,z)$. Now,
if~$a_q(z)=0$---which is implied
by~$z\in\scrS_q$---then~$K_q'(Y,z)=K_q(Y,z)$ for
any~$q$-contour~$Y$ by Theorem~\ref{thm4.2}.
But $\partial_{\bar z}K_q(Y,z)=0$ by the fact
that $\rho_z(Y)$, $Z_q(\Int_m Y,z)$ and $Z_m(\Int_m Y,z)$
are holomorphic and $Z_q(\Int_m Y,z)\neq 0$.
Since~$s_q$ is given in
terms of an absolutely converging power series in the~$K_q$'s, we
thus also have that~$\partial_{\bar z}e^{s_q(z)}=0$.
Hence $\partial_{\bar z}\zeta_q(z)=0$ for all~$z\in\scrS_q$.

To prove part~(3), let~$z\in\scrS_m\cap\scrS_n$ for some distinct
%
%
indices~$m,n\in\RR$.
Using
Lemma~\ref{lemma4.1}
we then have
\begin{equation}
\theta_m(z)\ge \theta(z) e^{-2e^{-\tau/2}}
\end{equation}
and similarly for~$n$.
Since~$\alpha\ge2e^{-\tau_0/2}\ge2e^{-\tau/2}$,
we thus have
$z\in\scrL_\alpha(m)\cap\scrL_\alpha(n)$.
Using the first bound in \eqref{s-bounds}, we further have
\begin{equation}
\biggr|\frac{\partial_z\zeta_m(z)}{\zeta_m(z)}-
\frac{\partial_z\zeta_n(z)}{\zeta_n(z)} \biggl|
\ge\bigl|\partial_ze_m(z)-\partial_ze_n(z)\bigr|-2e^{-\tau/2}.
\end{equation}
Applying Assumption~C3, the right hand side is not less
than~$\alpha-2e^{-\tau/2}$. Part~(4) is proved analogously; we
leave the details to the reader.
\end{proofsect}

Before proving Theorem~\ref{thmB}, we prove the following lemma.
\begin{lemma}
\label{lemCauchyDerBounds}
Let $\epsilon>0$, let $\tau_1$ be the
constant from Proposition~\ref{prop4.3}, and let
\begin{equation}
\label{sqLdef}
s_q^{(L)}(z)=\frac 1{|\Lambda|}\log\ZZ_q'(\T_L,z)
\end{equation}
and
\begin{equation}
\label{zetaqLdef}
\zeta_q^{(L)}(z)=\theta_q(z)e^{s_q^{(L)}(z)}.
\end{equation}
Then there exists a constant $M_0$ depending only on $\epsilon$
and $M$
such that
\begin{equation}
\label{logzetaL-der}
{\bigl|\partial_z^\ell\zeta_q^{(L)}(z)\bigr|}
\leq (\ell!)^2 (M_0)^\ell{\bigl|\zeta_q^{(L)}(z)\bigr|}
\end{equation}
holds for all $q\in\CalS$, all $\ell\geq 1$,
all $\tau\geq\tau_1$,
all $L\geq\tau/2$ and all $z\in\tilde\scrO$
with $a_q(z)\leq \tau/(4L)$ and $\dist(z,\tilde\scrO^\cc)\geq \epsilon$.
\end{lemma}
\begin{proofsect}{Proof}
We will prove the lemma withe the help of Cauchy's theorem.   Starting with
the derivatives of $\theta_q$, let
$\epsilon_0=\min\{\epsilon,1/(4\widetilde M_1)\}$
where $\widetilde M_1=1+4  M$ is the constant from Lemma~\ref{lemLipf},
and let $z'$ be a point in the disc $\D_{\epsilon_0}(z)$
of radius $\epsilon_0$ around $z$.
Using the bounds \eqref{thetazeta}
and \eqref{fbound}, we now bound
\begin{equation}
\bigl|\theta_q(z')\bigr|
\leq e^{\tilde\epsilon-f(z')}
\leq e^{\tilde\epsilon+\widetilde M_1\epsilon_0}e^{-f(z)}
\leq
e^{\tilde\epsilon+\widetilde M_1\epsilon_0+a_q(z)}e^{-f_q(z)}
\leq |\theta_q(z)| e^{2\tilde\epsilon+\widetilde M_1\epsilon_0+a_q(z)}.
\end{equation}
With the help of Cauchy's theorem and the estimates
$\tilde\epsilon\leq 1/8$, $\widetilde M_1\epsilon\leq 1/4$
and $a_q(z)\leq 1/2$, this implies
\begin{equation}
\label{logthetader}
\frac{\bigl|\partial_z^\ell\theta_q(z)\bigr|}
{\bigl|\theta_q(z)\bigr|}
\leq \ell! \epsilon_0^{-\ell} e^{1/4+1/4+1/2}
\leq \ell! (2\epsilon_0^{-1})^\ell
.
\end{equation}
In order to bound the derivatives of $s_q^{(L)}$,
let us consider a multiindex $\ssX$ contributing
to the cluster expansion of $s_q^{(L)}$, and let
$k=\max_{Y:\ssX(Y)>0}\diam Y$.
Defining
\begin{equation}
\epsilon_k=\min\{\epsilon,(20 e \widetilde M_1k )^{-1}\},
\end{equation}
where
$\widetilde M_1=1+4 M$ is the constant from Lemma~\ref{lemLipf},
we will show that the weight $K_q'(Y,\cdot)$
of any contour $Y$ with $\ssX(Y)>0$ is analytic inside the disc
$\D_{\epsilon_k}(z)$
of radius $\epsilon_k$
about $z$.
Indeed, let $|z-z'|\leq \epsilon_k$.  Combining the assumption
$a_q(z)\leq \tau/(4L)\leq 1/2$ with Lemma~\ref{lemLipf}, we have
\begin{equation}
\label{aqzz'}
\begin{aligned}
e^{-a_q(z')}
&\geq e^{-a_q(z)}-2e\widetilde M_1\epsilon_k
\geq 1-a_q(z)-2e\widetilde M_1\epsilon_k
\\
&\geq1-\frac 65\max\{a_q(z),10e\widetilde M_1\epsilon_k\}
\geq  e^{-2\max\{a_q(z),10e\widetilde M_1\epsilon_k\}}.
\end{aligned}
\end{equation}
Here we used the fact that $x+y\leq \frac 65\max\{x,5y\}$
whenever $x,y\geq 0$ in the last but one step, and the fact
that $e^{-2x}\leq 1-(1-e^{-1})2x\leq 1-\frac 65 x$ whenever $x\leq 1/2$
in the last step.  We thus have proven that
\begin{equation}
\label{aqz'k}
a_q(z')\leq \max\{2a_q(z),20e\widetilde M_1\epsilon_k\}
\leq\max\bigl\{\tfrac\tau{2L},\tfrac1k\bigr\}\leq\frac\tau{4k},
\end{equation}
so by Theorem~\ref{thm4.2}, $K_q'(Y,z')=K_q(Y,z')$
and $Z_q(\Int_mY,z')\neq 0$ for all $m\in\CalS$ and
$z'\in \D_{\epsilon_k}(z)$.
As a consequence, $K_q'(Y,\cdot)$ is analytic inside the disc
$\D_{\epsilon_k}(z)$, as claimed.

At this point, the proof of the lemma is an easy exercise.  Indeed,
combining Cauchy's theorem with the bound
$|K_q'(Y,z')|\leq e^{-(\tau/2+c_0)|Y|}\leq
e^{-c_0|Y|}e^{-(\tau/2)\diam Y}$, we get the
estimate
\begin{equation}
\left|\partial_z^\ell\prod_Y K_q'(Y,z')^{\ssX(Y)}\right|\leq
\ell!\epsilon_k^{\ell}
\prod_Y e^{-(c_0+\tau/2)|Y|\ssX(Y)}
\leq\ell!\epsilon_k^{-\ell}e^{-(\tau/2)k}
\prod _Y e^{-c_0|Y|\ssX(Y)}.
\end{equation}
Bounding $\epsilon_k^{-\ell}e^{-(\tau/2)k}$
by $\epsilon_1^{-\ell} k^\ell e^{-k}\leq
(\ell e^{-1}\epsilon_1^{-1})^{\ell}$, we conclude that
\begin{equation}
\Bigl|\partial_z^\ell\prod_Y K_q'(Y,z')^{\ssX(Y)}\Bigr|\leq
\ell!(\ell e^{-1}\epsilon_1^{-1})^{\ell}
\prod _Y e^{-c_0|Y|\ssX(Y)}.
\end{equation}
Inserted into the cluster expansion for
$s_q^{(L)}$, this gives the bound
\begin{equation}
\left|\partial_z^\ell
s_q^{(L)}(z)
\right|\leq
\ell!(\ell e^{-1}\epsilon_1^{-1})^{\ell}
,
\end{equation}
which in turn implies that
\begin{equation}
\bigl|\partial_z^\ell
e^{s_q^{(L)}(z)}
\bigr|\leq
\ell!(\ell e^{-1}\epsilon_1^{-1})^{\ell}2^\ell
\bigl|e^{s_q^{(L)}(z)}\bigr|.
\end{equation}
Combining this bound with the bound \eqref{logthetader}, we obtain
the bound \eqref{logzetaL-der} with a constant $M_0$ that depends
only on $\epsilon$ and $\widetilde M_1$, and hence only on
$\epsilon$ and $M$.
\end{proofsect}

Next we will prove Theorem~\ref{thmB}. Recall the definitions of
the sets~$\scrS_\epsilon(m)$ and~$\scrU_\epsilon(\QQ)$
from
\eqref{2.7a} and
\eqref{2.7}
and the fact that in Theorem~\ref{thmB}, we set~$\kappa=\tau/4$.

\begin{proofsect}{Proof of Theorem~\ref{thmB}(1--3)}
Part~(1) is a trivial consequence of the fact that
$\theta_m(z)$, $\rho_z(\eusmN)$ and~$\rho_z(Y)$
are analytic functions of~$z$
throughout~$\tilde\scrO$.

In order to prove part~(2), we note that
$z\in\scrS_{\kappa/L}(q)$ implies that
$a_q(z)\le\kappa/L=\tau/(4L)$ and hence by
Theorem~\ref{thm4.2}(ii) we have that~$K_q'(Y,z)=K_q(Y,z)$
for
any~$q$-contour contributing
to~$\ZZ_q(\T_L,z)$.
This immediately implies that the functions
$s_q^{(L)}$ and $\zeta_q^{(L)}(z)$
defined in \eqref{sqLdef} and \eqref{zetaqLdef}
are analytic function in $\scrS_{\kappa/L}(q)$.
Next we observe that $\tau\geq 4\tilde c_0+16$ implies
that $\tau L/8\geq \tau/8\geq \log 4$ and hence
$4e^{-\tau L/4}\le e^{-\tau L/8}$.  Since
$z\in\scrS_{\kappa/L}(q)$ implies
$a_q(z)<\infty$ and hence $\theta_q(z)\neq 0$,
the bounds \twoeqref{zetamL}{der-zetamL} are then direct consequences of
Lemma~\ref{lemma4.5} and the fact that~$\partial\T_L=\emptyset$.

The bound \eqref{uplogderL} in part (3) finally is nothing but the bound
\eqref{logzetaL-der} from Lemma~\ref{lemCauchyDerBounds},
while he bound \eqref{nondegL} is proved exactly as for
Theorem~\ref{thmA}.
Note that so far, we only have used that $\tau\geq\tau_0$,
except for the proof of \eqref{uplogderL}, which through
the conditions from Lemma~\ref{lemCauchyDerBounds} requires
$L\geq \tau/2$, and give a constant $M_0$ depending on
$\epsilon$ and $M$.
\end{proofsect}

\begin{proofsect}{Proof of Theorem~\ref{thmB}(4)}
We will again rely on analyticity and Cauchy's Theorem.
Let~$\QQ\subset\RR$ and let~$\QQ'\subset\CalS$ be the set of
corresponding interchangeable spin states. Clearly, if~$m$ and~$n$
are interchangeable, then~$\zeta_m^{(L)}=\zeta_n^{(L)}$ and,
recalling that~$q_m$ denotes the set of spins corresponding
to~$m\in\RR$, we have
\begin{equation}
\varXi_\QQ(z)=Z_L^\per(z)
-\sum_{n\in\QQ'}\bigl[\zeta_n^{(L)}(z)\bigr]^{L^d}
=Z_L^\per(z)-\sum_{n\in\QQ'} Z_n'(\T_L,z).
\end{equation}
Pick a~$z_0\in\scrU_{\kappa/L}(\QQ)$.
For $n\in \QQ'$, we then have $a_n(z_0)\leq \tau/(4L)$,
and by the argument leading to
\eqref{aqz'k} we have that $a_n(z)\leq \tau/(2L)$
provided $\tau/(4L)\leq 1/2$ and
$2e\widetilde M_1|z-z_0|\leq \frac 15\frac\tau{4L}$.
On the other hand, if $m\in\CalS\setminus\QQ'$, then
$a_m(z_0)\geq \tau/(8L)$, and by a similar argument, we get
that $a_m(z)\geq \tau/(16L)$ if $\tau/(8L)\leq 1$ and
$2e\widetilde M_1|z-z_0|\leq \frac 1{10}\frac\tau{8L}$.
Noting that $\tau\geq\tilde\tau_0$ implies $\tau\geq 4c_0+16\geq 16$,
we now set
\begin{equation}
\epsilon^{(L)}=\min\{\epsilon, (10e\widetilde M_1L^d)^{-1}\}.
\end{equation}
For $z\in\D_{\epsilon^{(L)}}(z_0)$ and $n\in\QQ'$, we then have
$a_n(z)\frac L2\leq \tau/4$ and hence $Z_n'(\T_L,z)=Z_n(\T_L,z)$,
implying in particular that
\begin{equation}
\label{Xibd1}
\varXi_\QQ(z)=Z_L^{\text{\rm big}}(z)
+\sum_{m\in\CalS\smallsetminus\QQ'}Z_m(\T_L,z).
\end{equation}
Note that this implies, in particular, that $\varXi_\QQ(\cdot)$
is analytic in $\D_{\epsilon^{(L)}}(z_0)$.

Our next goal is to prove a suitable bound on the right hand side
of \eqref{Xibd1}. By Lemma~\ref{lemma5.1}, the first term
contributes no more than $2L^d\zeta(z)^{L^d}e^{-\tau L/4}$,
provided $\tau\geq 4\tilde c_0+16$ and $L$ is so large that
$5L^de^{-\tau L/4}\leq \log 2$. On the other hand,
since~$z\in\D_{\epsilon^{(L)}}(z_0)$ implies that
that~$a_m(z)\ge\tau/(16L)$ for all $m\not\in\QQ'$, the bound
\eqref{thm4.6.iii} implies that each $Z_m(\T_L,z)$ on the right
hand side of \eqref{Xibd1} contributes less than
$2\zeta(z)^{L^d}e^{-\tau L^{d-1}/32}$ once $L$ is so large that
$4L^de^{-\tau L/4}\leq \log 2$. By putting all of these bounds
together and using that
$\zeta(z)^{L^d}\leq\zeta(z_0)^{L^d}e^{\widetilde M_1|z-z_0|L^d}
\leq e^{1/(10e)}\zeta(z_0)^{L^d}$ by the bound \eqref{fbound} and
our definition of $\epsilon^{(L)}$, we get that
\begin{equation}
\label{Xibd2}
\left|\varXi_\QQ(z)\right|
\leq 5 |\CalS| L^d \zeta(z_0)^{L^d}e^{-\tau L^{d-1}/32}
\end{equation}
whenever
$z\in\D_{\epsilon^{(L)}}(z_0)$
and $L$ is so large that $L\geq \tau/2$ and $5L^de^{-\tau L/4}\leq
\log 2$.  Increasing $L$ if necessary to guarantee that
$\epsilon^{(L)}=(10e\widetilde M_1 L^{d})^{-1}$
and applying Cauchy's theorem to bound the derivatives of $\varXi_\QQ(z)$,
we thus get
\begin{equation}
\label{Xibd3}
\left|\partial_z^\ell\varXi_\QQ(z)\right|_{z=z_0}
\leq
\ell!(10e\widetilde M_1 )^{\ell}
5 |\CalS| L^{d(\ell+1)}
\zeta(z_0)^{L^d}e^{-\tau L^{d-1}/32}
\end{equation}
provided $L\geq L_0$, where $L_0=L_0(d,M,\tau,\epsilon)$ is chosen
in such a way that for $L\geq L_0$, we have $L\geq \tau/2$,
$5L^de^{-\tau L/4}\leq \log 2$ and $(10e\widetilde M_1
L^{d})^{-1}\leq\epsilon$.
Since $z_0\in\scrU_{\kappa/L}(\QQ)$ was arbitrary and
$|\CalS|=\sum_{m\in\RR}q_m$, this proves the desired bound
\eqref{error}
with
$C_0=10e\widetilde M_1 =10e(1+4M)$.
\end{proofsect}

\section*{Acknowledgements}
\noindent M.B.~would like to acknowledge the hospitality of
Microsoft Research in Redmond, where large parts of this paper
were written. The research of~R.K. was partly supported by the
grants GA\v{C}R~201/00/1149, 201/03/0478, and MSM~110000001. The
research of~M.B. was partly supported by the
grant~NSF~DMS-0306167.

\end{document}